\documentclass[useAMS,usenatbib]{mn2e}
\usepackage{epsfig}
\usepackage{graphicx}
\usepackage{times}
\usepackage[usenames,dvipsnames]{color}
\usepackage{amsmath}

\def\cm3{cm$^{-3}$}
\def\kms{km~s$^{-1}$}
\def\msunyr{M$_{\odot}$\,yr$^{-1}$}
\def\lsun{L$_{\odot}$}

\def\msun{M$_{\odot}$}

\def\one{\ts {\,\sc i}}
\def\two{\ts {\,\sc ii}}
\def\three{\ts {\,\sc iii}}

\def\six{\ts {\sc vi}}
\def\beq{\begin{equation}}
\def\eeq{\end{equation}}

\def\lesssim{\mathrel{\hbox{\rlap{\hbox{\lower4pt\hbox{$\sim$}}}\hbox{$<$}}}}
\def\gtrsim{\mathrel{\hbox{\rlap{\hbox{\lower4pt\hbox{$\sim$}}}\hbox{$>$}}}}

\def\lesssim{\mathrel{\hbox{\rlap{\hbox{\lower4pt\hbox{$\sim$}}}\hbox{$<$}}}}
\def\gtrsim{\mathrel{\hbox{\rlap{\hbox{\lower4pt\hbox{$\sim$}}}\hbox{$>$}}}}

\def\one{{\,\sc i}}
\def\two{{\,\sc ii}}
\def\three{{\,\sc iii}}

\def\six{{\sc vi}}

\def\xii{{\sc xii}}

\def\rphot{$R_{\rm phot}$}

\def\cmfgen{{\sc cmfgen}}
\def\heracles{{\sc heracles}}

\newcommand{\iso}[2]{\ensuremath{^{#1}\rm{#2}}}

\def\aj{AJ}

\def\apj{ApJ}

\def\apjl{ApJL}
\def\aap{A\&A}

\def\mnras{MNRAS}
\def\nat{Nature}

\def\jqsrt{JQSRT}

 \def\rphot{$R_{\rm phot}$}
 \def\tphot{$T_{\rm phot}$}
 \def\edphot{$N_{\rm e,phot}$}

 \def\rcds{$R_{\rm cds}$}
 \def\tcds{$T_{\rm cds}$}
 \def\edcds{$N_{\rm e,cds}$}

 \def\rbb{$R_{\rm bb}$}
 \def\tbb{$T_{\rm bb}$}
 \def\lbb{$L_{\rm bb}$}

\def\foe{10$^{51}$\,erg}

\title[Super-luminous Type IIn SNe]{Numerical simulations of super-luminous supernovae of type IIn}

\author[Luc Dessart, Edouard Audit, and D.J. Hillier]
{Luc Dessart,$^{1}$ Edouard Audit,$^{2}$ and D. John Hillier$^{3}$\\ \\
$^{1}$: Laboratoire Lagrange, UMR7293, Universit\'e Nice Sophia-Antipolis, CNRS,
Observatoire de la C\^{o}te d'Azur, 06300 Nice, France. \\
$^{2}$: CEA, Maison de la Simulation, USR 3441, CEA-CNRS-INRIA- Univ. Paris-Sud -
Univ. de Versailles, 91191, Gif-sur-Yvette Cedex, France. \\
$^{3}$:
Department of Physics and Astronomy \& Pittsburgh Particle Physics, Astrophysics,
and Cosmology Center (PITT PACC), \\
University of Pittsburgh, 3941 O'Hara Street, Pittsburgh, PA 15260, USA.\\
}

\begin{document}

\date{Accepted 2015 March 18. Received 2015 March 09; in original form 2015 January 06}

\pagerange{\pageref{firstpage}--\pageref{lastpage}} \pubyear{2011}

\maketitle
\label{firstpage}

\begin{abstract}
We present numerical simulations that include 1-D Eulerian multi-group radiation-hydrodynamics,
1-D non-Local-Thermodynamic-Equilibrium (non-LTE) radiative transfer,
and 2-D polarised radiative transfer for super-luminous interacting supernovae (SNe).
Our reference model is a $\sim$10\,\msun\ inner shell with \foe\ ramming into a
$\sim$3\,\msun\ cold outer shell (the circumstellar-medium, or CSM)
that extends from 10$^{15}$ to 2$\times$10$^{16}$\,cm and  moves at 100\,\kms.
We discuss the light curve evolution, which cannot be captured adequately with a grey approach.
In these interactions, the shock-crossing time through the optically-thick CSM is much
longer than the photon diffusion time. Radiation is thus continuously leaking from the shock
through the CSM, in disagreement with the shell-shocked model that is often invoked.
Our spectra redden with time, with a peak distribution in the near-UV during the first month
gradually shifting to the optical range over the following year.
Initially Balmer lines exhibit a narrow line core and the broad line wings
that are characteristic of electron scattering in the SNe IIn atmospheres (CSM). At later
times they also exhibit a broad blue shifted component which arises from the cold dense shell.
Our model results are broadly consistent with the bolometric light curve and
spectral evolution observed for SN\,2010jl.
Invoking a prolate pole-to-equator density ratio in the CSM, we can also reproduce the
$\sim$2\% continuum polarisation, and line depolarisation, observed in SN\,2010jl.
By varying the inner shell kinetic energy and the mass and extent of the outer shell,
a large range of peak luminosities and durations, broadly
compatible with super-luminous SNe IIn like 2010jl or 2006gy, can be produced.
\end{abstract}


\section{Introduction}

Since the original identification of interacting supernovae (SNe; \citealt{dopita_etal_84}; \citealt{niemela_etal_85})
and the creation of the SN IIn class \citep{schlegel_90}, the sample of such peculiar Type II SNe has grown from
a few to a few tens of events. It has also revealed an intriguing diversity, spanning a range of luminosities, duration,
line profile morphology, with events like SNe 1988Z
\citep{stathakis_sadler_91,turatto_etal_93}, 1994W \citep{sollerman_etal_98,chugai_etal_04}, 2006gy \citep{smith_etal_07a},
and the enigmatic SN\,2009ip \citep{mauerhan_etal_13,pastorello_etal_13,margutti_etal_14}.
The origin of Type IIn SNe is associated with the interaction of fast material with
previously-expelled slowly-expanding material \citep{grasberg_nadezhin_86}.
The resulting shock leads to deceleration of the inner shell, acceleration of the outer shell,
and conversion of kinetic energy into internal and radiative energy.
Because (standard) non-interacting SNe radiate merely 1\% of their total energy (the rest being kinetic),
interacting SNe can easily reach extraordinary luminosities by tapping this abundant reservoir of kinetic energy.
For example, an interaction extracting 30\% of a \foe\ ejecta could produce a SN IIn with a luminosity
of 10$^{10}$\,\lsun\ for 3 months.

Super luminous interacting SNe require both a high kinetic energy in the inner shell
and a large mass reservoir in the outer shell (see, e.g., \citealt{moriya_etal_13a}).
Two good examples of such SNe IIn are SN\,2006gy \citep{smith_etal_07a} and more recently
SN\,2010jl \citep{stoll_etal_11}. The latter has extensive photometric and spectroscopic observations 
\citep{stoll_etal_11,zhang_etal_12,ofek_etal_14,fransson_etal_14}
and also has spectropolarimetric data \citep{patat_etal_11,williams_etal_14}.
How stars produce the outer shell is unclear -- it may
be  material ejected by the pair-production instability \citep{barkat_etal_67,woosley_etal_07}
or it may be material ejected via a super-Eddington wind \citep{owocki_etal_04}.
Here, we simply consider the interaction
between a fast moving inner shell and a slowly-moving massive outer shell, and
leave aside speculations concerning their origin.

    Numerical simulations of interacting SNe are generally undertaken with radiation hydrodynamics
    \citep{chugai_etal_04,woosley_etal_07,moriya_etal_13a,whalen_etal_13},
    hydrodynamics combined with a parameterised cooling function \citep{van_marle_etal_10,chen_etal_14},
    or pure hydrodynamics (by assuming adiabaticity; see, e.g., \citealt{blondin_etal_96}).
    When studying super-luminous SNe IIn, radiation hydrodynamics with a non-grey (multi-group)
    treatment is a must because the radiation contribution is a sizeable fraction of the pressure/energy
    in the problem, the characteristic temperatures of the radiation and of the gas may differ strongly,
    and the emerging radiation provides essential signatures characterising the event.

    The radiative transfer modelling for SNe IIn comes at present in two flavours.
    It is either based on the results of radiation hydrodynamics simulations
    but limited to a single line, e.g., H$\alpha$ \citep{chugai_etal_04}, or
    based on an ad-hoc atmospheric structure and the assumption of steady-state
    radiation through a diffusive optically thick inner boundary \citep{dessart_etal_09}.

   In this work, we discuss numerical simulations for super-luminous SNe of type IIn,
   including studies with radiation hydrodynamics, radiative transfer, and polarised radiative transfer codes.
   In Section~\ref{sect_her}, we first present the multi-group 1-D Eulerian radiation hydrodynamics
   simulations we perform with \heracles, including the numerical approach, the standard interaction
configuration we adopt as initial conditions, and the results for both the dynamics and the radiation.
We then present in Section~\ref{sect_cmfgen} non-LTE radiative-transfer simulations with \cmfgen, which we compute
for a number of snapshots using the results from the radiation-hydrodynamical modelling.
We discuss line profile broadening, and the relative roles of expansion and electron scattering.
In Section~\ref{sect_pol}, we present polarisation calculations based on these \cmfgen\ simulations but
artificially distorted along an axis of symmetry, in order to produce prolate density configurations.
We assess the level of asymmetry needed to explain the observed continuum polarisation
(and line depolarisation) reported for SN\,2010jl.
In Section~\ref{sect_dep}, we discuss the dependency of our results on the interaction properties,
varying the kinetic energy of the inner shell and the mass/extent of the circumstellar medium (CSM).
Finally, in Section~\ref{sect_conc}, we present our conclusions.

\begin{figure*}
\epsfig{file=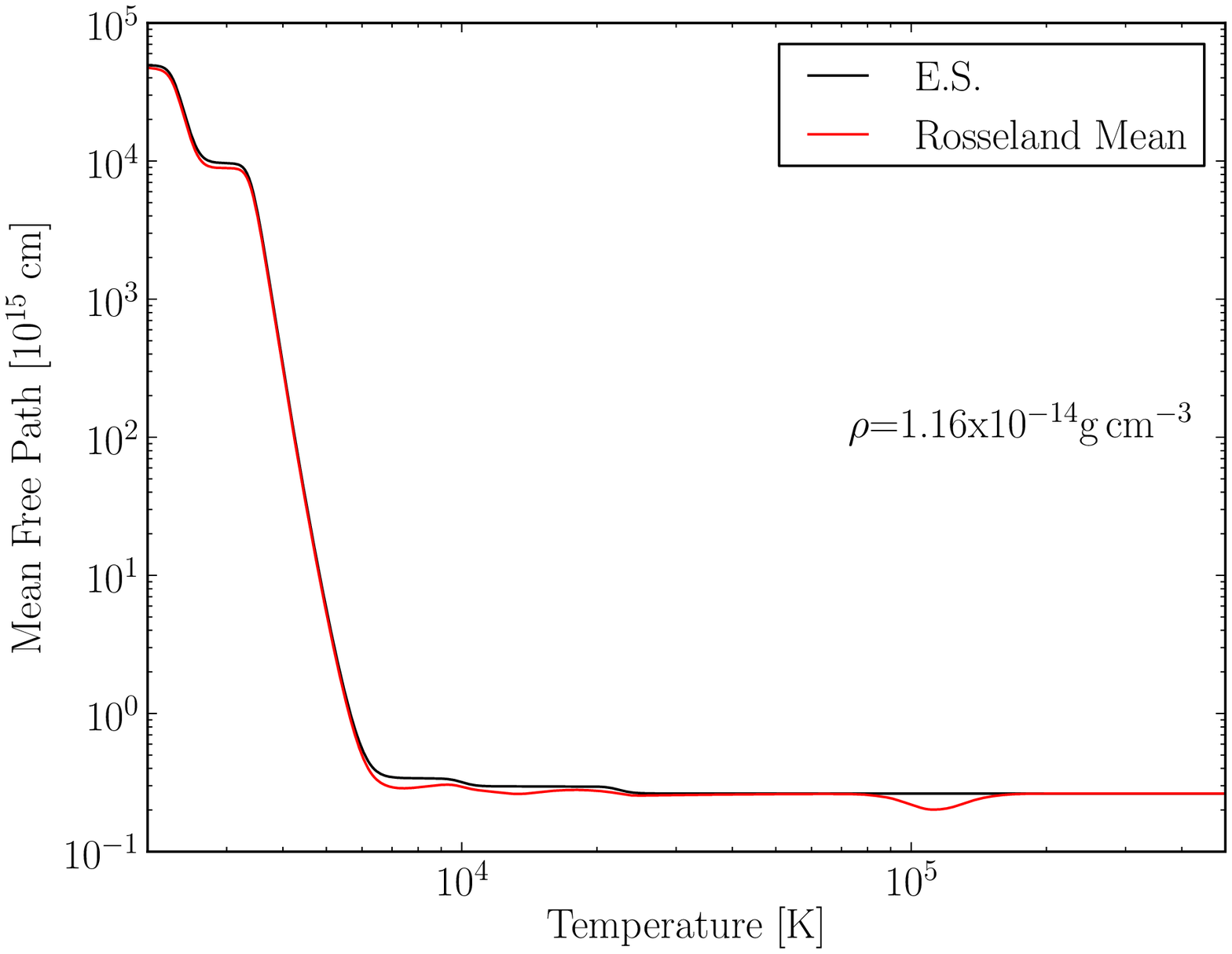,width=8.5cm}
\epsfig{file=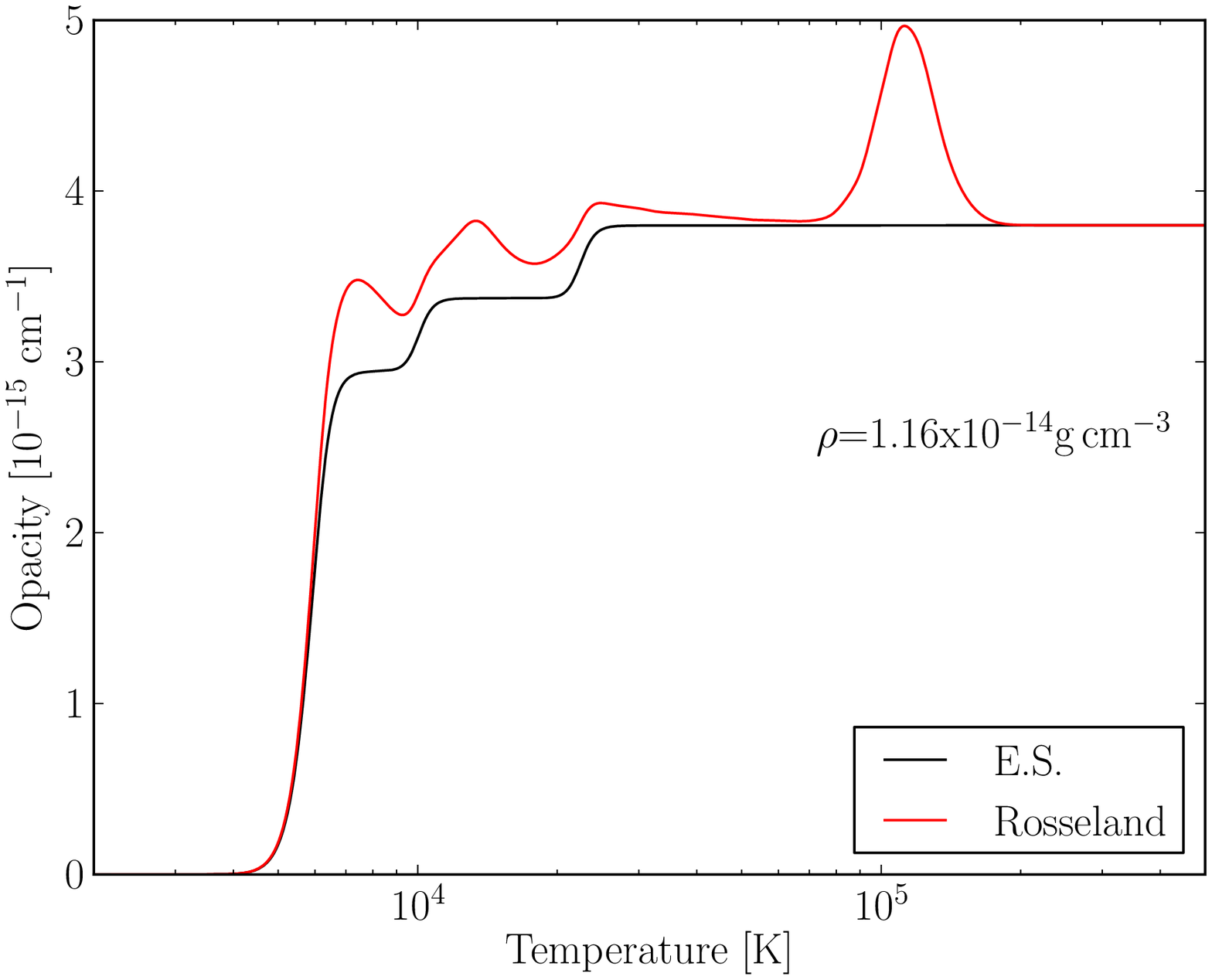,width=8.5cm}
\epsfig{file=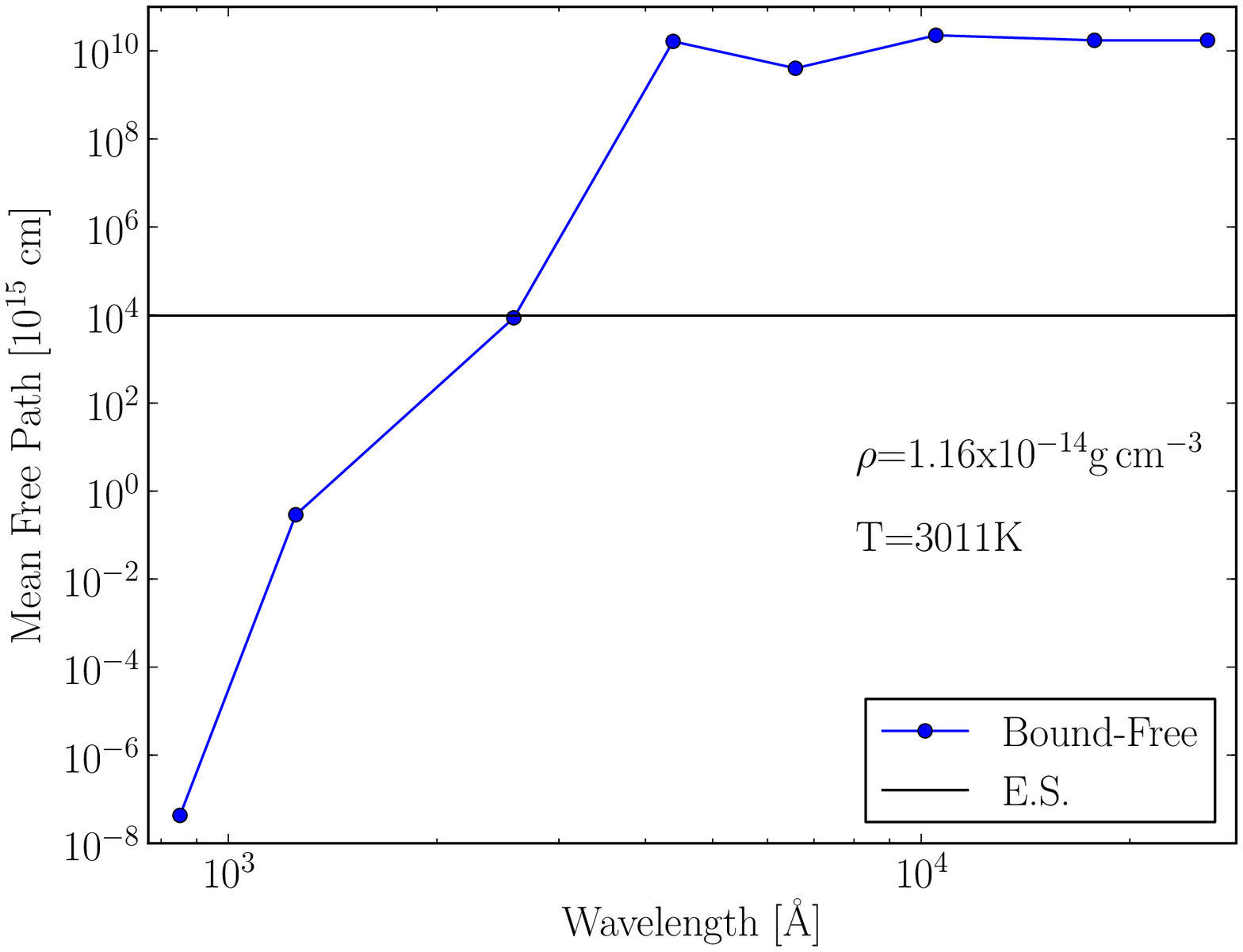,width=8.5cm}
\epsfig{file=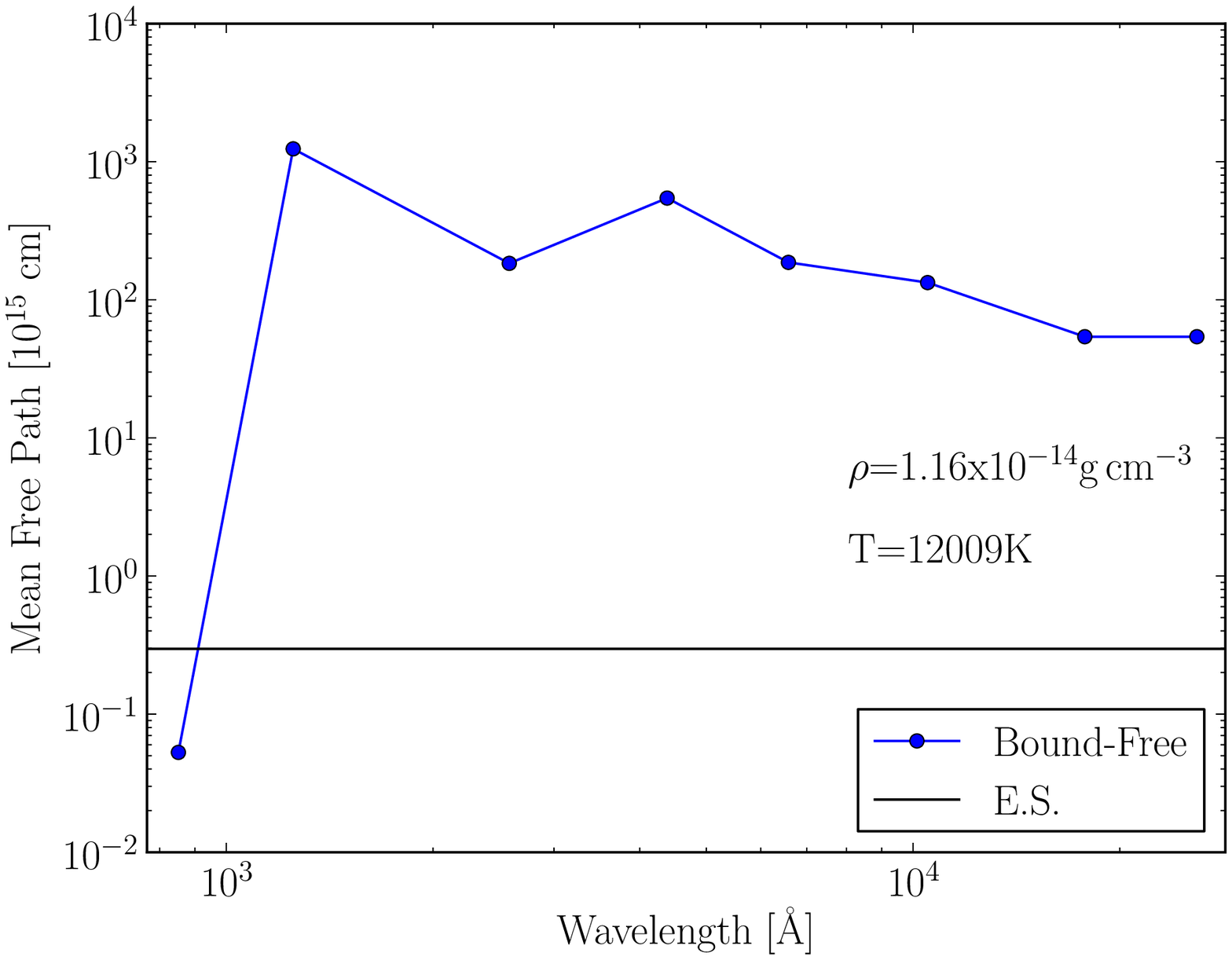,width=8.5cm}
\caption{
{\it Top row:} Illustration of the variation of the photon mean free path (left) and inverse mean
free path (right) at a mass density of 1.16$\times$10$^{-14}$\,g\,cm$^{-3}$.
The electron-scattering opacity (black) is the main
contribution to the Rosseland-mean opacity (red; which accounts for bound-bound and bound-free processes),
except in the regions of partial ionisation of H and He at 8000-20000\,K and in the iron-opacity bump in the
range 100000-200000\,K.
The precise temperature regions where these offsets occur depend on density.
The most critical variation in the opacity is associated with the recombination of H and He, leading to a
very small opacity at low temperature.
{\it Bottom row:} Illustration of the wavelength dependence of the photon mean free path associated
with absorptive processes (blue) at
$\sim$\,3000\,K (left) and $\sim$\,12000\,K (right), corresponding to conditions in which hydrogen is
neutral and ionised, respectively. The filled dots correspond to the central wavelength of each energy group
used for both the opacity table and the \heracles\ simulations.
\label{fig_chi}
}
\end{figure*}

\section{Radiation hydrodynamics with \heracles}
\label{sect_her}

\subsection{Numerical approach}
\label{sect_her_setup}

  \heracles\ is a Eulerian multi-dimensional radiation-hydrodynamics code \citep{gonzalez_etal_07},
  with the possibility for multi-group radiation transport \citep{vaytet_etal_11}.
The hydrodynamics is treated using a standard second order Godunov scheme. For the radiation transfer,
conduction, flux limited diffusion (FLD) and the M1 moment model \citep{m1_model} are implemented
in \heracles, the latter two with
the possibility of using a multi-group approach. Apart from a comparison test that uses a grey treatment,
the simulations presented in this paper are all done using the
multi-group M1 moment model. As explained below, a multi-group treatment is necessary in order to take into
account the large opacity variations with frequency and the strong imbalance between radiation and gas
temperature. The M1 model has also some strong benefit compared to the more standard FLD approach.
First of all the M1 model is exact in both the diffusion and the free-streaming limit and is rather accurate in
between. It is therefore well suited to model regions with low or intermediate optical depth. Furthermore,
the M1 method solves explicitly for both the energy and the flux through moments of the radiative transfer equations.
The M1 model can distinguish between absorption/emission and scattering opacities. The effect of scattering
in regions where the scattering opacity is dominant can then be taken into account. Finally, the radiative
transfer is formulated in the co-moving frame and as is shown in \citet{vaytet_etal_11} the Doppler and
aberrations effects are properly accounted for, which is important when large velocity gradients are present.

We use the multi-group solver because grey radiation transport, often
used for simulations of non-interacting Type II SNe, is inadequate in SNe IIn.
In non-interacting Type II SNe, the radiation and the gas
have similar temperatures everywhere (even identical at large optical depths), so that adopting
the Rosseland mean opacity, for example, is not a bad representation of  the effective opacity  of the material.
In contrast, the radiation produced from the interaction in SNe IIn has a characteristic temperature that can
vastly differ from the medium that it traverses, and the total continuum optical depth is not much in excess
of about ten. In the next section, we illustrate the impact on the light curve produced by assuming a
grey rather than an energy-dependent LTE opacity for the material.

In this work, we adopt a uniform composition, with a H mass fraction of 0.633,
He mass fraction of 0.36564, and an iron mass fraction of 0.00136 to reflect approximately
the near-solar composition of a blue or a red supergiant (hereafter BSG/RSG)
star (for SN\,2010jl, \citealt{stoll_etal_11})
argue for a sub-solar metallicity, which may be important for understanding how these
SNe IIn come about, but is irrelevant for the present radiation-hydrodynamics considerations).

\begin{figure*}
\epsfig{file=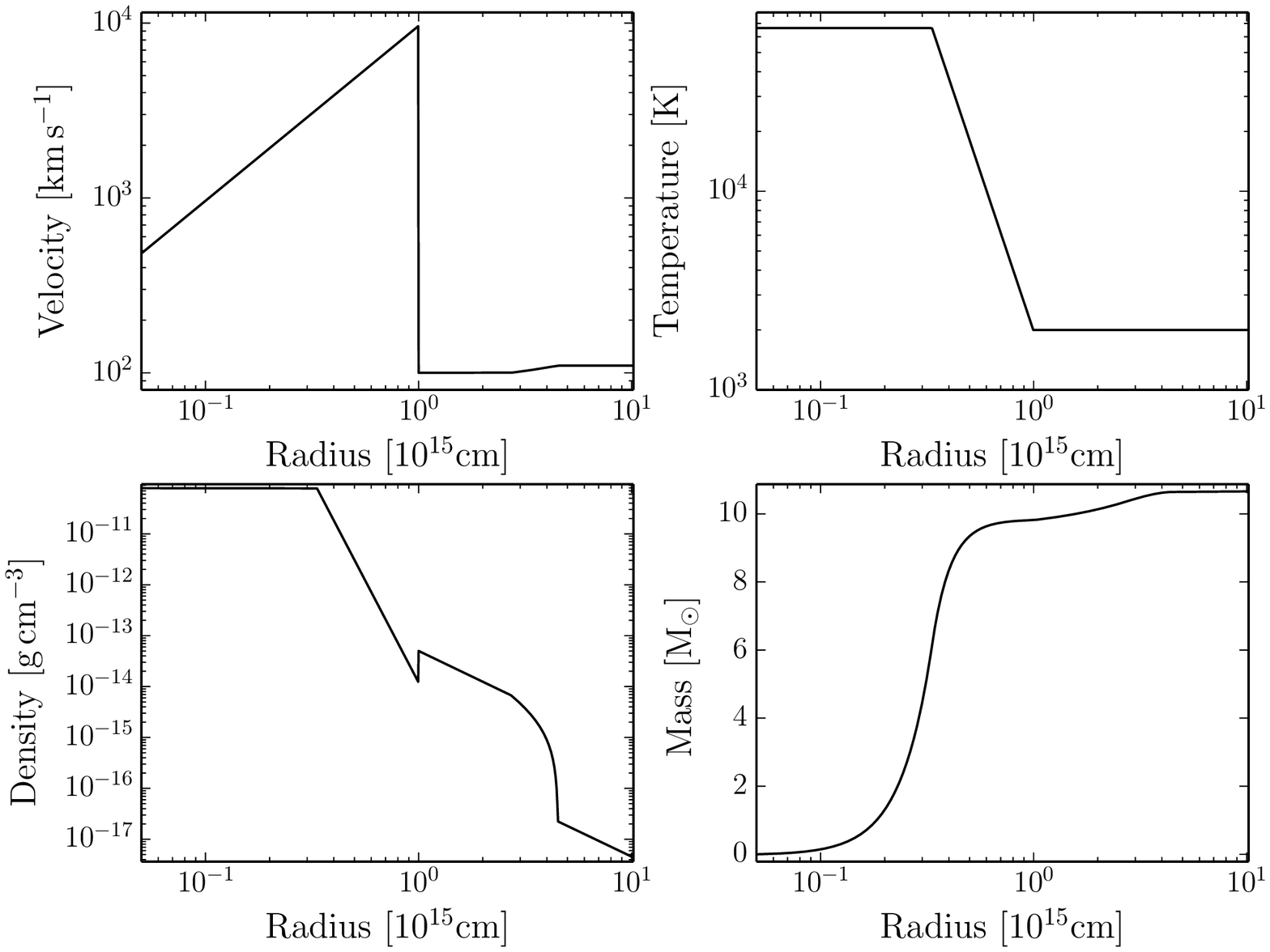,width=15cm}
\caption{Initial configuration for the reference interaction model X simulated with \heracles.
We show radial profiles of the velocity, the temperature, the mass density, and the Lagrangian mass.
Global quantities and additional details for this model are given in Table~\ref{tab_mod_sample}
and Section~\ref{sect_her_setup}.
\label{fig_init}
}
\end{figure*}

We supply the code with an opacity table for our adopted composition.
We compute our opacities as a function of density, temperature and energy group.
Energy groups are positioned at strategic locations to capture the strong variation
in absorptive opacity with wavelength. We use one group for the entire Lyman continuum
(including the X-ray range), two groups for the Balmer continuum, two for the Paschen continuum,
and three groups for the Brackett continuum and beyond.\footnote{We tried using several
energy groups to cover the X-ray and the EUV ranges better but this caused numerical problems
with the radiative transfer. The solver struggled converging when an exceedingly small energy/flux
was present in those energy groups, a situation encountered outside of the shocked region.
In practice, this should not impact our results sizeably because the optical depth to hard radiation
(all Lyman photons) is huge at all times considered here.}
Our opacity code computes the LTE level populations and ionisation state of the gas, and
then uses the atomic data available within \cmfgen\ to compute LTE opacities
and to make the opacity table.
The domain covered is 10$^{-20}$ to 10$^{-5}$\,g\,cm$^{-3}$ in 200 bins, and from 2000
to 500,000\,K in 500 bins --- we use the nearest edge for requests outside the table boundaries.
Here, we consider the contribution from electron scattering
and bound-free opacity for H\one, He\one-\two, and Fe\one-\xii.
A major limitation of our work is that we ignore line opacity and emissivity
in our \heracles\ simulations.  Metal line blanketing (in particular from Fe) would
enhance the opacity in the UV (inhibiting further the escape of UV photons).
We also neglect cooling due to CNO elements, which requires a non-LTE study with specific allowance
for enhanced photon escape due to the velocity field. Inclusion of these effects is left to a future study.

In Fig.~\ref{fig_chi}, we plot the photon mean free path and its inverse (i.e.,
the opacity) against wavelength for  different processes at different densities and temperatures.
In super-luminous SNe IIn the spectrum formation region is typically
optically thick in the Lyman continuum (and at X-ray wavelengths) and high-energy photons will
tend to thermalise, even at low electron-scattering optical depth.
Longward of the Lyman continuum, electron scattering opacity typically dominates
over absorptive opacities and so low-energy photons will thermalise inefficiently at
most times and locations.

The simulations in this work use
an ideal gas equation of state with $\gamma=5/3$.\footnote{We therefore ignore
the variations in thermodynamic quantities that arise from variations in ionisation.}
We also adopt a mean atomic weight of 1.38 in all simulations. In practice, this value
depends on the ionisation state of the gas, but in ways that need to be determined by using
a general equation of state --- this is left to a forthcoming study.

Our simulations are 1-D and use a uniform radial grid with 1600 points,
covering from 0.05 to 2$\times$\,10$^{16}$\,cm.
Degrading the resolution by a factor of four produces the same overall properties (e.g.,
the bolometric maximum is changed by $\sim$\,10\%), although
the shock is less resolved and more numerical diffusion occurs as the interaction region crosses grid zones
in the course of the simulation. This has a visible impact on the properties of the forward and reverse shocks
and the associated temperature jumps, but it does not influence significantly the energy solved for by \heracles.
Very high resolution simulations that yielded better resolution of the shock structure gave, for example, similar
results for the properties of the CSM, the shock propagation speed, and the emergent luminosity.

The initial configuration for the interaction is determined analytically.
We prescribe a density and a temperature structure for the inner shell (tagged as the
SN ejecta, but the results apply to any ejected shell with similar properties) and the outer shell
(we assume the CSM arises from pre-SN mass loss in the form of a wind).
We adopt a structure for the SN ejecta which is based on the simulations
of core-collapse SNe we perform \citep{DLW10a,DLW10b}, in practice comparable to the
formulation of \citet{chevalier_irwin_11} and \citet{moriya_etal_13a}.
In the present paper, we are only interested in studying the basic properties of super-luminous
SNe IIn and thus focus on one event (i.e., SN\,2010jl), for which we are guided by the parameters
of \citet{fransson_etal_14}, who argue for a SN explosion leading to interaction with a dense and
extended CSM.
We do not include any unstable nuclei in the simulation and thus ignore any contribution
from radioactive decay.

For our standard interaction model, named X,
we take a 9.8\,\msun\ inner ejecta (in homologous expansion) with \foe. Its density structure is given by a power law in radius
with exponent $N_{\rho}=$\,8 outside of $V_0\sim$\,3000\,\kms\ and constant within it. Its temperature
structure is given by a power law in density with exponent $N_{\rm T}=$\,0.4, rising from 2000\,K at the ejecta/CSM
interface radius located at $R_t$ (we enforce a maximum temperature of 65000\,K in the inner ejecta).
The SN ejecta is 11.6\,d old when the interaction starts.
For the outer shell, which starts at $R_t=$\,10$^{15}$\,cm, we adopt a wind
structure with a constant velocity of 100\,\kms\ (and constant temperature of 2000\,K),
but split that space into two regions of distinct density. Below 10$^{16}$\,cm,
we prescribe a mass loss rate of 0.1\,\msunyr, and beyond that radius we use
a mass loss rate of 10$^{-3}$\,\msunyr\ (i.e., two orders of magnitude smaller).
The motivation for this is three fold.
First, the CSM mass should not be unrealistically large. By truncating the high mass loss region,
we can  control the total CSM mass for any chosen mass loss. Second, the phase of high mass loss in massive stars
like $\eta$ Car lasts for a period of the order of ten years, not for centuries. With our choice of outer radius
and wind speed, the high mass loss phase has a duration of $\sim$\,28\,yr.
Third, the transition to a lower density outer region will facilitate a reduction in the
luminosity at later times, as observed in SN\,2010jl  past 300\,d (see below, and \citealt{fransson_etal_14}).
For convenience, we interpolate the density at the junction between the two shells
in order to smooth the profile and avoid an overly abrupt change in luminosity when the
shock reaches this region.
The inner edge of the outer shell has a kinematic age of 3.2\,yr.
The CSM has a total mass of 2.89\,\msun\ and a total kinetic energy of  5.2$\times$10$^{47}$\,erg.

\noindent
To summarise, for the SN ejecta ($R<R_t$), we adopt:
\begin{align*}
V(R) &= (R/ R_t) V_{\rm max}\, ,\\
V_0 &= V_{\rm max} / 3 \, ,\\
\rho(V) &= \rho_0 (V_0/V)^{N_{\rho}} \, {\rm for} \, V > V_0  \, ,\\
\rho(V) &= \rho_0 \, {\rm for} \, V < V_0 \, {\rm and} \\
T(V) &= 2000 \left( \rho(V)/\rho(V_{\rm max}) \right)^{N_T}\,{\rm K}\,.
\end{align*}
The proportionality constants $\rho_0$ and $V_{\rm max}$ are adjusted so that the total SN ejecta
mass and kinetic energy match the desired values (see Table~\ref{tab_mod_sample}).
For the CSM ($R>R_t$), we adopt:
\begin{align*}
\rho(R) &= \dot{M} / 4\pi R^2 V \, , \\
 V &= 100\, \hbox{\kms}\ {\rm and} \, , \\
  T &= 2000\,K
\end{align*}
where $ \dot{M} = \dot{M}_{\rm CSM, in}$ for $R<$\,10$^{16}$\,cm and $ \dot{M} = \dot{M}_{\rm CSM, out}$ for $R>$\,10$^{16}$\,cm.

The initial conditions for model X are shown in Fig.~\ref{fig_init}.
To explore some dependencies of our results, we vary these ejecta/CSM conditions
and discuss the implications in Section~\ref{sect_dep} (see also Table~\ref{tab_mod_sample}).

\begin{table*}
\caption{Summary of simulations performed in this work. The reference model X is discussed in detail
through the most part of the paper (Section~\ref{sect_her}), while additional simulations (Xe3 etc.) are discussed in Section~\ref{sect_dep}.
For all simulations, the minimum and maximum radii of the Eulerian grid are 5$\times$10$^{13}$ and 2$\times$10$^{16}$\,cm,
and the transition radius between the SN ejecta and the CSM lies at a radius $R_t$ of 10$^{15}$\,cm.
In all cases, the mass of the inner shell (the SN ejecta) is 9.8\,\msun, and the CSM wind velocity is 100\,\kms.
Results from the \heracles\ simulation for each model are given in the last three columns. The time to peak
is the time to reach maximum from the time when the rising bolometric luminosity is only 1\% of the value at peak
(this way, we cancel the light travel time to the outer boundary, where we record the flux).
We finally add results from Section~\ref{sect_cmfgen} for the bolometric correction and colour at peak.
Numbers in parenthesis are powers of ten.
\label{tab_mod_sample}}
\begin{tabular}{l@{\hspace{1mm}}c@{\hspace{2mm}}c@{\hspace{2mm}}c@{\hspace{2mm}}c@{\hspace{2mm}}c@{\hspace{2mm}}
c@{\hspace{2mm}}c@{\hspace{2mm}}c@{\hspace{2mm}}c@{\hspace{2mm}}c@{\hspace{2mm}}c@{\hspace{2mm}}c@{\hspace{2mm}}c@{\hspace{2mm}}}
\hline
model        &$E_{\rm kin, SN}$ &  $V_{\rm max, SN}$   & $E_{\rm kin, CSM}$ &  $M_{\rm CSM}$ & $\dot{\rm M}_{\rm CSM,in}$ &
$\dot{\rm M}_{\rm CSM,out}$ &     $L_{\rm bol, peak}$  &  B.C.$_{\rm @peak}$ & ($V-I$)$_{\rm @peak}$ & $t_{\rm peak}$  & $\int L dt$  \\
             & [\foe]       & [\kms]     &  [\foe] & [\msun] & \multicolumn{2}{c}{[\msunyr]}  &    [erg\,s$^{-1}$] &  [mag] & [mag] & [d]    & [\foe] \\
\hline
X            &    1   &       9608     & 5.17(-4)  &     2.89   &   0.1    &  0.001  & 3.024(43)  &   -1.06 &   0.15 & 19.4 &  0.32    \\
Xe3          &    3   &      16642     & 9.70(-4)  &     2.89   &   0.1    &  0.001  & 1.204(44)  &   -1.35 &   0.11 & 15.7 &  0.88    \\
Xe3m6        &    3   &      16642     & 5.15(-3)  &    17.31   &   0.6    &  0.006  & 2.080(44)  &   -1.39 &   0.01 & 55.7 &  2.05    \\
Xe3m6r       &    3   &      16642     & 6.08(-3)  &    26.73   &   0.6    &  0.006  & 1.818(44)  &   -1.05 &   0.06 & 68.3 &  2.13    \\
Xe10         &    10  &      30384     & 2.55(-3)  &     2.89   &   0.1    &  0.001  & 6.399(44)  &   -1.46 &   0.13 & 12.7 &  2.92    \\
Xe10m6       &    10  &      30384     & 1.31(-2)  &    17.31   &   0.6    &  0.006  & 1.091(45)  &   -1.80 &  -0.04 & 34.2 &  6.89    \\
Xm3          &    1   &       9608     & 1.46(-3)  &     8.66   &   0.3    &  0.003  & 3.906(43)  &   -0.84 &   0.13 & 47.9 &  0.49    \\
Xm6          &    1   &       9608     & 2.87(-3)  &    17.31   &   0.6    &  0.006  & 4.751(43)  &   -0.80 &   0.27 & 77.5 &  0.63    \\
\hline
\end{tabular}
\end{table*}

As the code is Eulerian, the flow constantly leaves the grid at the outer boundary.
Conditions are flow out, for both the gas and the radiation.
At the inner boundary, we use a reflecting condition for the radiation (zero flux).
For the gas, we use an inflow condition dynamically consistent with the properties of the inner shell.
% Ideally, we should inject vacuum, but this is not practical.
We  assign to this injected material a low temperature of 2000\,K, a velocity that
preserves the homologous expansion of the inner shell
material ($V \propto R$), and a density $\rho_{\rm ib} = \rho_1 x^3$, with $x = t/ (t+dt)$, and where
$\rho_1$ is the density in the first active zone,  $t = R_{\rm ib}/V_{\rm ib}$, and $dt$ is the time step.
 The injected material has no dynamical or radiative influence on the  physics of the interaction and our results.
 We continue the simulations until the interaction region moves out of the grid, until
the luminosity becomes about 100 times smaller than at maximum, or when 2 years
have passed, whatever comes first.

\begin{figure}
\epsfig{file=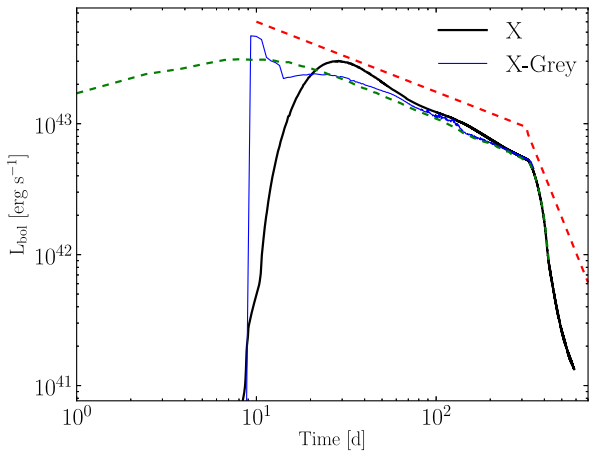,width=8.5cm}
\caption{Bolometric light curve evolution for the SN IIn model X (multi-group; black)
and its grey counterpart (blue).
We add the bolometric luminosity at the shock (green dashed curve), which differs at times $\lesssim$200\,d
from the recorded bolometric luminosity because of optical depth and light travel time effects
(there is an 8\,d light travel time to the outer boundary from the initial location of the interaction region
and a $\sim$\,35\,d diffusion time through the CSM once it has become ionised).
For comparison, we show the analytical model light curve (red dashed curve) of \citet{fransson_etal_14} for SN\,2010jl.
\label{fig_lbol}
}
\end{figure}

\begin{figure*}
\epsfig{file=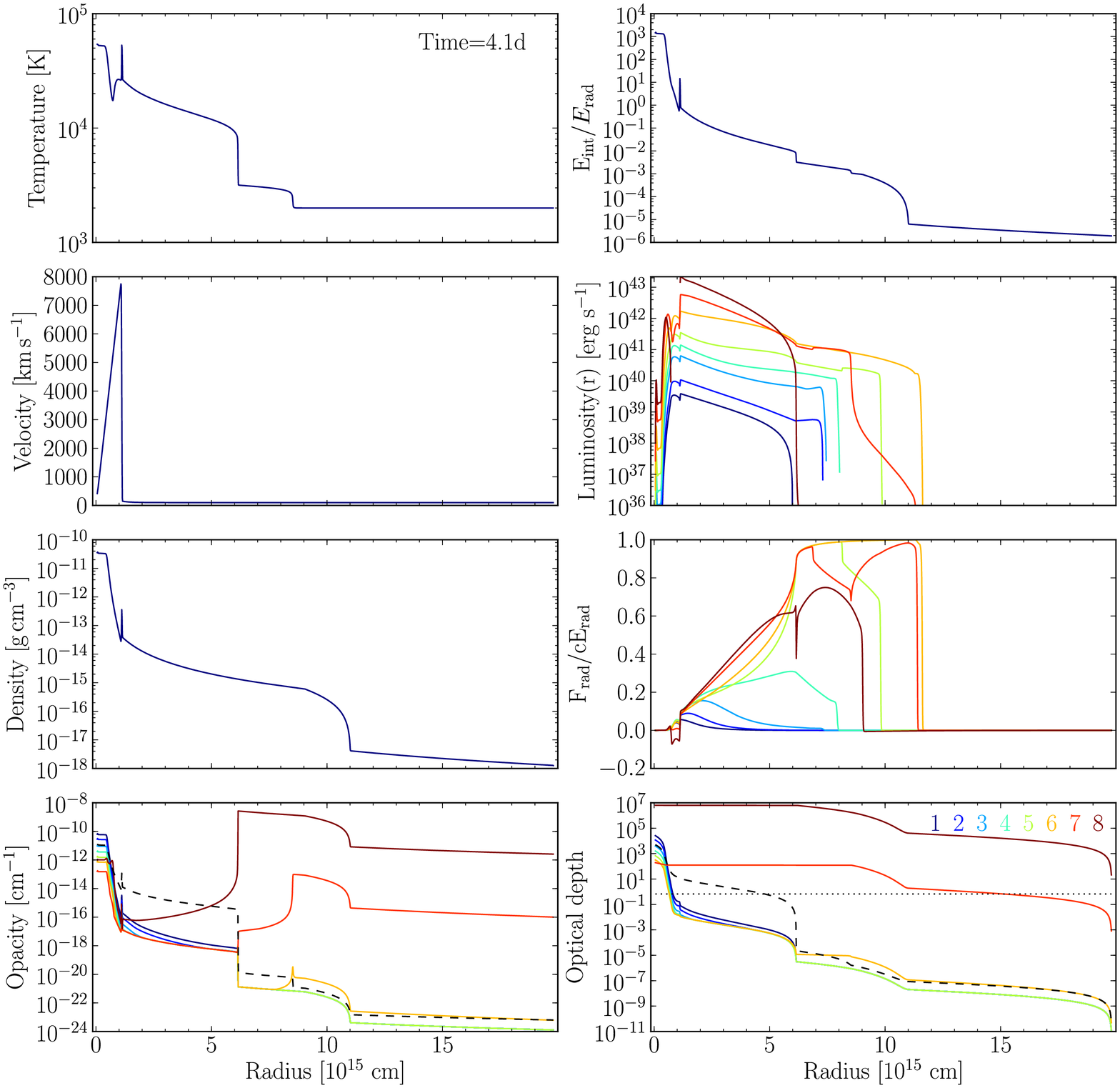,width=17cm}
\caption{
Radial slices through the grid showing the temperature, the ratio of the internal (i.e., gas)
energy and the radiation energy (i.e., the mean intensity), the velocity, the luminosity, the
mass density, the reduced flux, the opacity (inverse mean free path) for
each group (coloured curves) and for electron scattering (dashed black curve) together with
the associated optical depth (integrated inwards from the outer boundary; the dotted black line corresponds to $\tau=2/3$).
The coloured curves correspond to each one
of the eight energy groups (ordered from low to high energy/frequency;
see labels in bottom right panel and discussion in Section~\ref{sect_her_setup}).
We use 3 groups longward of the Paschen jump, two groups in the Paschen
continuum, two groups in the Balmer continuum, and one group for the Lyman continuum.
The time is 4.1\,d after the onset of the interaction. The electron scattering contribution
is not included in these groups.
\label{fig_hydro1}
}
\end{figure*}

\begin{figure*}
\epsfig{file=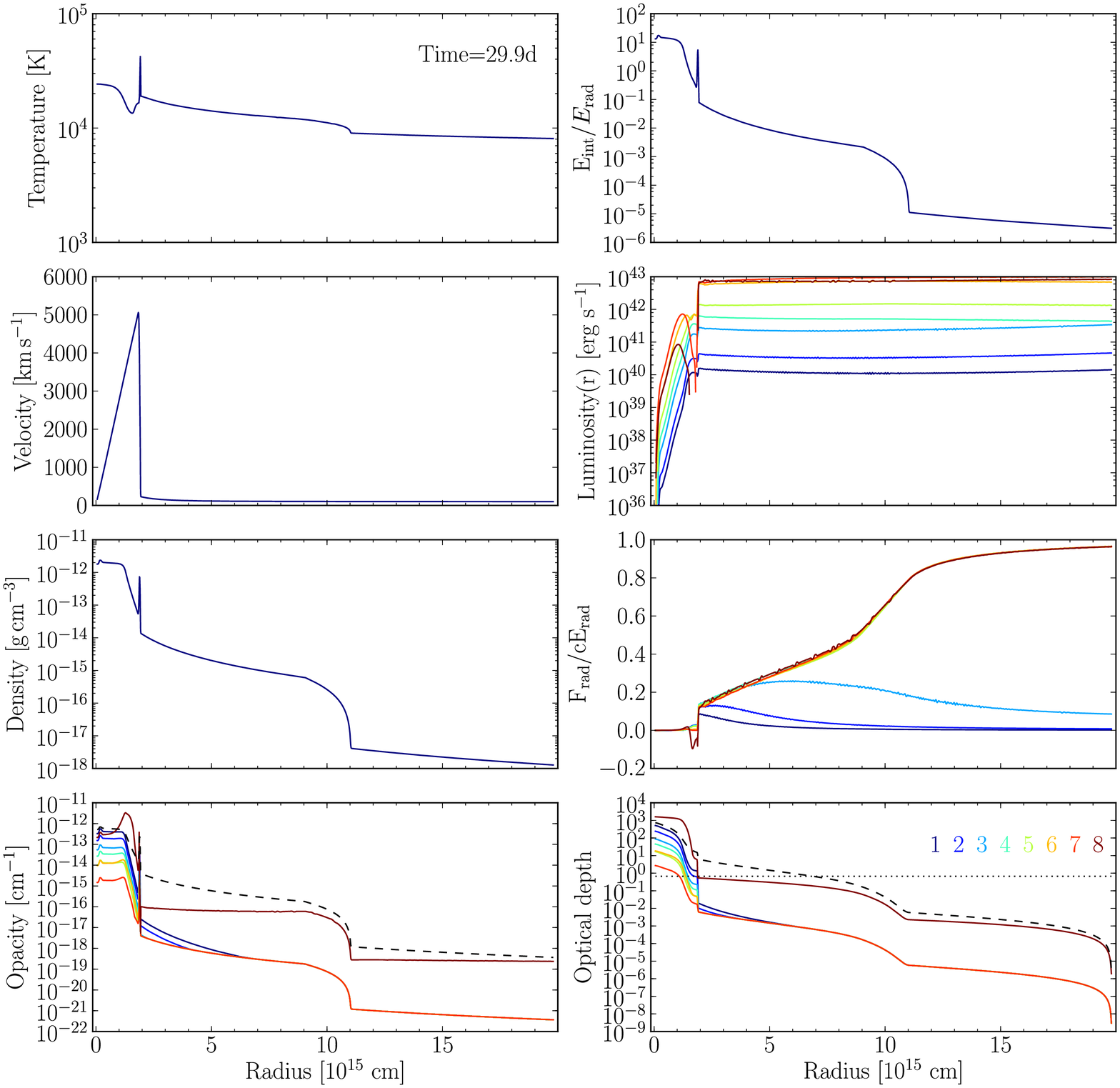,width=17cm}
\caption{Same as Fig.~\ref{fig_hydro1}, but now for a time of 29.9\,d after the onset of the interaction.
\label{fig_hydro2}
}
\end{figure*}

\begin{figure*}
\epsfig{file=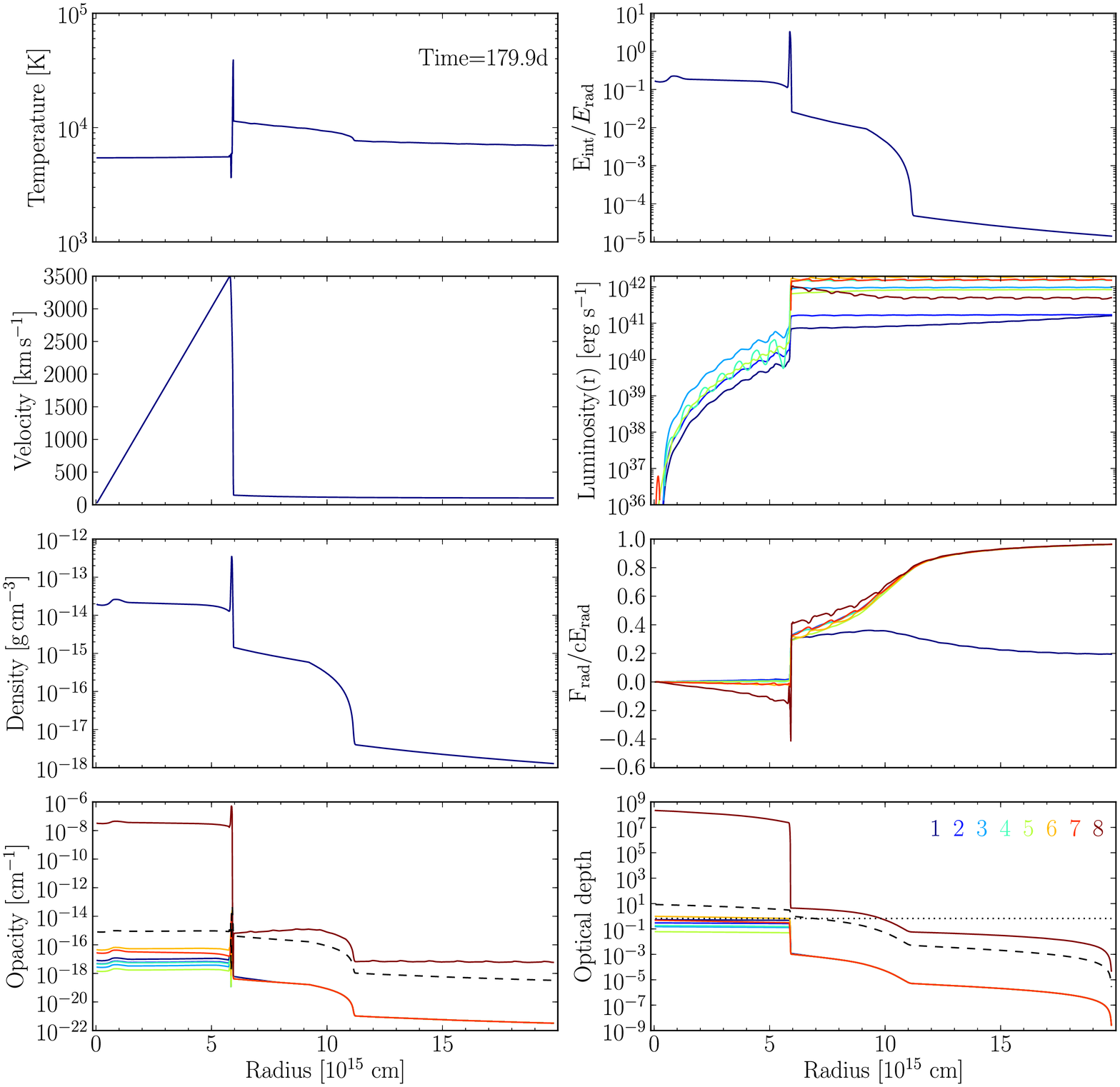,width=17cm}
\caption{Same as Fig.~\ref{fig_hydro1}, but now for a time of 179.9\,d after the onset of the interaction.
\label{fig_hydro3}
}
\end{figure*}

\begin{figure*}
\epsfig{file=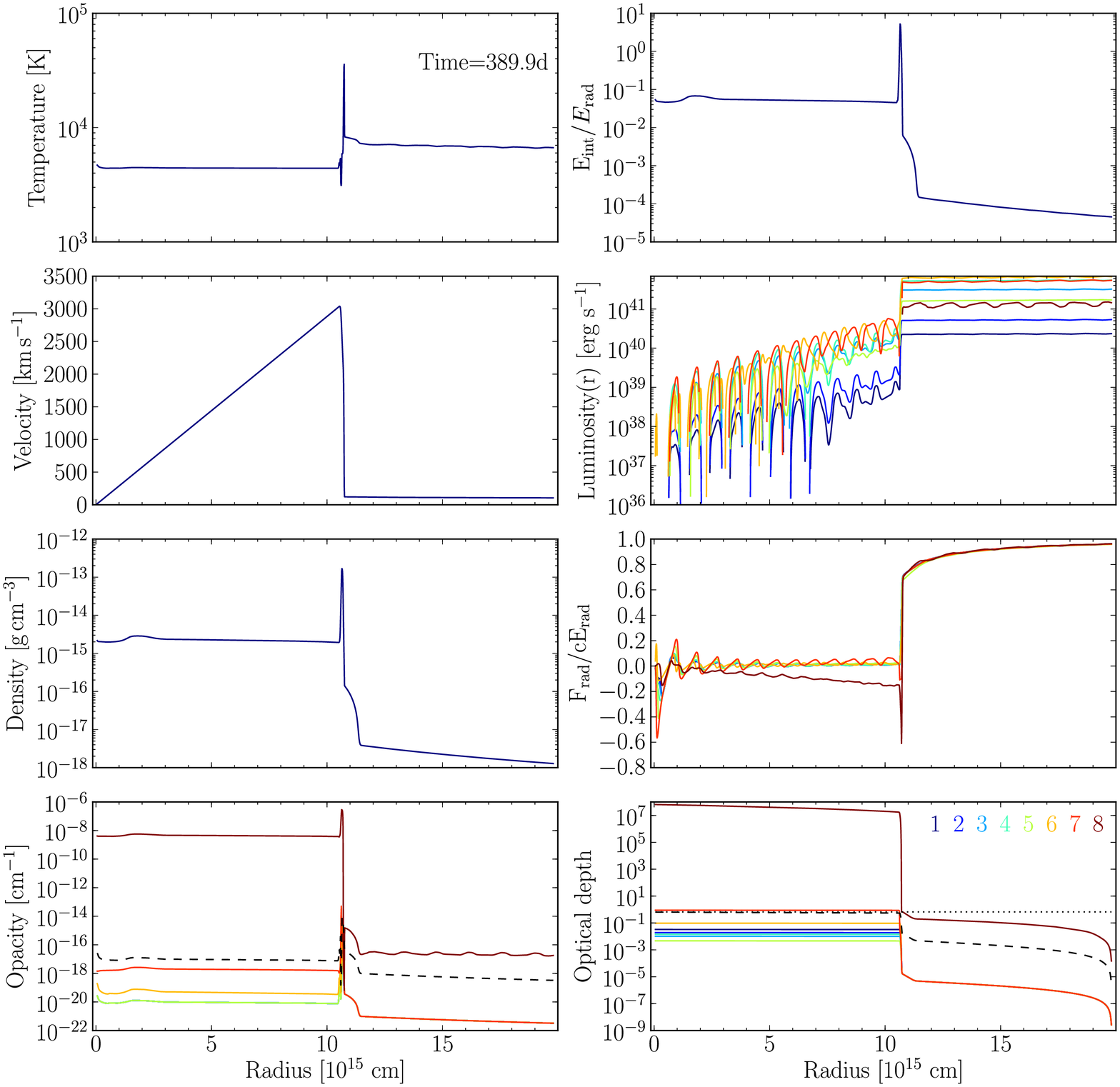,width=17cm}
\caption{Same as Fig.~\ref{fig_hydro1}, but now for a time of 389.9\,d after the onset of the interaction.
\label{fig_hydro4}
}
\end{figure*}

\subsection{Results from the \heracles\ simulation}

\begin{figure*}
\epsfig{file=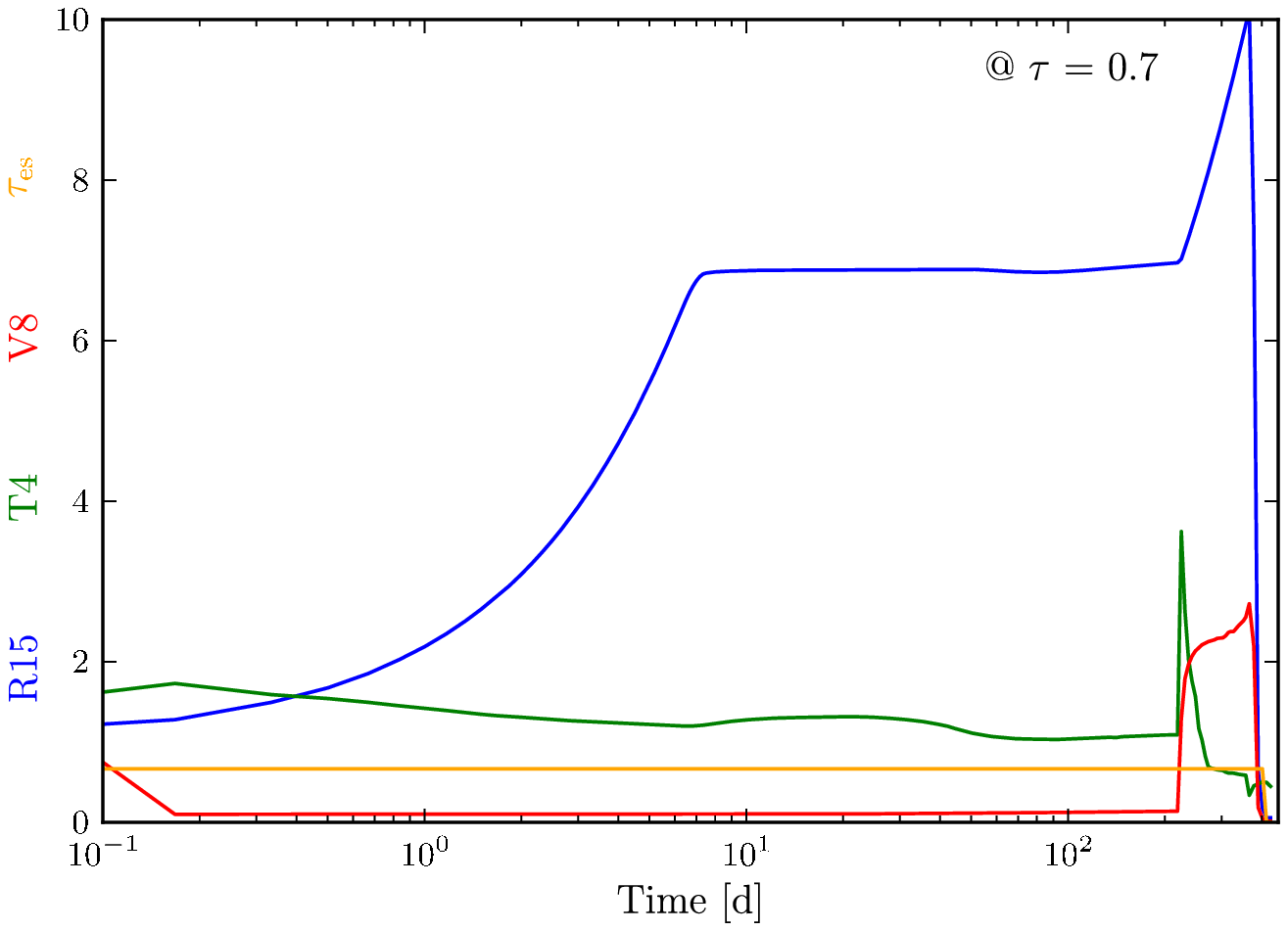,width=8.5cm}
\epsfig{file=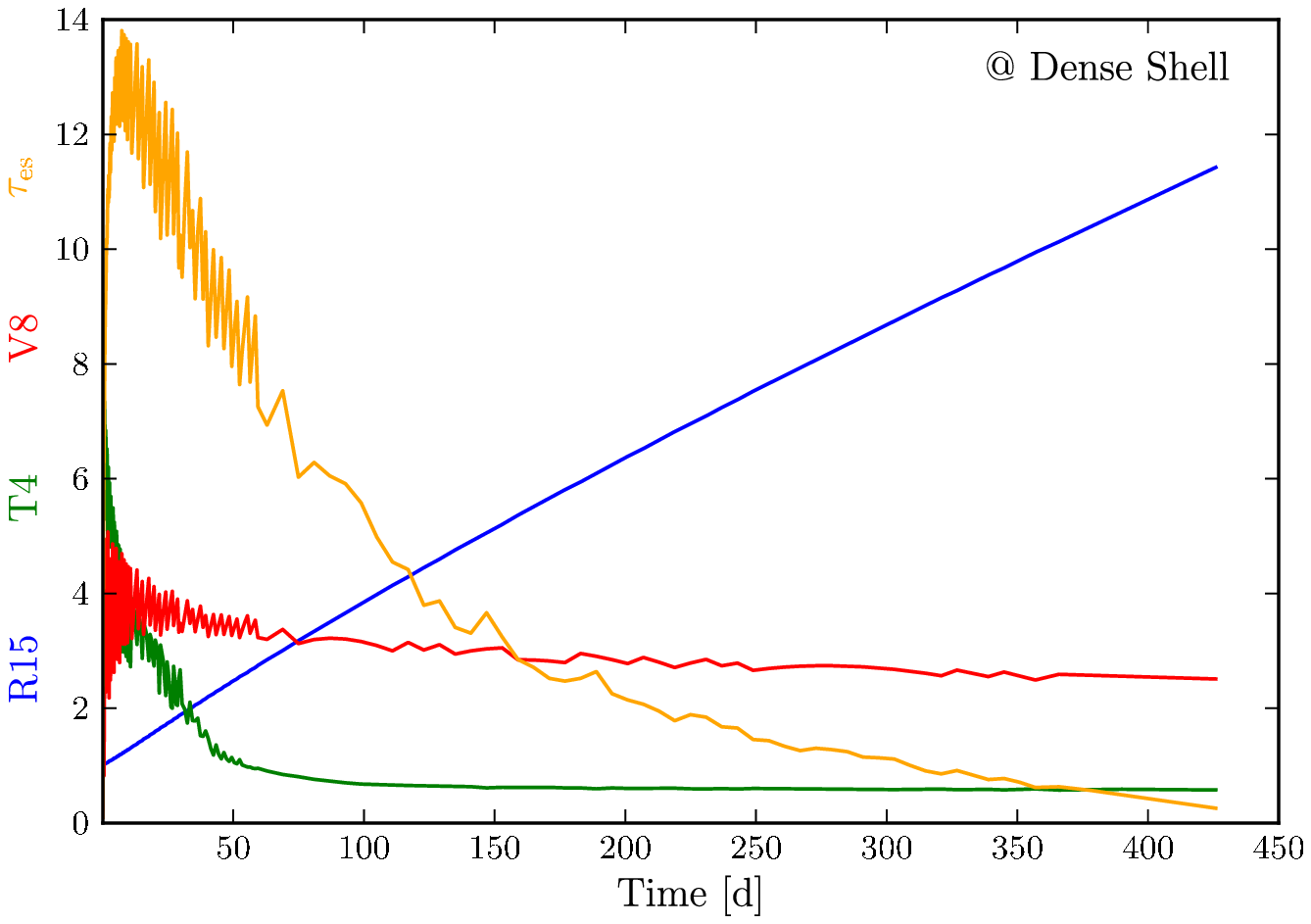,width=8.5cm}
\caption{Illustration of the evolution of the material properties at the photosphere (left; logarithmic scale for abscissa).
and at the location of maximum density within the dense shell (right).
Note the migration of the photosphere to larger radii during the first 8 days, as well as the
sudden jump in photospheric properties at $\sim$\,220\,d, when the CSM material exterior
to the CDS is no longer optically thick (because the CDS has swept up the dense part of the CSM).
\label{fig_prop_loc}
}
\end{figure*}

\begin{figure*}
\epsfig{file=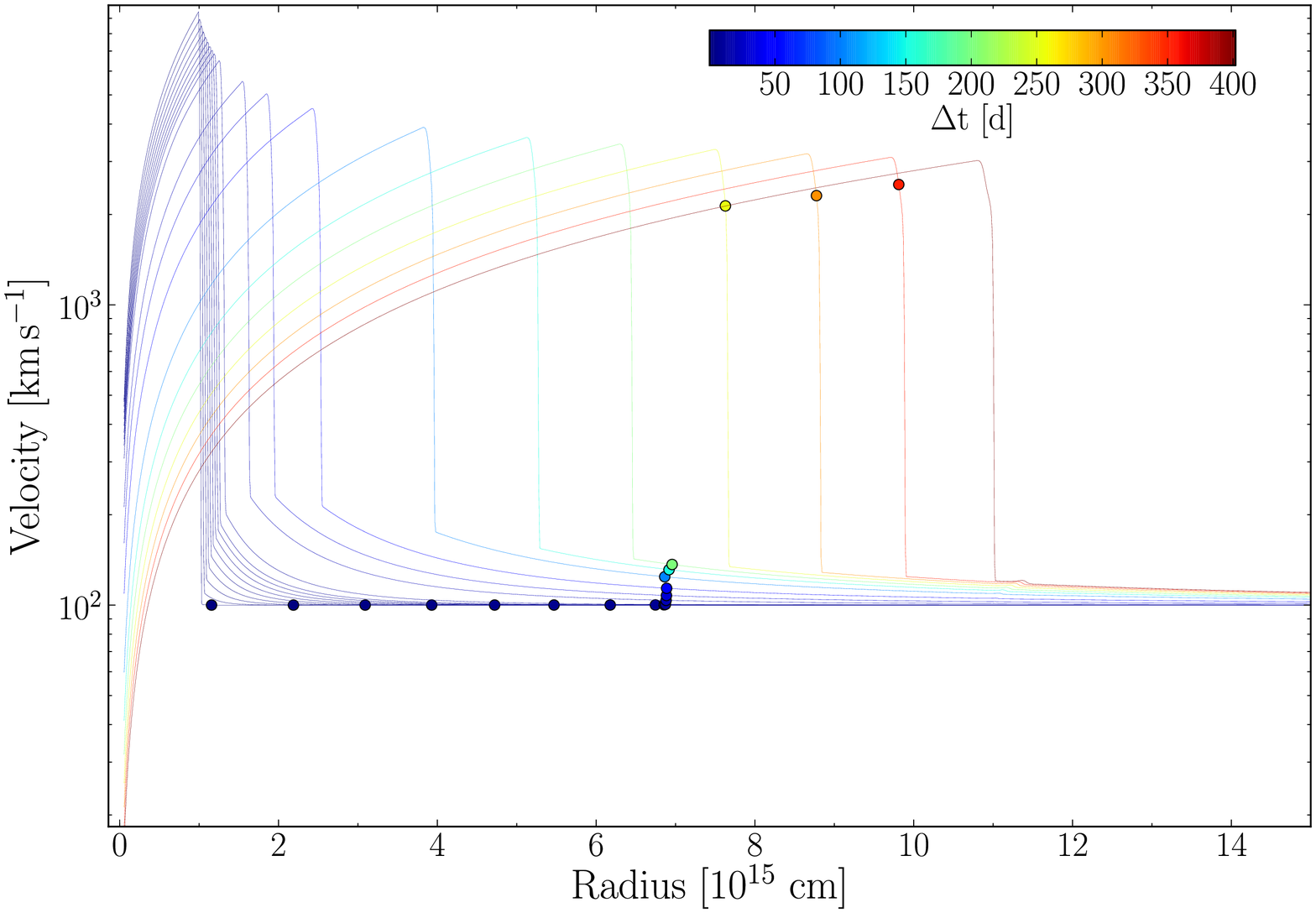,width=8.5cm}
\epsfig{file=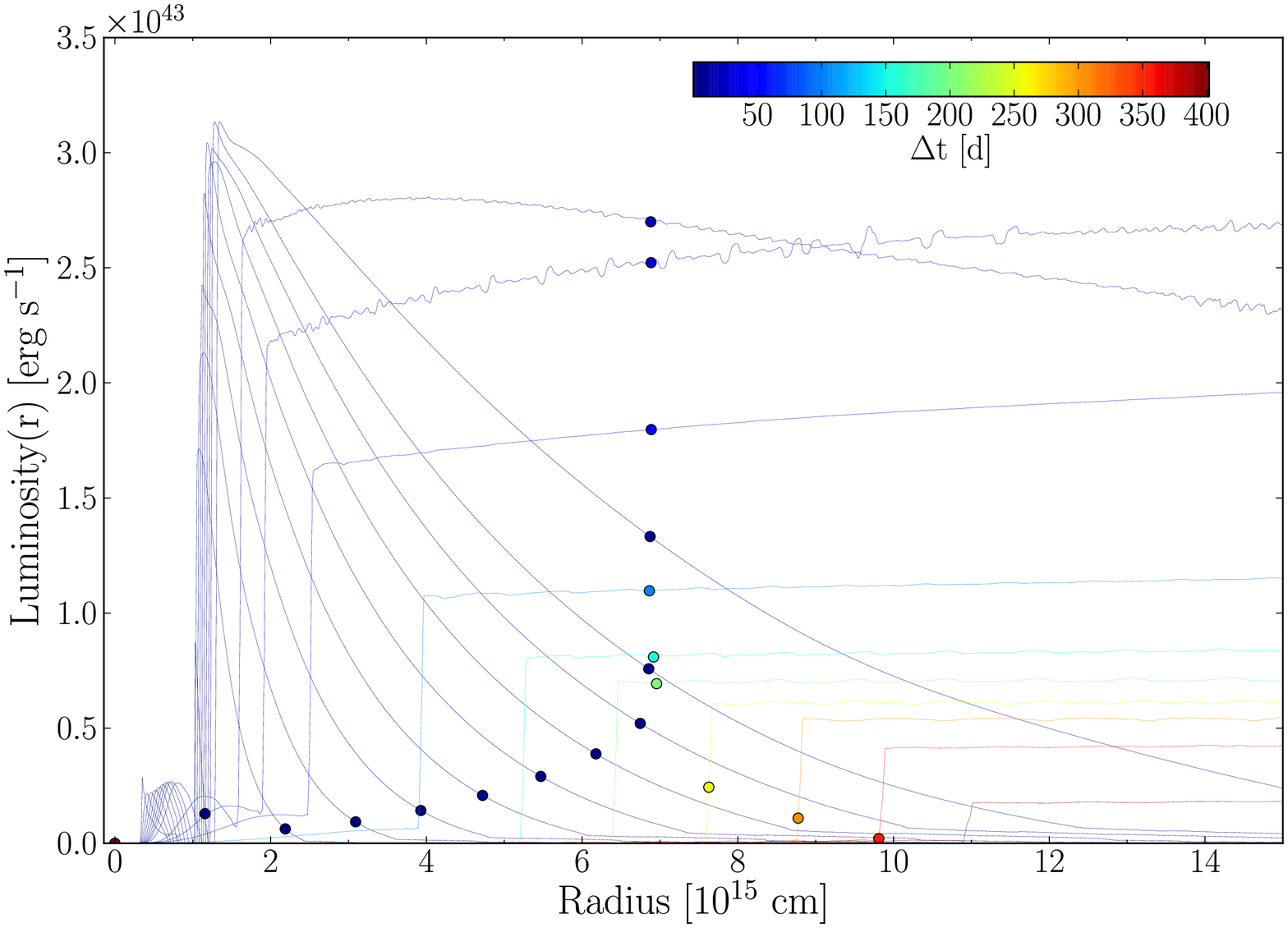,width=8.5cm}
\epsfig{file=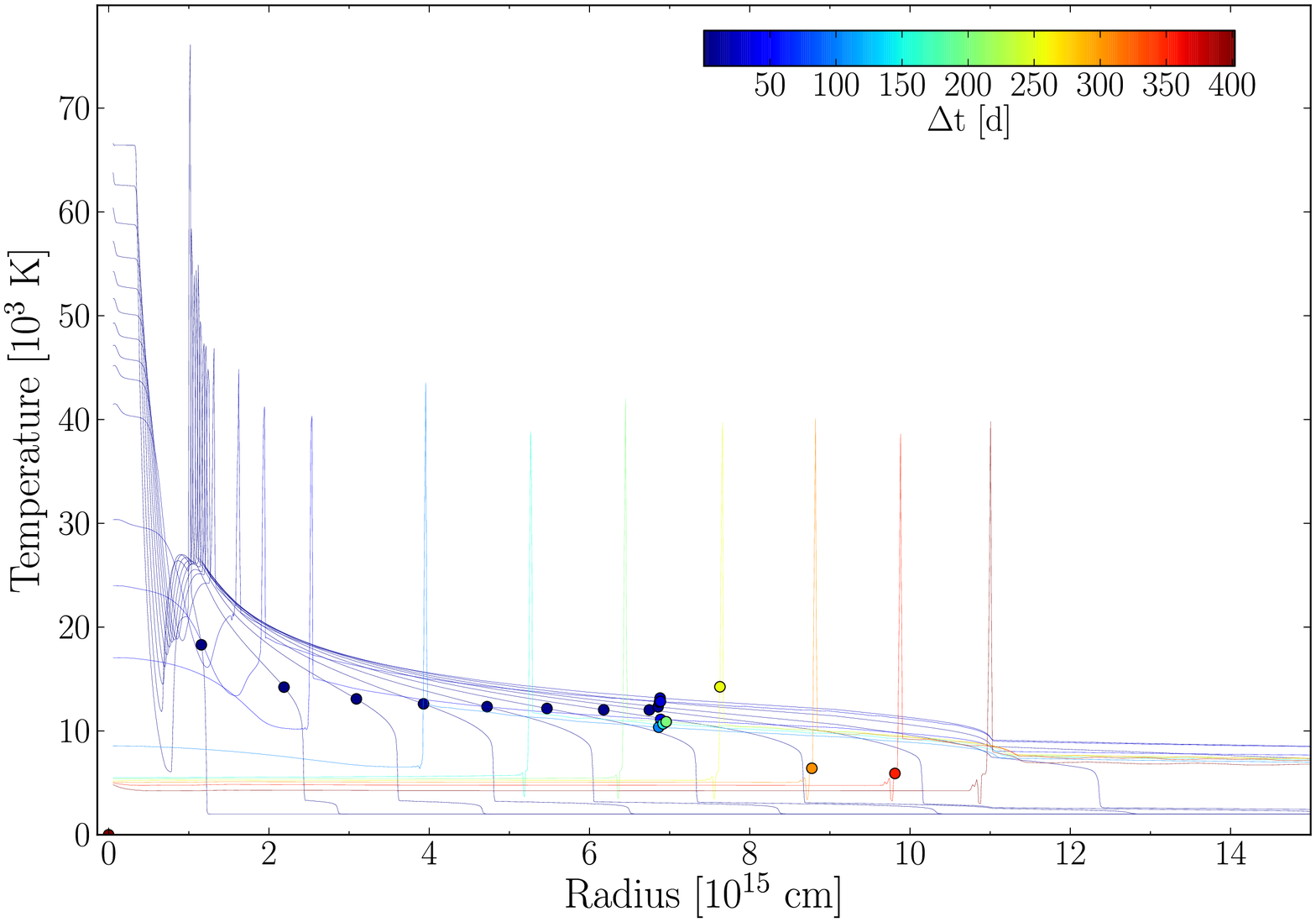,width=8.5cm}
\epsfig{file=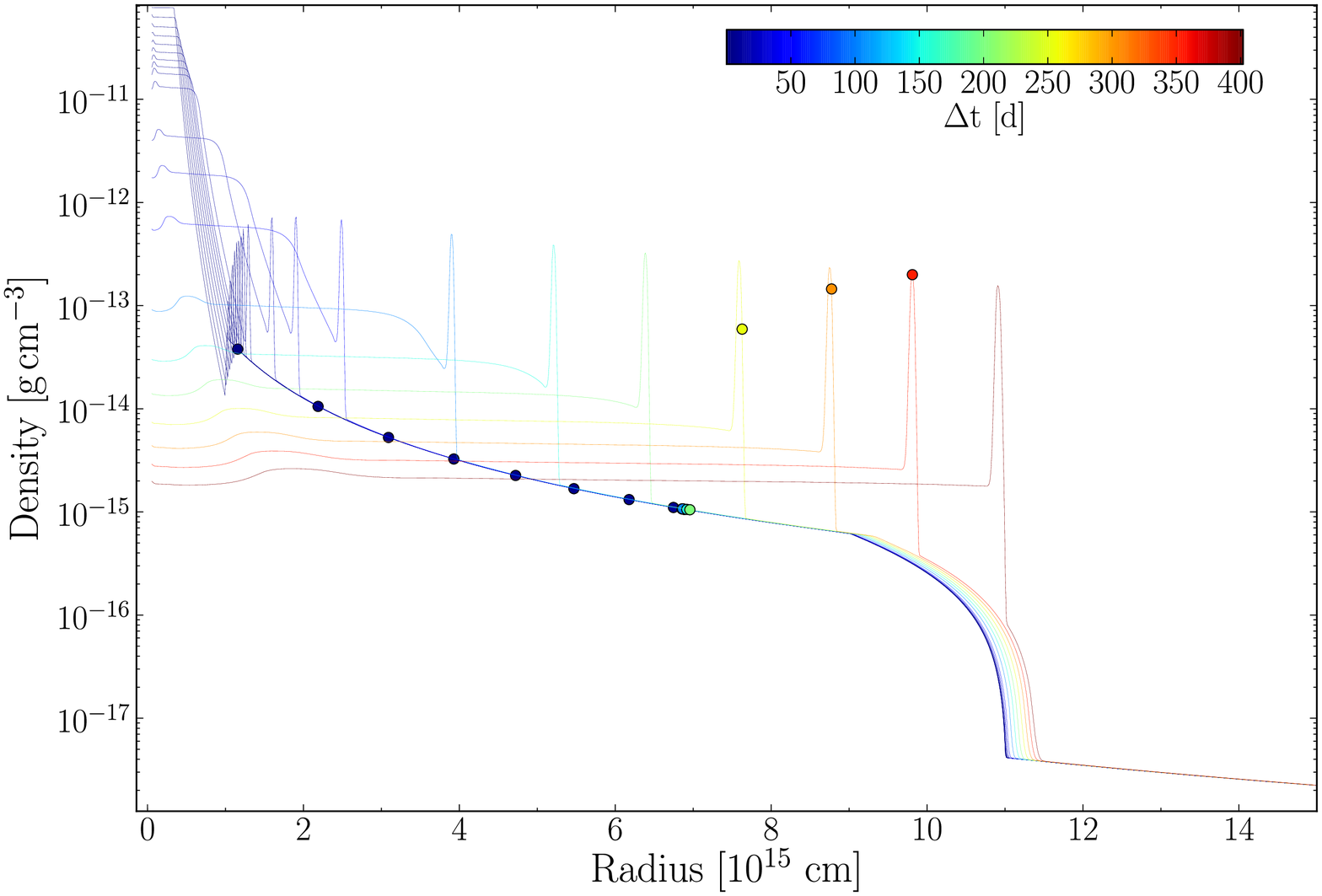,width=8.5cm}
\caption{
Evolution of the radial profiles for the velocity, luminosity, temperature, and mass density for the reference model X.
The epochs shown are 0.01,1, 2, 3, 4, 5, 6, 7, 8, 10, 20, 30, 50, 100, 150, 200, 250, 300, 350, 400\,d after the onset
of the interaction.
Dots refer to the location of the photosphere --- such a location exists for about 350\,d,
first located in the CSM and then in the CDS.
Past 350\,d, the total electron scattering optical depth is less than 2/3, making electron
scattering a poor frequency-redistribution mechanism at such late times.
\label{fig_evol}
}
\end{figure*}

\begin{figure*}
\epsfig{file=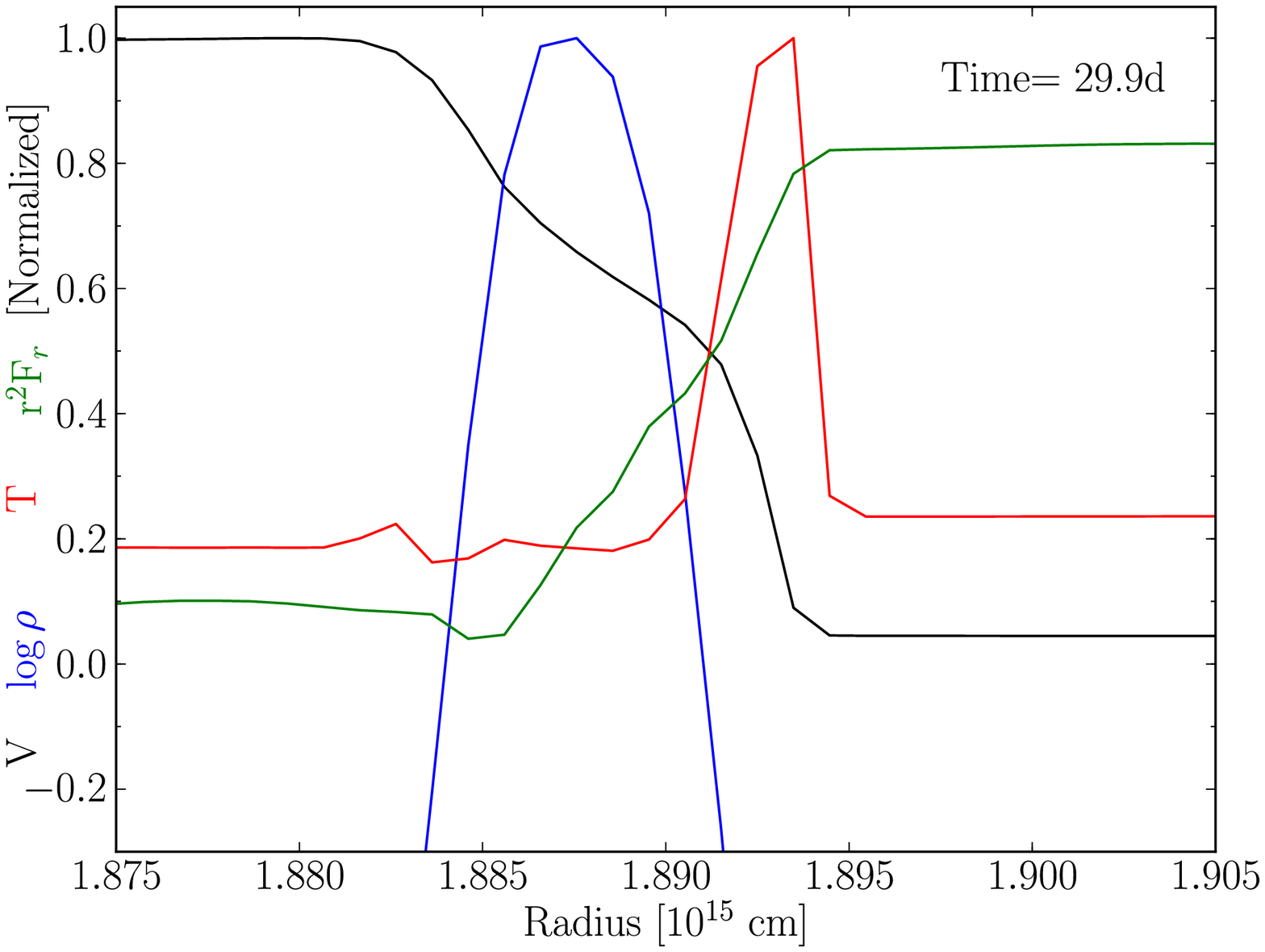,width=8.5cm}
\epsfig{file=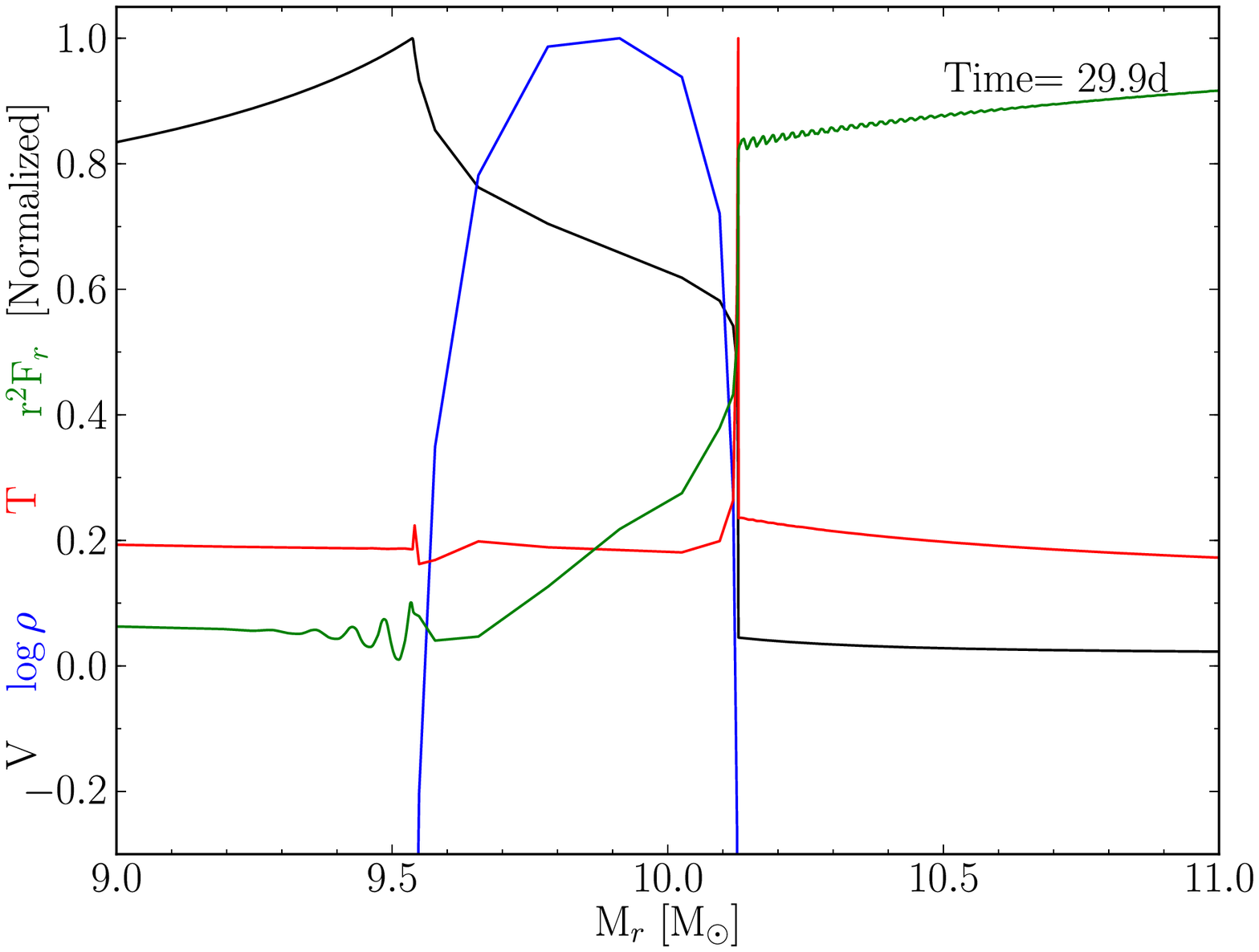,width=8.5cm}
\caption{Profiles for the density, temperature, velocity, and bolometric luminosity in
the shock region at maximum light (all normalised for visibility) as
a function of radius (left) and mass (right) in the reference model X.
The resolution is somewhat too small to fully resolve the shock structure, a problem
that arises from the Eulerian approach.
\label{fig_snap}
}
\end{figure*}

Below, we present the salient features from the interaction simulation, together
with illustrations of the bolometric light curve (Fig.~\ref{fig_lbol}),
the interaction evolution (Figs.~\ref{fig_hydro1}--\ref{fig_hydro4}), the properties
of the photosphere and of the dense shell (Fig.~\ref{fig_prop_loc}),
the evolution of the fluid and of the radiation properties versus radius (Fig.~\ref{fig_evol}),
and the shock structure (fig.~\ref{fig_snap}).
We define the photosphere as the location where the optical depth integrated inwards
from the outer boundary is 2/3. For convenience, we consider only the electron-scattering
opacity when computing the photosphere location. This choice is suitable because the bulk
of the emerging radiation emerges in the optical spectral region where electron scattering
is the dominant opacity source. At X-ray and UV wavelengths energy-dependent optical depths
would need to be used.

Initially the CSM, by construction, is at 2000\,K and neutral. It has a huge
opacity to Lyman-continuum photons but is transparent at longer wavelengths (Fig.~\ref{fig_chi}).
However, after about a week, this CSM is ionised (the ionisation is nearly complete for hydrogen,
but only partial for helium). This rapid change occurs as an ionisation front sweeps the CSM
(see the early migration of the photosphere, shown as a filled dot, in Fig.~\ref{fig_prop_loc}),
triggered by the huge luminosity arising at the shock.
At the onset of interaction, both the CSM density and the ejecta velocity are maximum,
so the shock luminosity is at its maximum.
The luminosity is dominated by the contribution from the forward shock and
can be estimated with $L_{\rm shock} \sim 2 \pi r^2 \rho_{\rm csm} v_{\rm shock}^3$
(see, e.g., \citealt{chugai_danziger_94}),
which is of the order of 3$\times$10$^{44}$\,erg\,s$^{-1}$.
As time progresses, the fast ejecta is decelerated by the CSM and
the shock luminosity decreases.
Within 10\,d, the maximum velocity is $\sim$\,6000\,\kms\ and the shock luminosity
$\sim$3$\times$10$^{43}$\,erg\,s$^{-1}$. These luminosities are extreme.
Adopting a representative radiative luminosity of a few 10$^{43}$\,erg\,s$^{-1}$ crossing the CSM over
the first ten days, this amounts to a total radiative energy on the order of 10$^{49}$\,erg, which is
100-1000 times the energy required to ionise the hydrogen atoms in that H-rich shell.
Only a small fraction of this energy is absorbed but this is more than sufficient
to cause rapid ionisation of the CSM.

Once the CSM is ionised, the minimum continuum opacity is set by electron
scattering, and is moderate at all wavelengths. Photons, which are injected at the shock,
are thermalised due to the huge X-ray/UV opacity and converted to UV/optical photons.
These photons
are trapped and escape after diffusing for  a time $t_{\rm diff} \sim \tau_{\rm csm} \Delta R/c$,
where $\Delta R = R_{\rm phot} - R_{\rm cds}$, and $R_{\rm phot}$ is the
photospheric radius and $R_{\rm cds}$ is the Cold-Dense-Shell (CDS) radius.
This CDS is very narrow (i.e., $10^{13}$\,cm) and is bounded by the reverse shock and the forward shock.
As can be seen in Figs.~\ref{fig_hydro3}--\ref{fig_hydro4} and Fig.~\ref{fig_snap},
the forward shock exhibits a Zel'dovich spike \citep{ZR67}
where the gas temperature becomes very high  before relaxing to its post-shock value.
In practice, the CDS is not cold --  its temperature at these early times is in excess of 20000\,K,
having, however, efficiently cooled from X-ray shock temperatures by radiative
emission.\footnote{We obtain higher spike temperatures of $>$10$^5$\,K  with higher resolution.
However, it is in practice difficult to resolve the spike structure given the scale of the problem.
Shock temperatures of 10$^7$-10$^8$\,K are obtained if we simulate the same interaction
configuration without radiation transport (i.e., equivalent to hydrodynamics only).
When radiation is included, radiation mediates the properties of the shock. The large
heat capacity of the photon gas reduces the shock temperature.
The diffusion and the escape of radiation from the shocked region exacerbates the reduction
in temperature of the shocked region compared to a pure hydrodynamics configuration.}
In our reference simulation (i.e., for this adopted CSM structure) at 10 days after the onset
of the interaction, $R_{\rm phot}\sim$\,7$\times$10$^{15}$\,cm,  the CSM (electron scattering) optical depth
is $\sim$15, and the diffusion time from the shock to the photosphere is $\sim$35\,d.

The light curve we obtain if we assume grey radiation transport (within the M1 model) is quite
different (blue curve in Fig.~\ref{fig_lbol}), especially at early times. The luminosity is
maximum at the beginning,
because the grey opacity is very low for cold gas. In contrast, the opacity to
X-rays and UV photons is huge at low temperature. Hence, in the grey case, the photons initially
produced at the shock travel without being absorbed and reach the outer boundary unadulterated.
It is only subsequently that the streaming radiation raises the CSM temperature and increases
its optical depth, causing a dip and a bump in the light curve, but the corresponding opacity in the
X/UV ranges is still much lower than what it is in Nature. The radiation properties (bolometric luminosity,
but also colours etc.) are thus affected. This test clearly shows that multi-group
radiation transport (with groups that span from short to long wavelength, even coarsely) is important
when modelling SNe IIn.

The evolution of the photospheric radius is quite complicated. Initially it progresses to
larger radii as the CSM becomes more ionised. Once the medium is fully ionised, the
photospheric radius remains almost constant (Fig.~\ref{fig_evol}). Depending
on circumstances, the photospheric radius can then either decrease or increase.
First, the CSM may recombine and its opacity to radiation decrease, leading to a recession
of the photosphere. This is not seen in our simulations because the large and sustained supply of
radiation from the shock maintains a high ionisation in the CSM. The other circumstance,
which inevitably occurs, is that the shock (and the CDS),  eventually overtakes  $R_{\rm phot}$
causing the photospheric radius to increase again.
This occurs in this simulation at $\sim$\,200\,d, which
implies that the average shock velocity over the first 200 days is $\sim$3000\,\kms\ (Fig.~\ref{fig_prop_loc}).
Because of expansion, the electron-scattering optical depth of the CDS decreases with time.
\footnote{This holds once the CSM is fully ionised. Radial compression of CSM material by the
shock does not change the CSM electron-scattering optical depth; this would require lateral compression,
inexistent in a spherically-symmetric simulation. Absorptive optical depths may increase (because such
process are often proportional to the square of the density) but except for the optical depth of
 the Lyman continuum these also decline.}

The evolution of the light curve for the first $\sim$\,200\,d is essentially at fixed $R_{\rm phot}$.
The morphology of the light curve (Fig.~\ref{fig_lbol}) differs from the monotonically decreasing shock luminosity
because of optical depth effects, very much in the manner of \iso{56}Ni powered optically-thick
ejecta of standard SNe \citep{arnett_82,chevalier_irwin_11,moriya_etal_13a}.
At early times, the $L_{\rm shock}$ is huge but the optical depth
is large so radiation energy is stored and released on a diffusion time scale of about a month.
Bolometric maximum occurs at $t \sim t_{\rm diff}$.
Past maximum, the SN radiation exceeds for a while and eventually becomes equal to
the shock luminosity.
A steady-state configuration sets in because the shock luminosity evolves slowly and
the diffusion time is only a few days (because the CDS gets closer to $R_{\rm phot}$).
Thus, the bolometric evolution over the first 200\,d reflects primarily variations not in $R_{\rm phot}$
but in the radiation temperature, itself dependent on the amount of energy injected and stored
between the photosphere and the CDS. For times $t<t_{\rm diff}$, this energy accumulates
and the temperature of the optically-thick CSM gas goes up to a maximum around bolometric maximum.
For times $t>t_{\rm diff}$, less and less energy is stored until the steady-steady state regime is reached.
As the shock luminosity decreases, the radiation temperature decreases too and the SN luminosity
ebbs.\footnote{In theory,
the adopted value of the mean atomic weight should influence the gas temperature. For partial ionisation,
the mean atomic weight could be around 0.6-0.7, rather than 1.38, which corresponds to
the neutral H-rich gas adopted here.
We ran a simulation for the reference model X with a value of 0.6, but
found very similar results, probably because in super-luminous  SNe IIn, the entire material
is partially ionised throughout most of the evolution --- the ionisation
state of the plasma does not change.}

At the end of the simulation, a total of 0.3$\times$\foe\ has been radiated, drawn
from the \foe\ of kinetic energy stored in the inner ejecta.  There still remains,
untapped, 0.7$\times$\foe\ of kinetic energy in the system, contained in the un-shocked SN ejecta
(i.e., the SN ejecta material not affected by the interaction, which represents 50\% of the total SN ejecta mass)
and in the massive CDS,
which still advances through space at $\sim$3000\,\kms. So, for this configuration,
we obtain a conversion efficiency of kinetic to radiation energy of 30\% (little additional
kinetic energy will be extracted in this interaction because the remaining CSM is very low
density).
In a multi-dimensional configuration, some kinetic energy could also be stored in the lateral
direction. Hence, the present 1-D simulation likely overestimates what the equivalent 3-D simulation
would produce.
Unlike \citet{moriya_etal_13a}, we make no attempt to estimate this, in part because
these lateral motions will be much slower than the radial motion of the material
and thus cannot represent a large energy loss for the radiation.

 In the next section, we will discuss in more detail the radiative transfer properties
 of this interaction. One thing to note is that despite the significant optical depth
 of the configuration, the photon mean free path is non negligible. This is particularly
 true in the inner ejecta region once expansion has caused significant cooling because
 these regions do not benefit from the shock luminosity (the bulk of the radiation
 streams radially outwards from the interaction).
 There is a temperature jump ahead of the interaction (Fig.~\ref{fig_snap})
 caused by radiation leakage from
 the shock, as in the phenomenon of shock breakout in core-collapse SNe \citep{klein_chevalier_78}.
 This structure is comoving with the shock, since the conditions that cause it persist
 for as long as the shock remains optically-thick.

Our results are in agreement with the radiation hydrodynamics simulations
of \citet{moriya_etal_13a}. Our simulation also
emphasises that the shell shocked model \citep{smith_mccray_07} is not adequate
for optically-thick super-luminous SNe IIn, as also pointed out by \citet{moriya_etal_13b}.
The basic inconsistency with the shell-shocked model is that it is not possible for a
shock to cross a very extended
CSM, ionise it, store energy within it, and {\it subsequently} let this shocked material radiate
the deposited energy.
This scenario applies to shocks crossing the interior of a stellar envelope (as in successful
core-collapse SNe) only because the stellar interior has a huge optical depth. At every location
except for the outermost stellar layers, the radiation dominated shock progresses outward faster
than the photons it carries in its wake because the photon mean free path is exceedingly small
and photon diffusion times exceedingly long compared to the shock-crossing time.
In interacting SNe, this configuration does not hold at all.
Even for very high CSM densities, the shock-crossing time through the CSM will always
far exceed the diffusion time through the shell so that radiation deposited by the shock is
{\it continuously} being radiated by the shocked material.
The present simulation shows that the basic light curve properties of
super luminous SNe can be explained with a simple and physically consistent setup.
It also illustrates how the simplistic ``diffusion model'' can be misleading, and how an argument
based on dimensional analysis can sometimes lead to wrong conclusions.

In our simulations spectrum formation
takes place primarily in the slow moving CSM for about 200\,d, but past that the spectrum
formation region will be tied to the much faster moving CDS.
The CDS optical depth will be too small to ensure full thermalisation within the slow
unshocked material even if the location where the optical depth is 2/3
(i.e., the photosphere), lies there.
This implies that line profiles should evolve and become more
Doppler-broadened as time passes in super-luminous SNe,
while line broadening by electron scattering would dominate only at early times
\citep{dessart_etal_09}. This behaviour is important to study as it carries
information on the potentially large residue of kinetic
energy left untapped in the interaction. For example, this matters when estimating
the kinetic energy in the inner shell and whether this inner shell originates in an energetic
explosion that requires the gravitational collapse of the iron core of a massive star.

The radiation injected by the shock and crossing the CSM does work on the material
and accelerates it. This radiative acceleration is however modest, and limited to the peak
time of the light curve, leading to an increase of the CSM velocity from 100 to $\sim$\,200\,\kms\
(Fig.~\ref{fig_evol}).
At early times, the material that receives a larger acceleration, just ahead of the shock, lies at
a large optical depth and is thus less visible that the photons emitted further out, in regions
where the acceleration is more modest.
At late times, the material ahead of the shock is too optically-thin to receive a significant
acceleration.
The radiative acceleration in the \heracles\ simulation is likely underestimated
because of the neglect of line opacity and the potential boost from line desaturation caused
by velocity gradients \citep{cak,owocki_14}.
However, our simulations suggest that the radiation arising
at the shock only cause a modest acceleration of the unshocked CSM.
This is compatible with the near-constant width of the narrow P-Cygni profile seen in SN\,2010jl
\citep{fransson_etal_14}. As we discuss in the next section, this modest acceleration seems an unlikely
origin of the peak blueshift seen in the broad emission component of H$\alpha$.

\section{Spectroscopic modelling with \cmfgen}
\label{sect_cmfgen}

\subsection{Numerical approach}

At selected epochs during the evolution, from 20 to $\sim$\,200\,d after the onset of the interaction,
we post-process the \heracles\ simulation with the non-LTE radiative transfer code \cmfgen\
\citep{HM98_lb,HD12}.
While there is no explicit time-dependence in the \cmfgen\ simulations undertaken here,
time dependence is implicitly taken into account since we use the density, temperature, and velocity
structures computed with \heracles. We thus retain a much higher physical consistency
than in \citet{dessart_etal_09}.
We use an adaptive grid to resolve carefully the strongly varying density and temperature profiles,
as well as the optical depth.
The grid is truncated at small radii when the inward integrated Rosseland-mean optical depth
reaches 30.
The final models discussed here use 100 grid points and adopt a turbulent velocity of 20\,\kms.

We use the same H, He, and Fe mass fractions as for the radiation-hydrodynamical simulations.
We limit the model atom to H\one, He\one, He\two, Fe\one--\six\ with the same super-level
assignment as in \citet{dessart_etal_13b}.
Higher ionisation stages are unnecessary since the gas temperature in the simulations at the times
considered is below $\sim$\,80000\,K (this maximum temperature is limited to a few zones
at the shock and to early times).
In the present simulations, the spectrum formation region has a characteristic temperature of about
10000\,K or less.
A discussion of the signatures of ions from CNO and other intermediate mass elements
(which are merely trace elements compared to H and He here) is deferred to a subsequent study.

The current version of \cmfgen\ does not allow us
to calculate non-LTE line blanketed model atmospheres for non-monotonic velocity flows.
This property of interacting SNe is however fundamental and should be taken into account.
In this work, to overcome this limitation and take into account the non-monotonic velocity structure,
we ignore lines when treating  the continuous spectrum, but model the lines
for the non-LTE analysis using the Sobolev approximation. Non-monotonic velocity flows do not
cause major issues with the Sobolev approximation\footnote{One problem
we needed to address is that the Sobolev optical depth becomes infinite for one particular direction
when the radial velocity gradient is not positive.
This occurs because in a spherical flow the Sobolev optical depth is proportional to
$1/\lvert1+\sigma\mu^2\rvert$, where $\sigma \equiv d\ln V/d\ln R-1$. The angular factor will be infinite
when $\mu=\pm 1/\sqrt{-\sigma}$.
%
% In practical terms this does not affect the
% escape probability very much due to the small solid angle subtended by the singularity. However, in the
% standard computational approach the angle grid is not well sampled, and numerical instabilities occurred,
% which we solved by limiting the increase in optical depth.
%
In practical terms this does not affect the escape probability very much due to the small solid angle subtended
by the singularity. However, in the standard computational approach the angle grid is not well sampled
(the sampling becomes worse as we move away from $\mu=1$), and numerical instabilities occurred
at one or two depths, which we overcame by limiting the increase in optical depth. In this work
we used the simple approach of choosing
$\max(|1+\sigma\mu^2 |,-0.2(1+\sigma))$ for $|1+\sigma\mu^2|$ when $\sigma < 0$.}
since it is
a local approximation that uses the absolute value of the velocity gradient to compute the optical depth
\citep{castor_70}.
Since the density and temperature structures are taken from the hydrodynamical simulation
we believe this to be a very reasonable first approach.

To compute the observed spectrum we wrote a new routine capable of computing J
in the comoving frame  in the presence of non-monotonic velocity fields. With this
routine we were able to allow for the influence of (non-coherent) electron scattering which was
iterated to convergence using a simple lambda-iteration. Relatively minor changes
were then made to {\sc cmf\_flux} \citep {BH05_2D} to facilitate the computation of the observed
spectrum in the observer's frame.

In \citet{dessart_etal_09}, we used a prescription for the density structure of the emitting region
and the ansatz was that an optically-thick shell was radiating the flux. In that approach, the entire
grid was in radiative equilibrium, and the flux was forced to diffuse through the densest regions
 of the atmosphere (these regions were also the hottest by imposing the diffusion approximation
 at the base). The dynamical configuration presented in Section~\ref{sect_her} is obviously quite different, but
 this simplistic treatment is in fact not so bad at early times, when the interaction region
 and the inner CSM are optically thick because in this case the details of the inner boundary conditions are
 lost. Hence, for as long as the configuration is optically thick, the approach of \citet{dessart_etal_09}
 can provide a good match to the observations of some SNe IIn like 1994W without being dynamically
 consistent.

\begin{figure*}
\epsfig{file=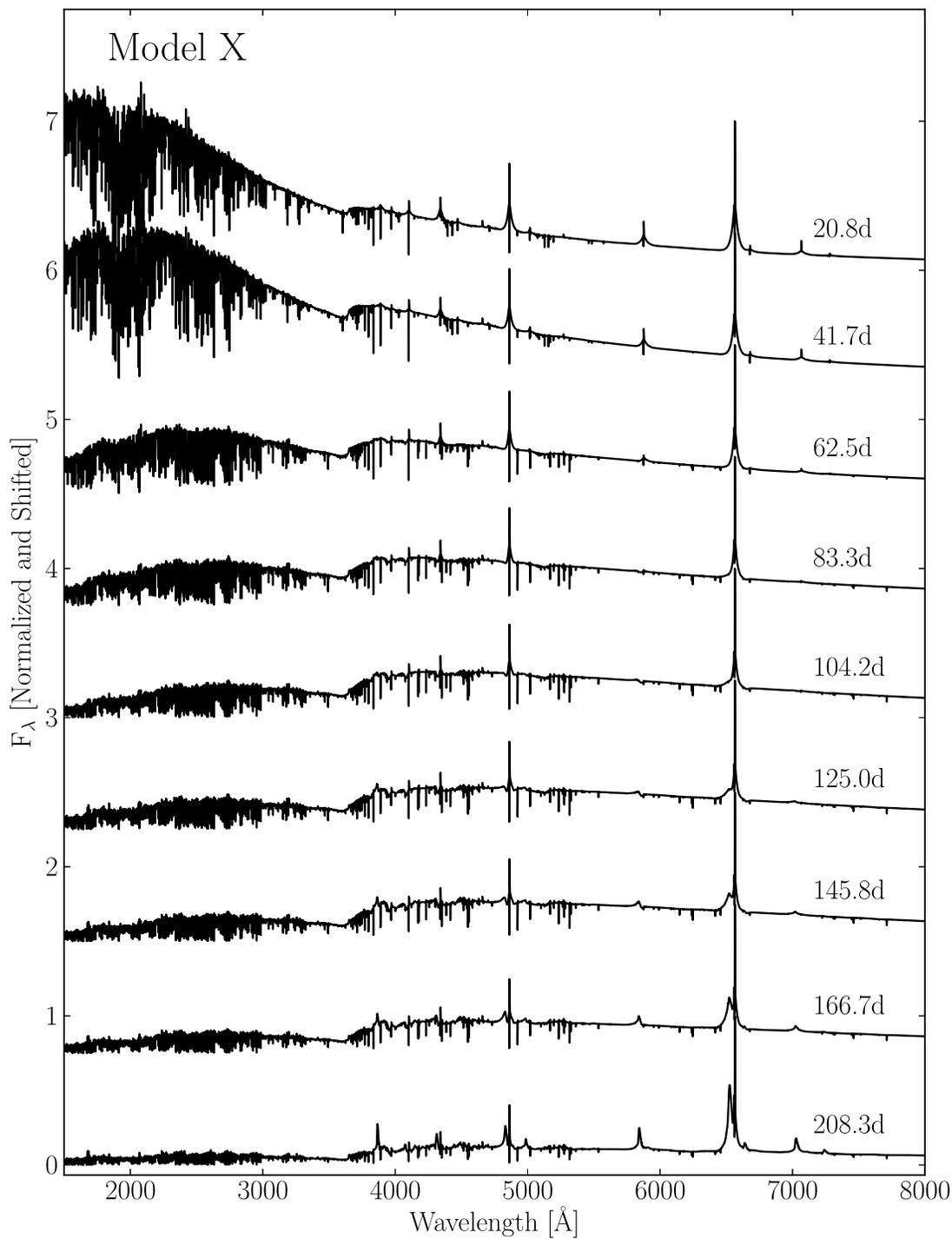,width=17cm}
\vspace{-2.cm}
\caption{
Montage of spectra at selected epochs (see right label) for our reference model X.
The high-frequency patterns in the UV are caused by line blanketing from Fe\two\ and Fe\three\
in this model, while the optical is composed of H\one, He\one, and Fe\two\ lines. At early times, lines
typically show a hybrid morphology with a narrow core and extended wings, up to about 2000\,\kms\ from
line centre. These wings are due to incoherent electron scattering.
As time progresses, a blue shifted component, associated with emission from the CDS,
grows in strength. Line broadening is then caused by electron scattering and expansion.
\label{fig_spec_evol}
}
\end{figure*}

\subsection{Spectral evolution}

We thus compute non-LTE spectra at 20.8, 41.7, 62.5, 83.3, 104.2,
125.0, 145.8, 166.7, 208.3\,d after the onset of the interaction.
The spectral evolution shown in Fig.~\ref{fig_spec_evol} reveals a striking similarity to observations of
numerous super-luminous SNe IIn, including SN\,2010jl \citep{zhang_etal_12,fransson_etal_14}.
The SED evolves very slowly with time, with a progression towards redder colours, a
diminishing flux shortward of 4000\,\AA, and a decreasing ionisation visible through the strengthening of
Fe\,\two\ lines. H\,\one\ and He\,\one\ lines behave in a more complicated way.

Line profiles prior to 100 days show the distinct symmetric morphology
resulting from frequency redistribution by scattering with thermal electrons \citep{C01_SN1998S,dessart_etal_09},
with a typical width of a few
1000\,\kms\ --- in earlier calculations, where we limited the velocities
to 100\,\kms\ on the \cmfgen\ grid, we obtained similar profiles. The width of this component only
reflects redistribution of photons in frequency space by electron scattering -- it provides no information
about the shock velocity.
Superposed on the broad wings, is a very narrow component with a P-Cygni profile morphology.
This narrow component  corresponds to line emission and absorption weakly affected by electron scattering,
and which  preferentially arises from the outer regions of the CSM (which are moving at $\gtrsim$100\,\kms\
in our model X) where line opacity may still be large but where the electron-scattering opacity is negligible.

At a qualitative level, the early-time spectra shown in Fig.~\ref{fig_spec_evol} are analogous to those computed in
\citet{dessart_etal_09} despite the very different approaches in the description of the density/temperature
structure.

\begin{figure*}
\epsfig{file=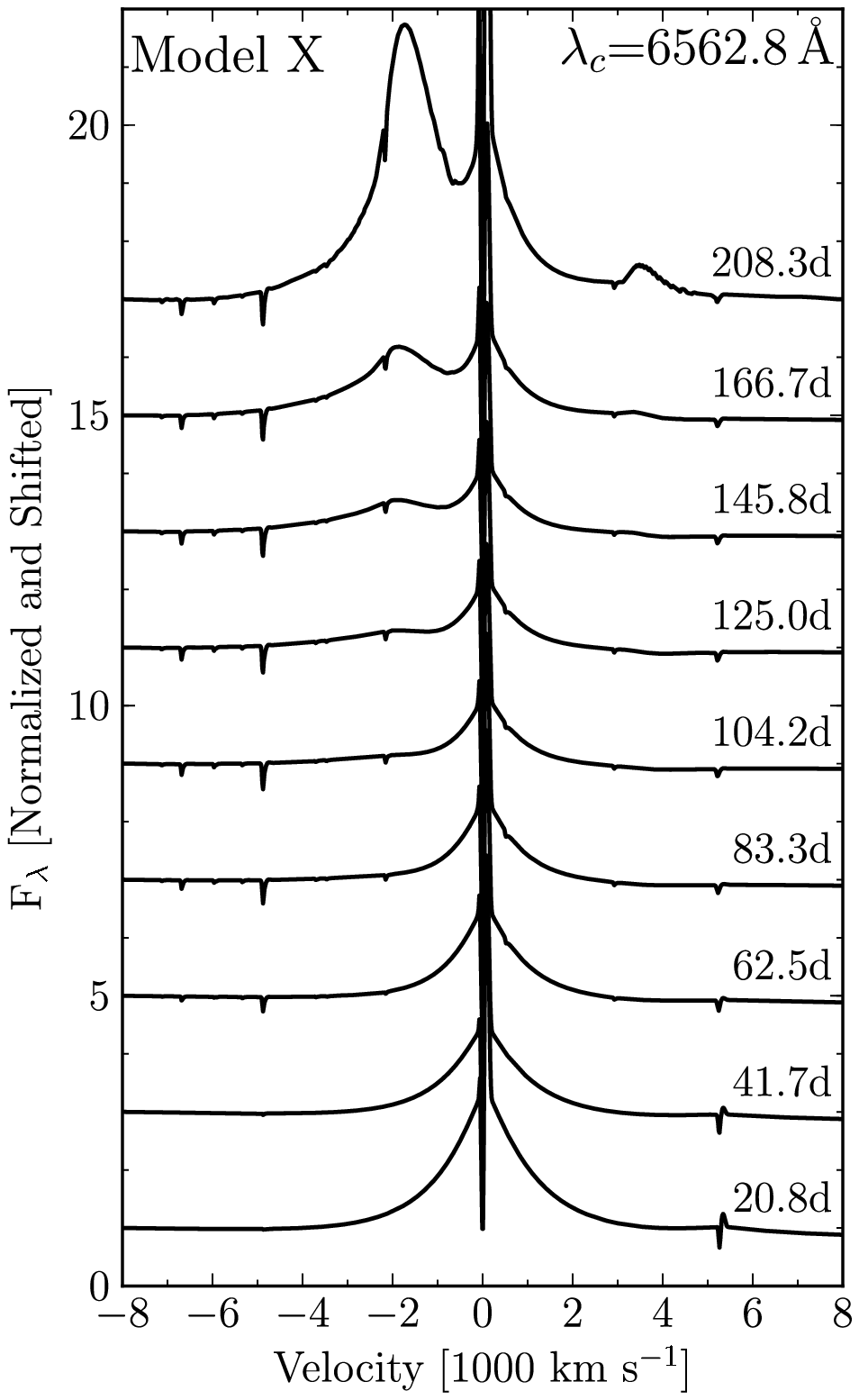,width=8.5cm}
\epsfig{file=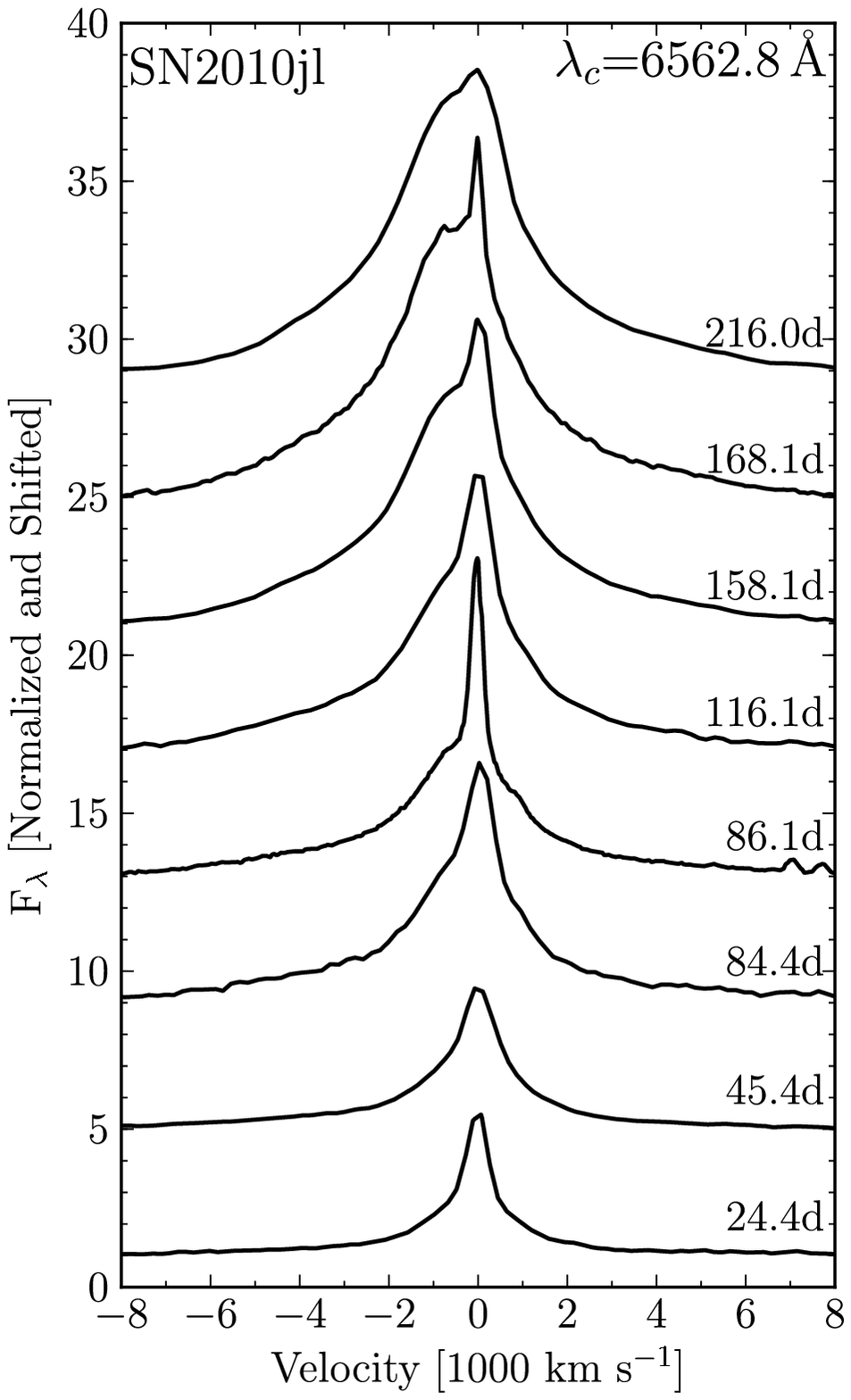,width=8.5cm}
\caption{
Montage of multi-epoch spectra for Model X (left) and SN\,2010jl (right; \citealt{zhang_etal_12}) shown
in velocity space with respect to the rest wavelength of H$\alpha$.
Qualitatively, the model reproduces well the strengthening of the blue-shifted emission component and the increasing
width of the observed H$\alpha$ line profile.
Quantitatively, the magnitude of the blueshift is somewhat too high while the width of the emission feature is underestimated.
\label{fig_spec_halpha}
}
\end{figure*}

After 100 days, the line profile morphology changes. In addition to the two components discussed above,
we see a blue shifted emission component that arises from the CDS. It is blue shifted
because the CDS is moving at a velocity of $\sim$3000\,\kms\ and is optically thick in the continuum
-- line photons emitted in the red are absorbed by the near side of the CDS.
The blueshift of peak emission seen in non-interacting Type II SNe \citep{DH05a} has a similar origin.
These H$\alpha$ properties agree well at the qualitative level with the observations of SN\,2010jl \citep{zhang_etal_12},
although the magnitude of the emission blueshift is overestimated and the magnitude of the line width
is underestimated (Fig.~\ref{fig_spec_halpha}).

% From ~/python/cmfgen_sniin/get_bb_prop.py
% From mkflc.pro
% From ~/python/cmfgen_sniin/prop_snapshots.py
\begin{table*}
\caption{
Summary of properties for the \cmfgen\ simulations based  on the reference interaction model X.
We include the properties at the photosphere, at the location
of maximum density in the interaction region (the CDS), and
the properties of the fitted blackbody (see Fig.~\ref{fig_comp_bb} for an example).
We select epochs when the CDS is well below the photosphere, so that the interaction region is optically thick.
\label{tab_mod_cmfgen}}
\begin{tabular}{l@{\hspace{2mm}}c@{\hspace{2mm}}c@{\hspace{2mm}}c@{\hspace{2mm}}c@{\hspace{2mm}}c@{\hspace{2mm}}c@{\hspace{2mm}}c@{\hspace{2mm}}
c@{\hspace{2mm}}c@{\hspace{2mm}}c@{\hspace{2mm}}c@{\hspace{2mm}}c@{\hspace{2mm}}c@{\hspace{2mm}}c@{\hspace{2mm}}}
\hline
\hline
      age     & $L_{\rm bol}$&      $M_B$  &      $M_V$  &      $M_R$ &      $M_I$  &   \rphot\   &  \tphot\  &   \edphot\    &   \rcds\    &    \tcds\  & \edcds\ &      \rbb\  &     \tbb\   &   \lbb\     \\
\hline
[d] & [erg\,s$^{-1}$] & [mag] &  [mag] &  [mag] &  [mag] & [10$^{15}$\,cm] & [10$^4$\,K] & [cm$^{-3}$] &  [10$^{15}$\,cm] & [10$^4$\,K] & [cm$^{-3}$] &   [10$^{15}$\,cm] & [10$^4$\,K] & [erg\,s$^{-1}$]  \\
\hline
20.8    &   2.15(43)  &    -18.61  &   -18.68  &   -18.85  &   -18.86  &  7.17 &    1.33 &    4.36(8)  &   1.25 &    1.66  &   1.76(11) &  1.172  &    1.084  &   2.259(43)  \\
41.7    &   2.23(43)  &    -18.84  &   -18.98  &   -19.14  &   -19.21  &  7.10 &    1.27 &    4.42(8)  &   1.78 &    1.18  &   1.33(11) &  1.448  &    1.033  &   2.371(43)  \\
62.5    &   1.21(43)  &    -18.40  &   -18.68  &   -18.90  &   -19.03  &  7.02 &    1.09 &    4.05(8)  &   2.86 &    1.01  &   1.05(11) &  1.616  &    0.879  &   1.289(43)  \\
83.3    &   9.88(42)  &    -18.16  &   -18.55  &   -18.82  &   -19.01  &  7.04 &    1.00 &    4.00(8)  &   3.37 &    0.79  &   3.93(10) &  1.848  &    0.788  &   1.050(43)  \\
104.2   &   8.65(42)  &    -17.95  &   -18.43  &   -18.74  &   -18.99  &  7.11 &    0.97 &    3.94(8)  &   3.91 &    0.76  &   2.25(10) &  2.083  &    0.715  &   9.068(42)  \\
125.0   &   7.99(42)  &    -17.74  &   -18.27  &   -18.62  &   -18.89  &  7.08 &    0.98 &    3.96(8)  &   3.64 &    0.61  &   9.24(9)  &  2.199  &    0.673  &   8.496(42)  \\
145.8   &   7.26(42)  &    -17.56  &   -18.12  &   -18.49  &   -18.77  &  7.15 &    1.00 &    3.90(8)  &   4.94 &    0.69  &   9.51(9)  &  2.199  &    0.646  &   7.503(42)  \\
% 166.7   &   5.77(42)  &    -17.22  &   -17.78  &   -18.18  &   -18.45  &  7.12 &    1.00 &    3.94(8)  &   5.47 &    0.66  &   7.62(9)  &  1.990  &    0.625  &   5.834(42)  \\
% 208.3   &   2.94(42)  &    -16.38  &   -16.92  &   -17.44  &   -17.59  &  7.16 &    1.04 &    4.08(8)  &   6.39 &    0.61  &   2.14(9)  &  1.391  &    0.609  &   2.768(42)  \\
\hline
\end{tabular}
\end{table*}

We give a summary of the \cmfgen\ model properties in Table~\ref{tab_mod_cmfgen}.
The \cmfgen\ bolometric luminosity agrees with the \heracles\ value to within 5-25\% at early times
(i.e. prior to $\sim$\,200\,d),
which is noteworthy since the two codes differ vastly in the treatment of the radiative transfer
(M1 method using eight energy groups versus a two-moment solver using 10$^5$
energy groups) and of the thermodynamic state of the gas (LTE versus non-LTE) ---
past 200\,d, the bolometric luminosity predicted by the two codes differs sizeably (about a factor of 2),
probably because non-LTE effects are stronger (the CDS is no longer embedded within the optically
thick CSM).
A mismatch is expected during the first 1-2 months because light-travel time effects
are ignored in the computation of the flux in \cmfgen\ (the time dependence of the model is
implicitly taken into account since we fix the temperature, but no explicit time dependence
is accounted for when computing the flux in the formal solution of the radiative transfer equation
in \cmfgen). This matters when the properties of the
interaction vary on a short time scale (i.e., comparable to the free-flight time of $\sim$8\,d
to the outer boundary, where the luminosity is recorded).

The slow evolution of the SED reflects the slowly varying shock luminosity
(which arises from the smooth variation in both the ejecta density/velocity
profile and from the slowly varying CSM density for a wind configuration),
leading to a decrease in radiation/gas temperature in the optically thick regions of the CSM.
Consequently, the electron scattering wings weaken as the ionisation level and the electron
scattering optical depth decrease (this is caused by the shrinking of
the spectrum formation region, which is bounded by the fixed photosphere and the
outward moving dense shell or shock).
For about 200\,d after the onset of the interaction, the spectrum formation process is
essentially unchanged in this simulation.

\begin{figure}
\epsfig{file=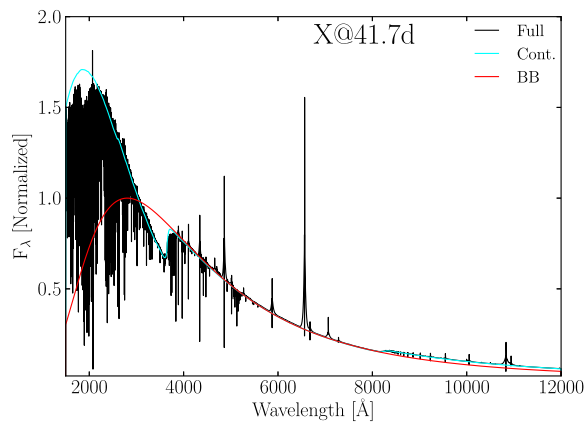,width=8.5cm}
\caption{
Comparison of the total synthetic flux (black), the continuum synthetic flux (blue)
and the best match blackbody flux to the optical range (red), for the reference model X
at 41.7\,d after the onset of interaction.
In the same order, the cumulative luminosities are 5.83$\times$\,10$^9$\,\lsun,
6.18$\times$\,10$^9$\,\lsun, and 4.45$\times$\,10$^9$\,\lsun.
The blackbody radius and temperature are 1.45$\times$\,10$^{15}$\,cm and 10331.6\,K.
The photospheric radius and temperature at that time are 7.10$\times$\,10$^{15}$\,cm and 12670.0\,K.
\label{fig_comp_bb}
}
\end{figure}

It is customary in the community to compare the SED with a blackbody. At each epoch,
we fit the \cmfgen\ SED flux ratio between 4800 and 8000\,\AA\  to infer $T_{\rm bb}$ --- we use
an iterative procedure to find the temperature for which the model and the blackbody have between
these two wavelengths the same flux ratio to within one part in 10000). We determine the corresponding blackbody
radius $R_{\rm bb}$ by matching the blackbody to the model luminosity $L_{\lambda}$ at $\lambda$
(i.e., through $R_{\rm bb}^2 = L_{\lambda} / 4\pi^2 B_{\lambda}(T_{\rm bb})$).
The total blackbody luminosity \lbb\ is then $4 \pi R_{\rm bb}^2 \sigma T_{\rm bb}^4$.
An example is shown in Fig.~\ref{fig_comp_bb}.
Although the match is excellent in the fitted range, the SED is not a blackbody.
We find non-planckian effects here associated with dilution due to electron scattering
(causing $R_{\rm bb} < R_{\rm phot}$; see, e.g., \citealt{DH05b}), and flux excess rather than blanketing
in the UV (as if multiple blackbody emitters with different temperatures were contributing).

\begin{figure}
\epsfig{file=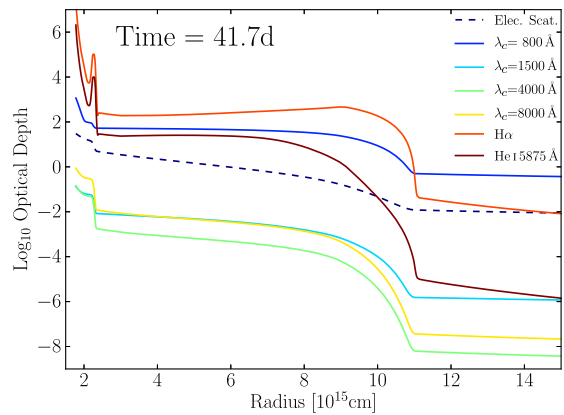,width=8.5cm}
\caption{
Radial variation of the optical depth associated with electron scattering, various continua,
and lines of H$\alpha$ and He\one\,5875\,\AA\ (for which we plot the Sobolev optical depth).
Thermalisation in the  Lyman continuum is ensured, but not in the Balmer and Paschen
continua. The large Sobolev length caused by the small velocity gradient in the CSM causes
the H$\alpha$ line optical depth to be very large, and to exceed that in the Lyman continuum.
\label{fig_tau}
}
\end{figure}

\begin{figure*}
\epsfig{file=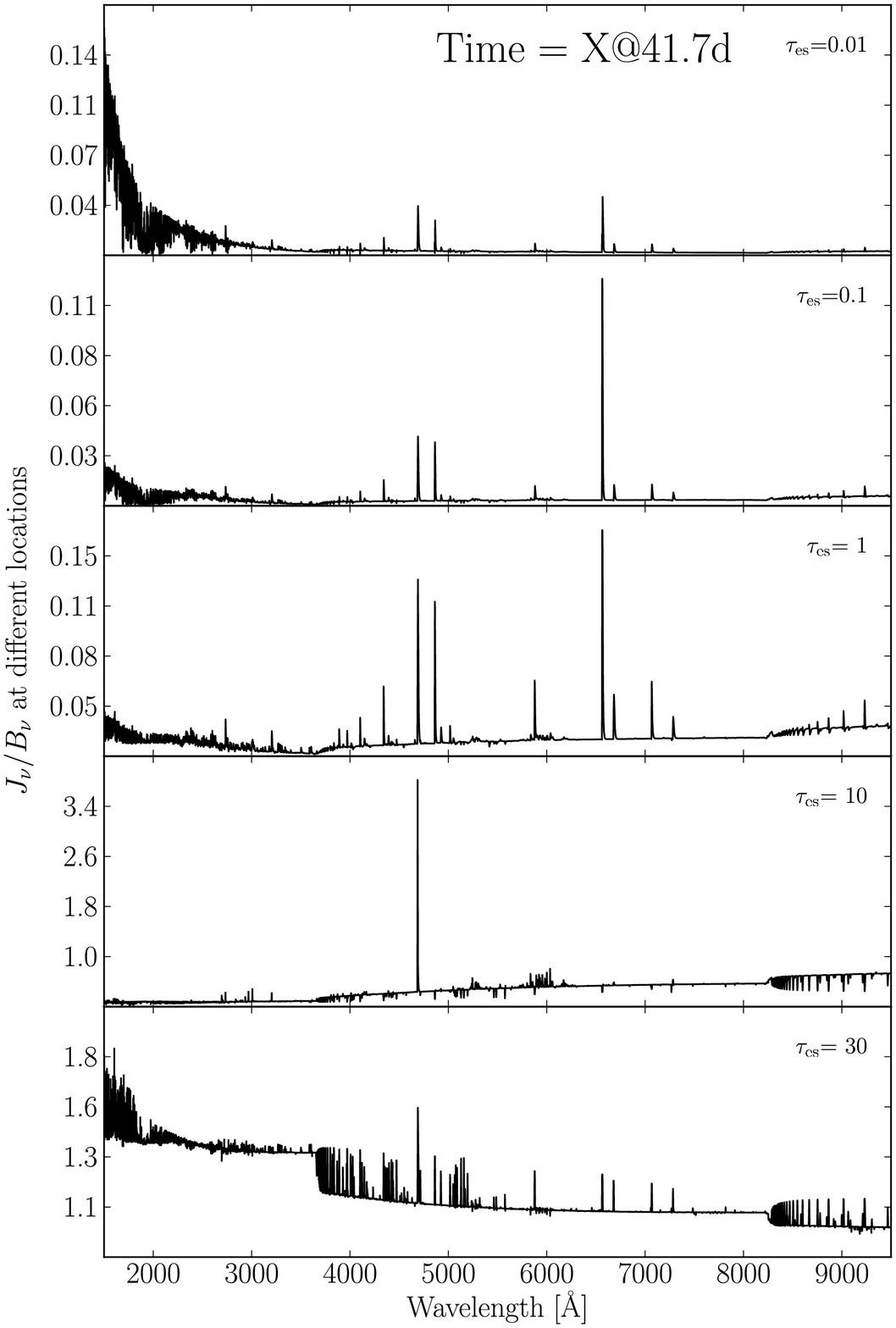,width=8.5cm}
\epsfig{file=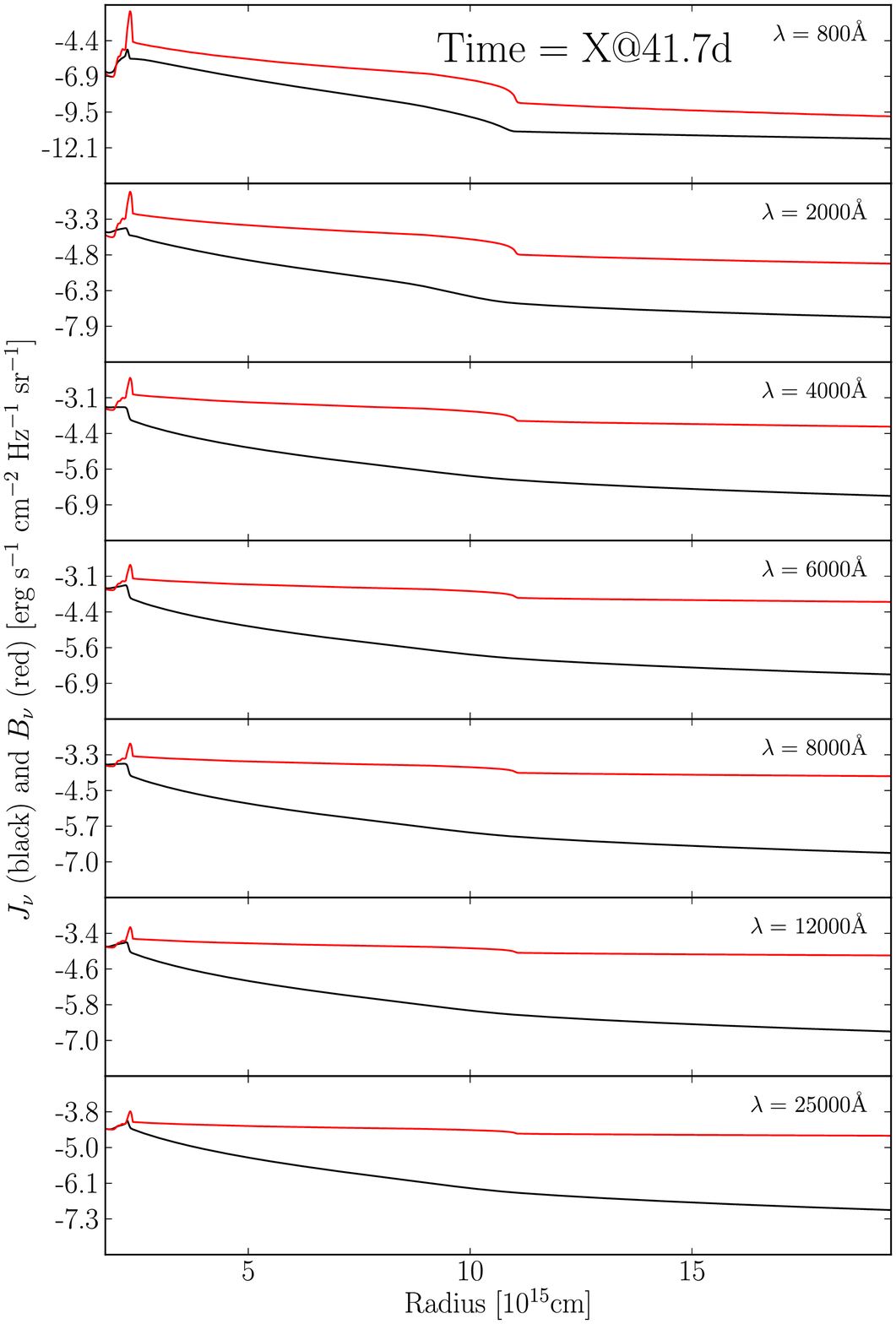,width=8.5cm}
\caption{{\it Left:} Variation of the ratio of the mean intensity $J_{\lambda}$ and the Planck function
$B_{\lambda}$ in the simulation at 41.7\,d, shown as a function of wavelength and
ejecta/CSM location. The label at right gives the electron-scattering optical depth $\tau_{\rm es}$
associated with each panel. At this time, the CDS is at $\tau_{\rm es}\sim$\,10.
{\it Right:} Variation of $J_{\lambda}$ and $B_{\lambda}$ versus radius and shown for various
continuum wavelengths.
\label{fig_j_and_b}
}
\end{figure*}

We show the optical depth for electron scattering, selected continuum wavelengths,
as well as selected lines in Fig.~\ref{fig_tau}. It is evident that thermalisation is complete for
UV photons in this model, but photons redder than the Lyman edge poorly thermalise,
even at an early epoch of 41.7\,d after the onset of interaction.
For these, electron scattering is the main source of opacity.
The blackbody radius underestimates the radius of the CDS at all times, although the bulk of the SN flux
arises from outside the CDS. Overall, the radiation field in this SN configuration is not well described
by the Planck distribution. These considerations are confirmed by Fig.~\ref{fig_j_and_b},
which illustrates how much smaller the mean intensity is compared to the Planck function, except at the location
of lines. Furthermore, in our reference model, the bulk of the flux arises at all times from layers
where the electron scattering optical depth is never greater than 10, and more typically a few only
(see also Fig.~\ref{fig_snap}).

\begin{figure*}
\epsfig{file=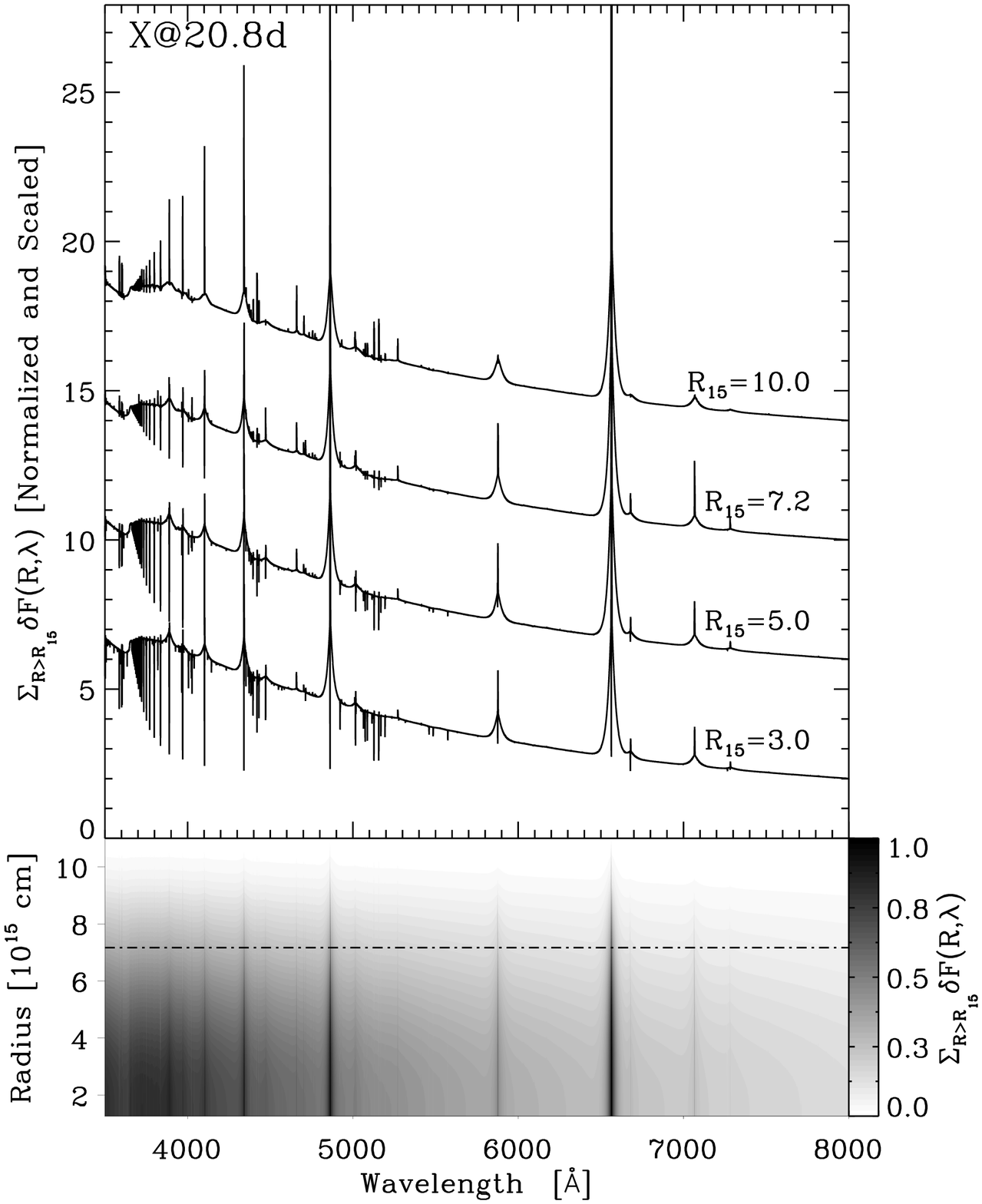,width=8.5cm}
\hspace{0.5cm}
\epsfig{file=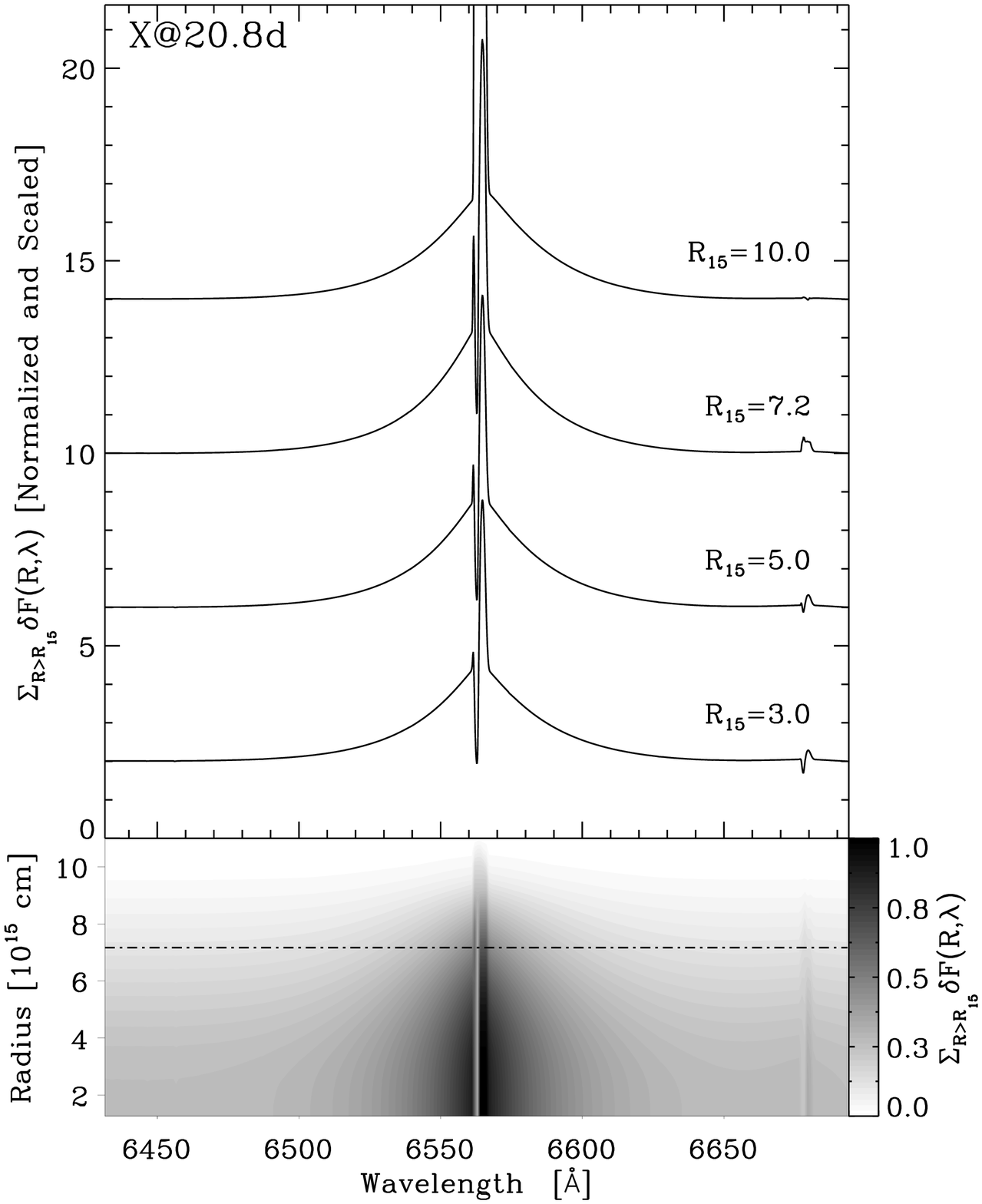,width=8.5cm}
\caption{{\it left:} Illustration of the wavelength ($\lambda$) and depth ($R$)
dependence of the quantity $\sum_{R>R_{15}} \delta F(R,\lambda)$, where
$\delta F(R,\lambda) = (2\pi/D^2) \int_{\Delta R} \, \Delta z  \, \eta(p, z,\lambda) \, e^{-\tau(p, z,\lambda)} p dp$,
and $R_{15}$ is $R$ in units of 10$^{15}$\,cm.
Here, $D$ is the distance; $\Delta R$ and $\Delta z$ are the shell thickness in the radial direction and along the ray with
impact parameter $p$, respectively; $\eta(p, z,\lambda)$  and $\tau(p, z,\lambda)$ are the emissivity and the
ray optical depth at location $(p,z)$ and wavelength $\lambda$.
The gray scale in the bottom panel shows how $\sum_{R>R_{15}} \delta F(R,\lambda)$ varies as we progress inwards
from the outer boundary, indicating the relative flux contributions of different regions.
If we choose $R_{15}$ as the minimum radius on the \cmfgen\ grid, we recover the total flux.
The grayscale also shows how lines increasingly broaden (and the peak line flux decreases) as we progress outwards.
The dash-dotted line corresponds to the radius of the electron-scattering photosphere,
and the dashed line (which overlaps with x-abscissa here) corresponds to the CDS radius.
The top panel shows selected cuts (see right label) of the quantity $\sum_{R>R_{15}} \delta F(R,\lambda)$.
From top to bottom ($R_{15}=$\,10, 7.2, 5.0, 3.0), $\sum_{R>R_{15}} \delta F(R,\lambda)$ represents 4, 45, 77, and 98
per cent of the total emergent flux at 6800\,\AA.
{\it Right:} Same as left, but now zooming in on the H$\alpha$ region. The depth variation of the line width, influenced by
electron scattering, is evident in the bottom panel.
\label{fig_dfr_d20}
}
\end{figure*}

\begin{figure*}
\epsfig{file=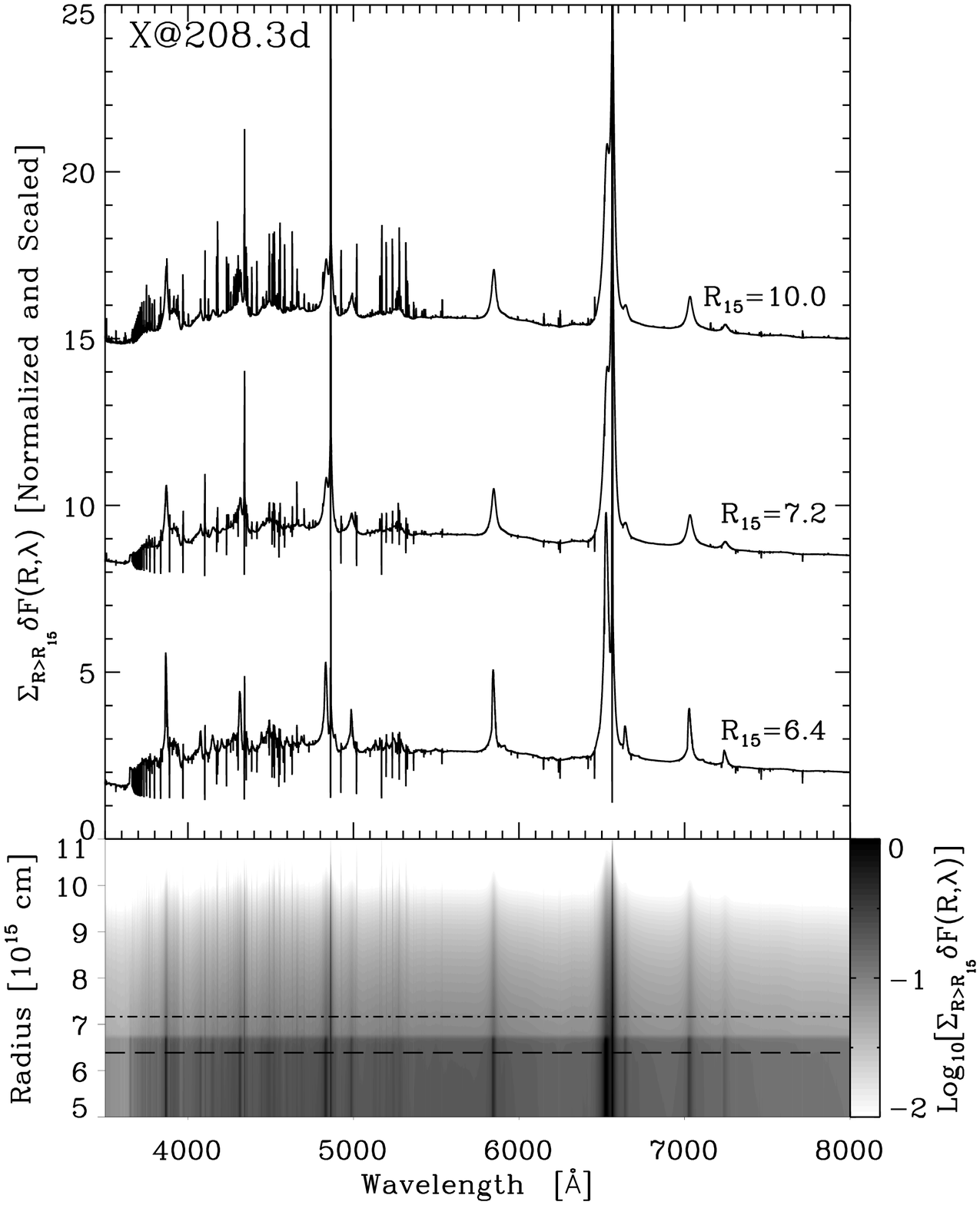,width=8.5cm}
\hspace{0.5cm}
\epsfig{file=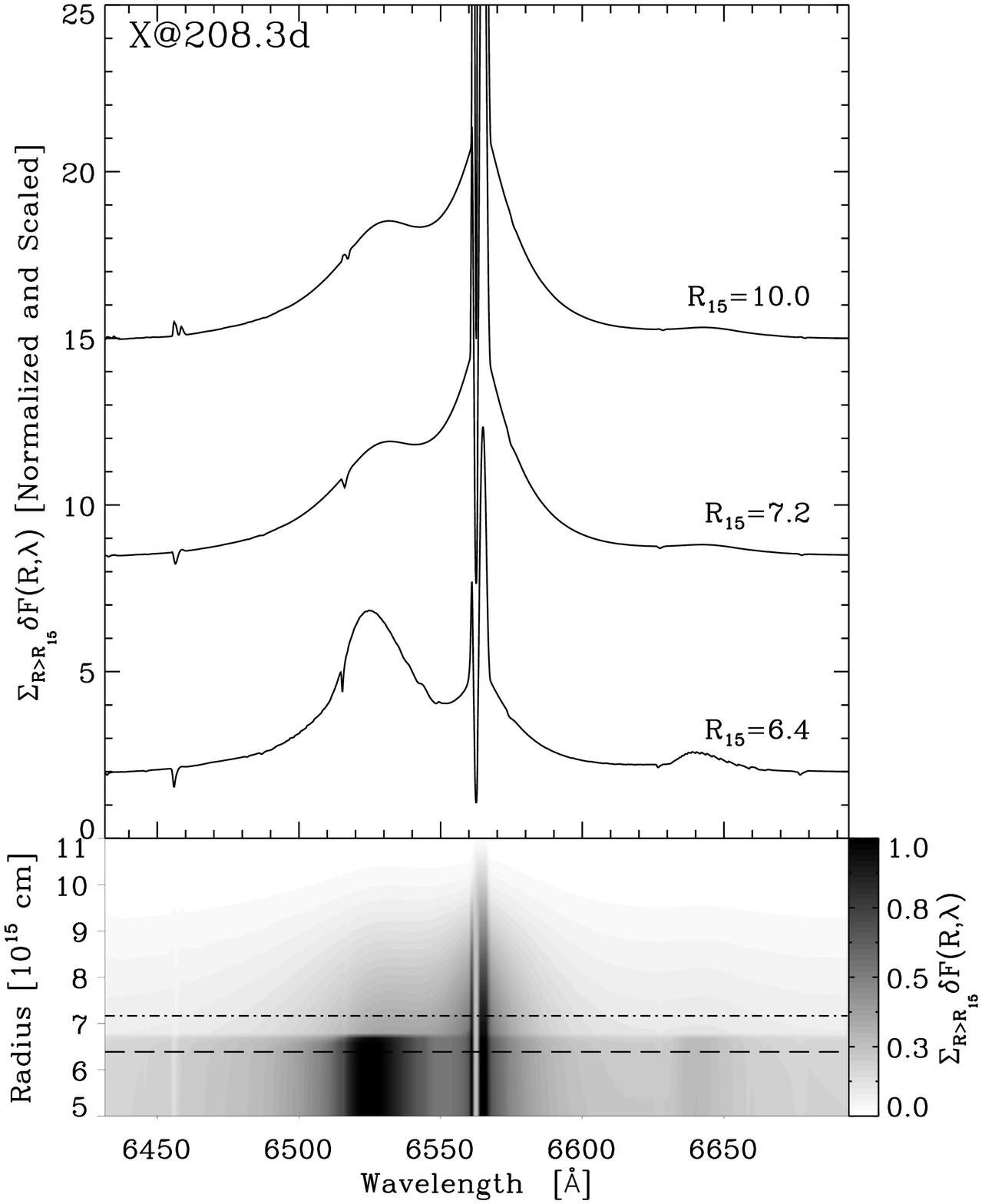,width=8.5cm}
\caption{Same as Fig.~\ref{fig_dfr_d20}, but now for a time of 208.3\,d after the onset
of interaction. In the left column, the greyscale is shown on a logarithmic scale.
From top to bottom ($R_{15}=$\,10, 7.2, 6.4), $\sum_{R>R_{15}} \delta F(R,\lambda)$ represents 6, 41, and 94
per cent of the total emergent flux at 6800\,\AA.
The flux contributions from the fast moving CDS and the slow moving CSM
are clearly visible. The emission from the CDS is blue shifted, while the CSM emission
is symmetric around line centre. Electron scattering affects the emitted photons from both components.
\label{fig_dfr_d208}
}
\end{figure*}

The shallow density structure, the wavelength-dependent opacities, and the strong albedo,
lead to a very spatially extended spectrum formation region. For example, the emission/absorption of
H$\alpha$ photons occurs throughout the grid, with the exception of the outermost regions of the CSM
where the density is very low.
At early times observed line photons primarily originate at larger radii because line photons emitted
deeper suffer more scatterings, and hence have a greater probability of being absorbed.
Line photons that could accrue very large frequency shifts
by multiple scattering with thermal electrons tend to be absorbed before they can escape.
In our reference model, we find that the bulk of H$\alpha$ photons arise from regions with an electron
scattering optical depth of a few, giving rise to emission in line wings extending to $\sim$\,3000\,\kms\ from
line centre, and characterised by a profile with a FWHM of $\sim$1500\,\kms. This is somewhat similar to
values in SN\,2010jl \citep{zhang_etal_12} at early epochs. However at later times the profiles
in SN\,2010jl are broader and blue shifted.  We also see this effect at late times, with the extra emission
associated with the CDS (Fig.~\ref{fig_spec_halpha}). The blueshift of the broad H$\alpha$ line is caused
by an optical depth effect that affects all SNe \citep{DH05a,anderson_etal_14}.
At early times, the peak blueshift is not seen because the H$\alpha$ photons emitted from the CDS
are unable to escape -- they undergo numerous electron scatterings and are eventually destroyed
by a continuum absorption.
All these properties are illustrated in Figs.~\ref{fig_dfr_d20}--\ref{fig_dfr_d208}.

 As found in earlier studies (e.g., \citealt{hillier_91}) the adopted turbulent velocity also affects the strength of the
electron scattering wings. This occurs since a larger turbulent velocity increases the Sobolev
length, and hence leads to an increased probability that a line photon scattering within the Sobolev
resonance zone will experience an electron scattering interaction, which, because of the large
Doppler shift, will cause it to be shifted out of resonance allowing it to escape.

\section{polarisation calculations}
\label{sect_pol}

  The dense shell that forms in the interaction is Rayleigh-Taylor unstable and should be simulated in 3-D --- this
  is left for future work. There is, however, observational evidence from spectropolarimetric observations that some
  SNe IIn are asymmetric on large scale \citep{leonard_etal_00,wang_etal_01,hoffman_etal_08,patat_etal_11}.
  This we try to address in this section by performing 2-D polarisation calculations.

We post-process the \cmfgen\ simulation for model X at 41.7\,d
by computing linear polarisation profiles in the same fashion as described
in \citet{DH11b}. We break spherical symmetry by enforcing a latitudinal
variation in density (thus preserving axial symmetry).
In practice, we use $\rho(r, \mu) = \rho(r, \mu=0) (1+A_1 \mu^2)$, where $\mu \equiv \cos\theta$,
and $\theta$ is the colatitude.
The magnitude of the asymmetry is controlled through the parameter $A_1$, which we
vary from  0.1, to 0.2, 0.4, 0.8, and 1.6.
The resulting axially symmetric ejecta has therefore a prolate morphology, with pole-to-equator density
ratios (at a given radius) of 1.1, 1.2, 1.4, 1.8, and 2.6.
This adopted type of asymmetry is chosen for its simplicity -- other choices might also yield similar
continuum and line polarisations.

\begin{figure}
\epsfig{file=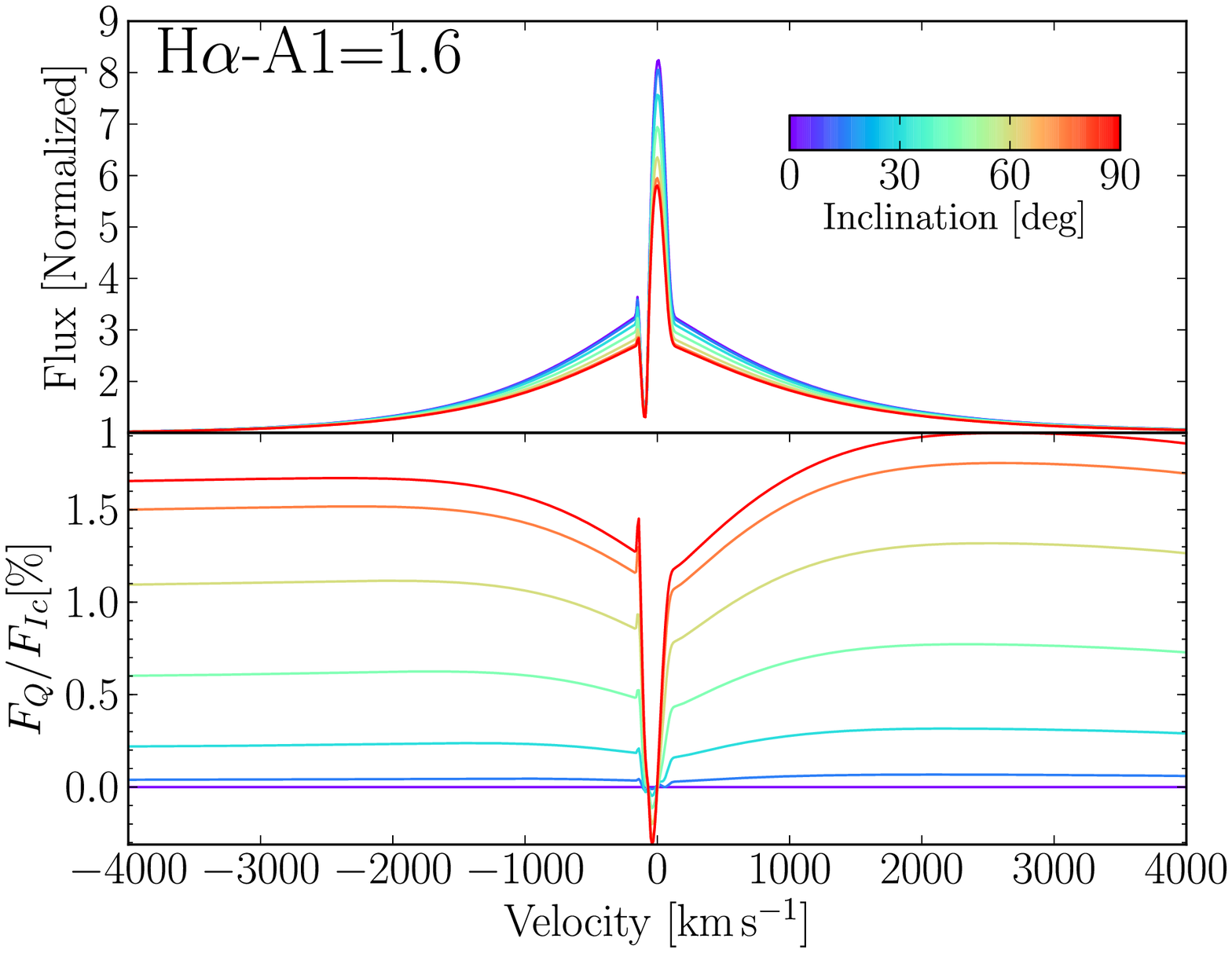,width=8.5cm}
\epsfig{file=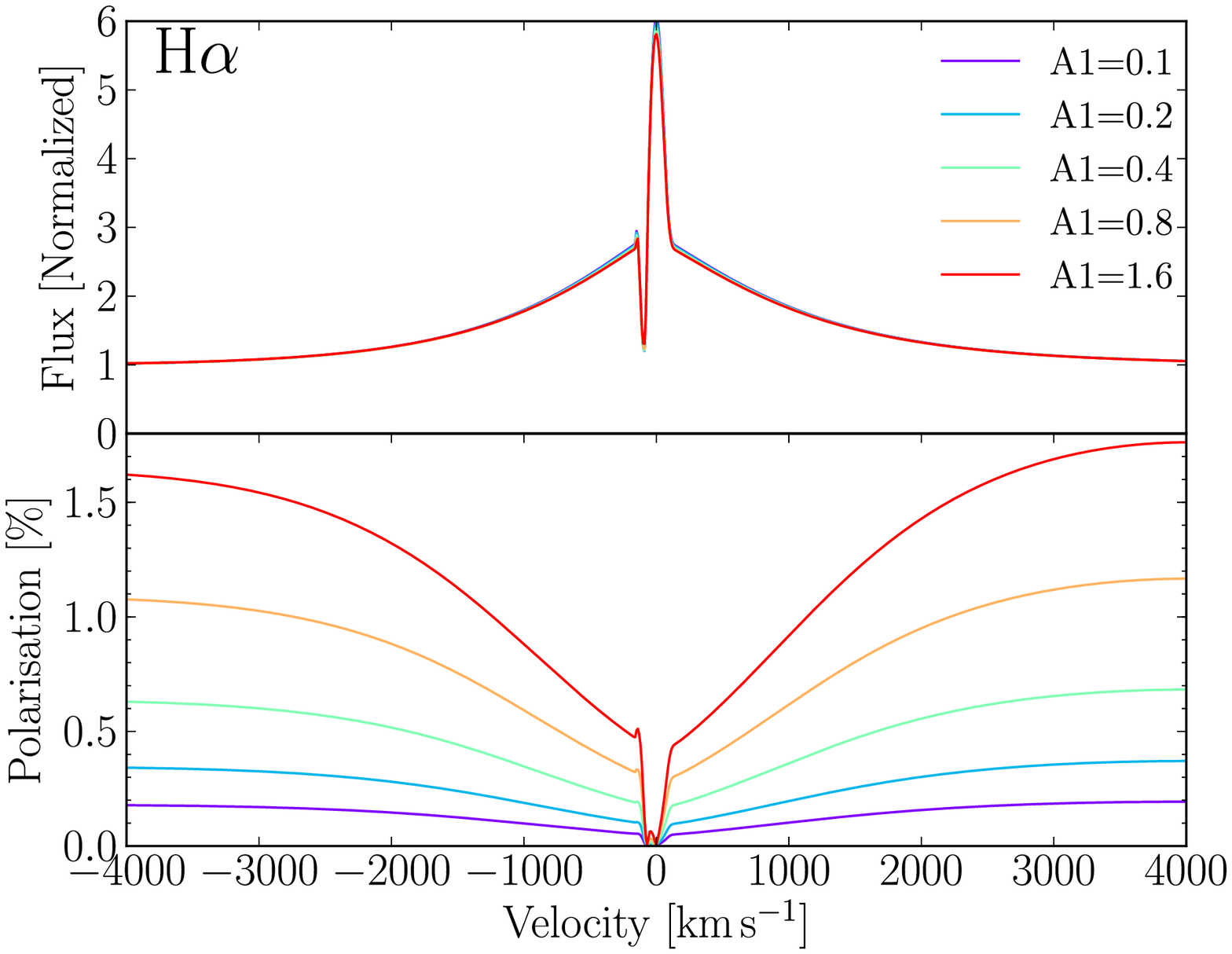,width=8.5cm}
\epsfig{file=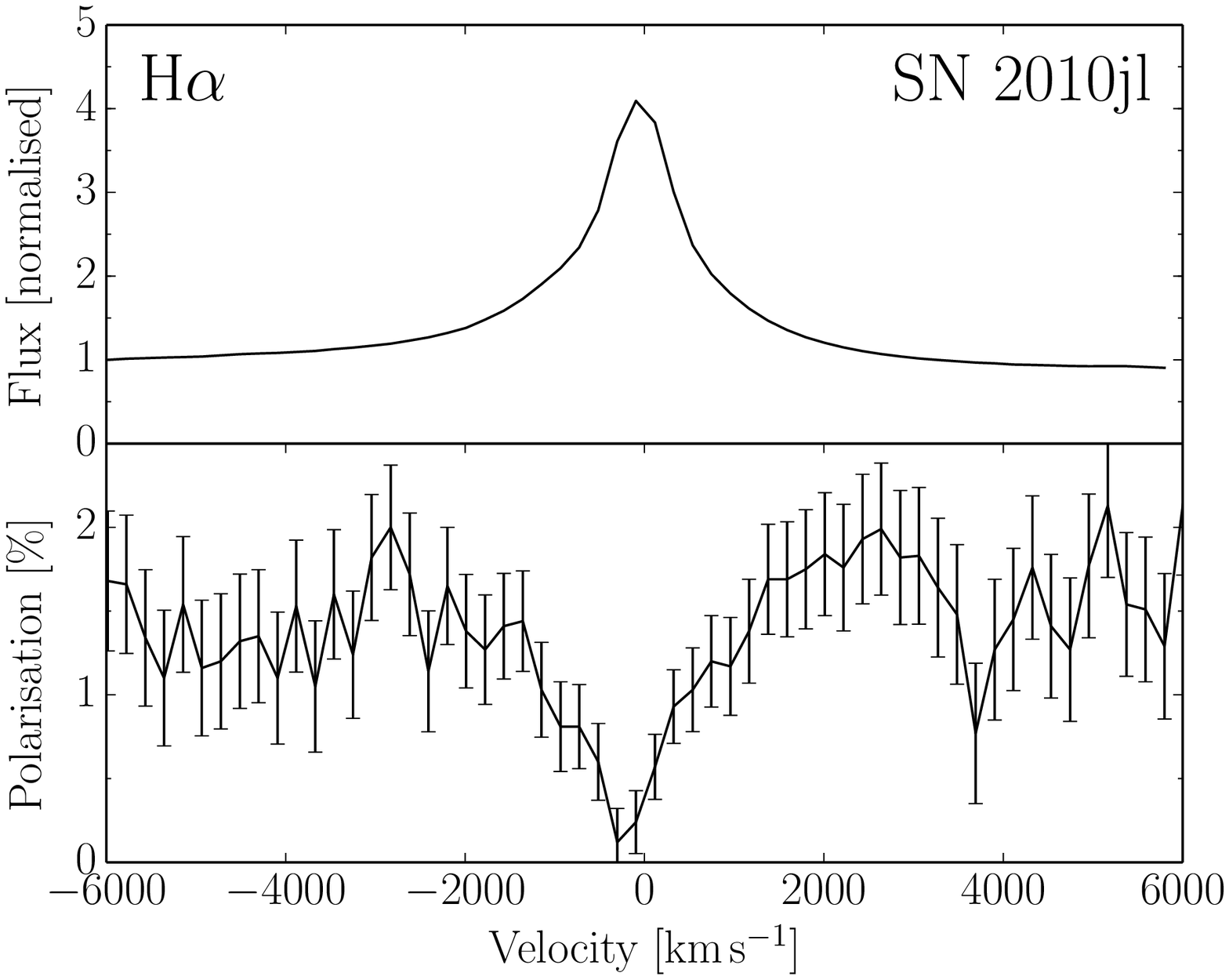,width=8.5cm}
\caption{
{\it Top:}  Total and polarised flux across the H$\alpha$ line and the overlapping continuum for
different inclinations (90$^\circ$ corresponds to an equatorial view), at 41.7\,d
after the onset of the interaction (hence close to the epoch of maximum light),
showing the larger continuum polarisation, the depolarisation across
the line profile, and the unpolarised line core. The prolate asymmetry is characterised by a pole-to-equator density
ratio of 2.6 ($A_1=$\,1.6).
{\it Middle:} Same as top, but now showing the percentage polarisation for different magnitudes of the asymmetry
($A_1$ values between 0.1 and 0.6) and for an equatorial view.
{\it Bottom:} Spectro-polarimetric observations of the H$\alpha$ line for SN\,2010jl
on November 18.2UT  \citep{patat_etal_11}, which is within a few days of our model time (left panel).
In the context of our model, the observed polarisation may be interpreted as arising from a prolate CSM with
a pole-to-equator density ratio of $\sim$\,3.
\label{fig_pol}
}
\end{figure}

\begin{figure}
\epsfig{file=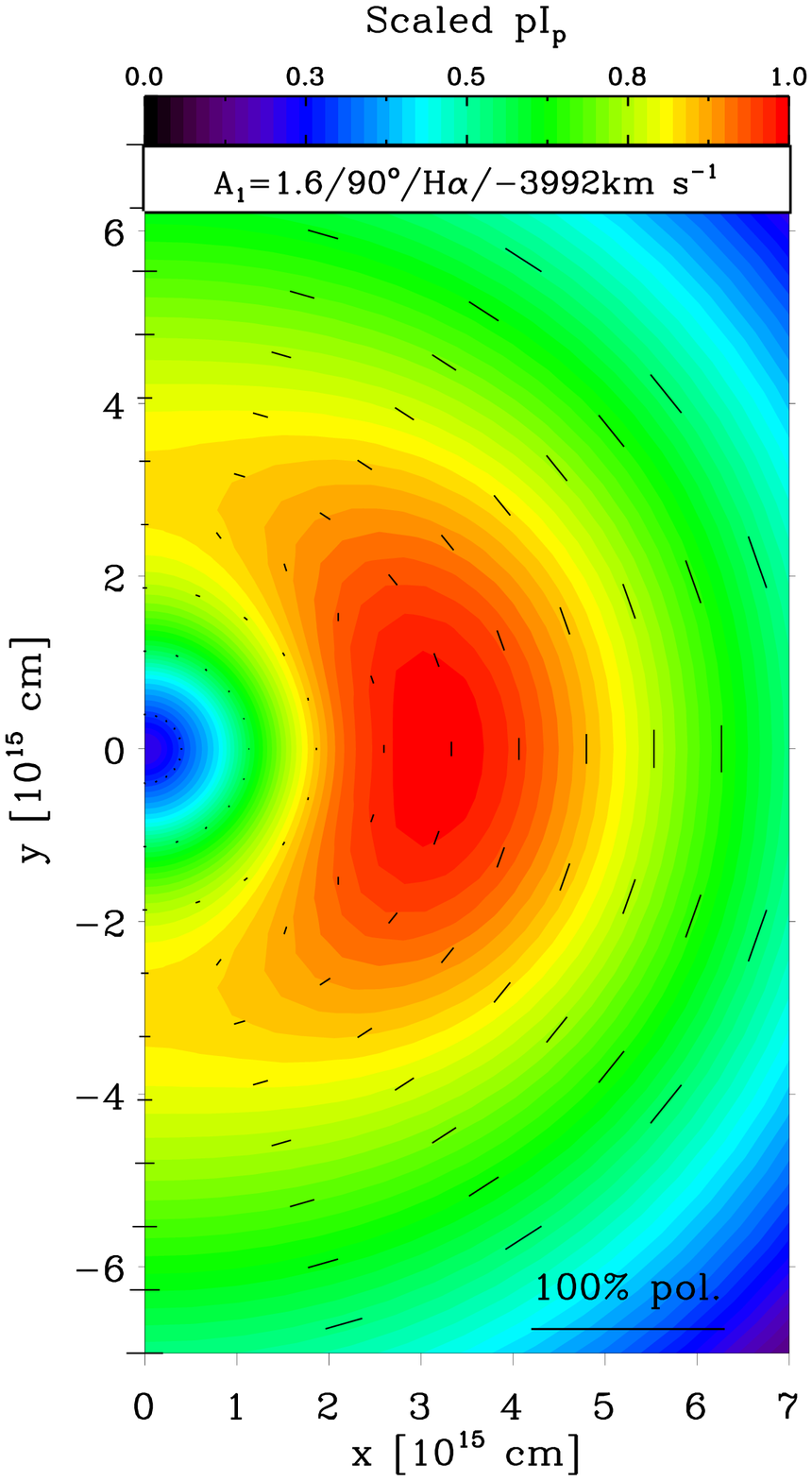,width=8.5cm}
\caption{
Colour map of the specific intensity (scaled by the impact parameter $p$) on the plane of the sky
at $-$3992\,\kms\ from the H$\alpha$ line centre (this Doppler velocity corresponds to a continuum region ---
see top panel of Fig.~\ref{fig_pol}),
for the model with $A_1=1.6$, at 41.7\,d after the onset of
the interaction, and seen at a 90$^\circ$ inclination (i.e. edge-on).
We overplot segments (black bars) showing the local polarisation strength and orientation
(the length corresponding to 100 per cent polarisation is shown at the bottom right).
\label{fig_pol_map}
}
\end{figure}

We show the flux in the H$\alpha$ line and in the overlapping continuum in Fig.~\ref{fig_pol} for
an observer along a line-of-sight perpendicular to the axis of symmetry. Along this direction, the
continuum polarisation has its maximum value, reaching about 1.6\% for $A_1=$\,1.6.
As we progress towards the H$\alpha$ rest wavelength from continuum regions (i.e., $\gtrsim$\,3000\,\kms\
from line centre), the polarisation decreases steadily, as observed in SN\,2010jl \citep{patat_etal_11,williams_etal_14}
and in some Wolf-Rayet stars \citep{schulte_ladbeck_etal_91,harries_etal_98}.
In our simulation, this reduced polarisation across the line is a signature of the distinct formation of the continuum
and the line. Because of the lower optical depth, continuum photon arise from
a much greater electron-scattering optical depth than do H$\alpha$ line photons (Fig.~\ref{fig_tau}).
Furthermore, line photons emitted in a recombination process start off with zero polarisation,
and therefore can only accumulate a lower polarisation through a reduced number of scatterings before escape.
This depolarisation is maximum within 100\,\kms\ of line centre since over this spectral range, one sees
photons that underwent essentially no scattering with free electrons
(photons scattered by free electrons receive a wavelength-shift and, statistically, end up somewhere in the wings;
see discussion in \citealt{dessart_etal_09}).
In this context, whatever the imposed deformation on the ejecta, the polarisation at line centre
is zero.\footnote{Equivalent simulations for Type II SNe may yield a residual polarisation at line centre
because these photons are emitted at a non negligible electron-scattering optical depth \citep{DH11b}.
Spectrum formation in SNe II-P and SNe IIn is drastically different.}

Surprisingly the polarisation is parallel to the major axis (i.e., positive; see top panel of Fig.~\ref{fig_pol})
--- this property is, naively, not expected  since scattering of the light from a point source by an optically thin prolate spheroid
has the polarisation at right angles to the major axis \citep{brown_mclean_77}. In our convention
(and that of \citealt{brown_mclean_77}), the polarisation is positive when it is parallel to the axis of symmetry
(for further details, see Section~2  of \citealt{DH11b}). The reason for this behaviour in the current simulation is that the CSM is
optically thick, and the observed (and hence scattered) flux predominantly escapes from equatorial regions,
where the densities are lowest (Fig~\ref{fig_pol_map}).

The correspondence  between our simulations and the morphology of the polarised and of the total fluxes of SN\,2010jl
suggests first that our model for super-luminous SNe IIn like 2010jl is adequate. Secondly, the observed polarisation
implies that the ionised CSM causing the polarisation is significantly asymmetric. If axially symmetric, a pole to equator
density ratio of $\sim$\,3 could explain the observed continuum polarisation. While the CSM is asymmetric, it is
however unclear whether the inner shell is also asymmetric --- it does not need to be.
Multi-dimensional radiation hydrodynamics simulations are needed to model the spatial properties of
the shocked CSM (for example, in the current context, the shock strength should also be latitude dependent, causing
variations in temperature, ionisation, velocities etc with angle), so that we can better characterise the nature
of the asymmetry at the origin of the observed polarisation in SNe IIn like 2010jl.

\section{Dependency on some model parameters}
\label{sect_dep}

   Varying the properties of either the CSM mass or the explosion energy can affect the
   emergent radiation. This has been discussed in the past for the bolometric luminosity or the colours
   (see, e.g., \citealt{chugai_etal_04,woosley_etal_07,van_marle_etal_10,moriya_etal_13a}).
   Here, we repeat such explorations to investigate the impact on spectral signatures as well.

   \subsection{Impact on the bolometric light curve}

   Taking model X as a reference, we produce models in which the SN ejecta kinetic energy
   is scaled by a factor of 3 (Xe3) or 10 (Xe10). We  produce models in which the CSM density
   is scaled by a factor of 3 (Xm3) or 6 (Xm6). We also consider combinations of both, with models Xe3m6
   or Xe10m6. Finally, we consider model Xe3m6r, where the high-density CSM extends to
   1.5$\times$10$^{16}$\,cm rather than 10$^{16}$\,cm only. Table~\ref{tab_mod_sample}
   summarises these model properties.

\begin{figure*}
\epsfig{file=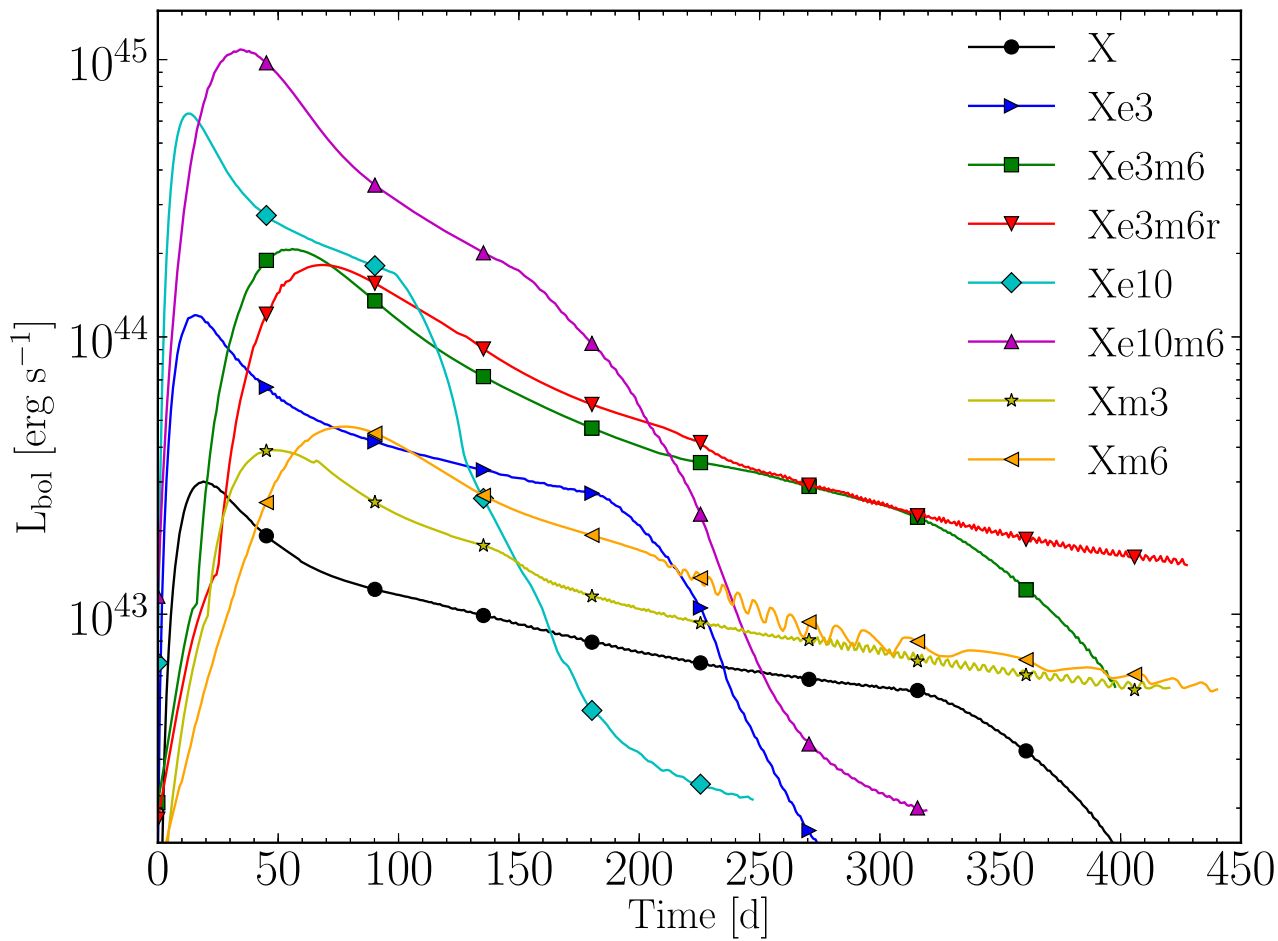,width=18cm}
\caption{Light curve diversity obtained by varying some of the parameters of our reference
model X (see details in Table~\ref{tab_mod_sample}). Models whose name contains e$n$
have a SN ejecta kinetic energy scaled by a factor $n$ relative to model X.
Models whose name contains m$n$ have a CSM mass loss rate scaled by a factor $n$ relative to model X.
Model Xe3m6r differs from the rest because the dense part of the CSM extends to 1.5$\times$10$^{16}$\,cm
rather than 10$^{16}$\,cm.
\label{fig_lc_dep}
}
\end{figure*}

   These alterations produce a considerable diversity of bolometric light curves (Fig.~\ref{fig_lc_dep}),
   in directions that are well understood.
   Increasing the kinetic energy causes a much greater shock luminosity at all times, but since the CSM
   structure is not changed, the diffusion time through the CSM is unchanged (the time to maximum is
   decreased by just a few days for higher SN ejecta kinetic energy). The bolometric light curves
   shift upward for increasing energy. The mean radiation-energy density in the CSM also increases
   with shock luminosity, which raises the radiation temperature.
   Colours are thus strongly affected, getting bluer for higher energy (see below and
   Table~\ref{tab_mod_sample} where we give the bolometric correction and the colour $V-I$
   at bolometric maximum).

   Varying the CSM mass (corresponding in our approach to a change in mass loss rate) modulates the impedance
   of the ``interaction engine". For larger CSM density and total mass, more kinetic energy is extracted from
   the SN ejecta. This energy is converted primarily into radiation.
   Consequently, varying the CSM mass affects the light curve properties. Higher CSM mass
   implies higher CSM optical depth and a longer diffusion time, hence a delayed
   time of peak. A larger CSM mass buffer implies more kinetic energy is tapped so the bolometric maximum
   is more luminous. In the model set X, Xm3, Xm6, the total energy radiated is 0.32, 0.49, and
   0.63$\times$\foe\ (for the same ejecta kinetic energy of \foe), so, the conversion
   efficiency can be considerably enhanced when the CSM mass exceeds significantly the SN ejecta mass,
   with values here as high as $\sim$\,70\% (models Xe3m6, Xe10m6, or Xm6).

   Combinations of higher CSM mass and higher ejecta kinetic energy (e.g., model Xe3m6) ) shift the reference
   model (quite suitable to SN\,2010jl) to the domain where SN\,2006gy lies \citep{smith_etal_07a},
   with peak luminosities of the order of a few 10$^{44}$\,erg\,s$^{-1}$ sustained for 2-3 months
   and a peak time that occurs later ($\sim$2 months after the onset of the interaction).

  Finally, extending the region of high density in the CSM from 10$^{16}$ to 1.5$\times$10$^{16}$ \,cm
  (models Xe3m6r versus model Xe3m6) leads to a sustained
  high luminosity beyond 300\,d. In fact, all the light curves in Fig~\ref{fig_lc_dep} show a break after 150-300\,d, except for
  model Xe3m6r. This light curve break in our simulations is associated with the shock (or the CDS) reaching
  the outer edge of the denser parts of the CSM (the shock in model Xe3m6r has not yet reached 1.5$\times$10$^{16}$ \,cm
  at the last time shown in Fig.~\ref{fig_lc_dep}) and starting to encounter lower density material (see Table~\ref{tab_mod_sample}
  for details).

\begin{figure}
\epsfig{file=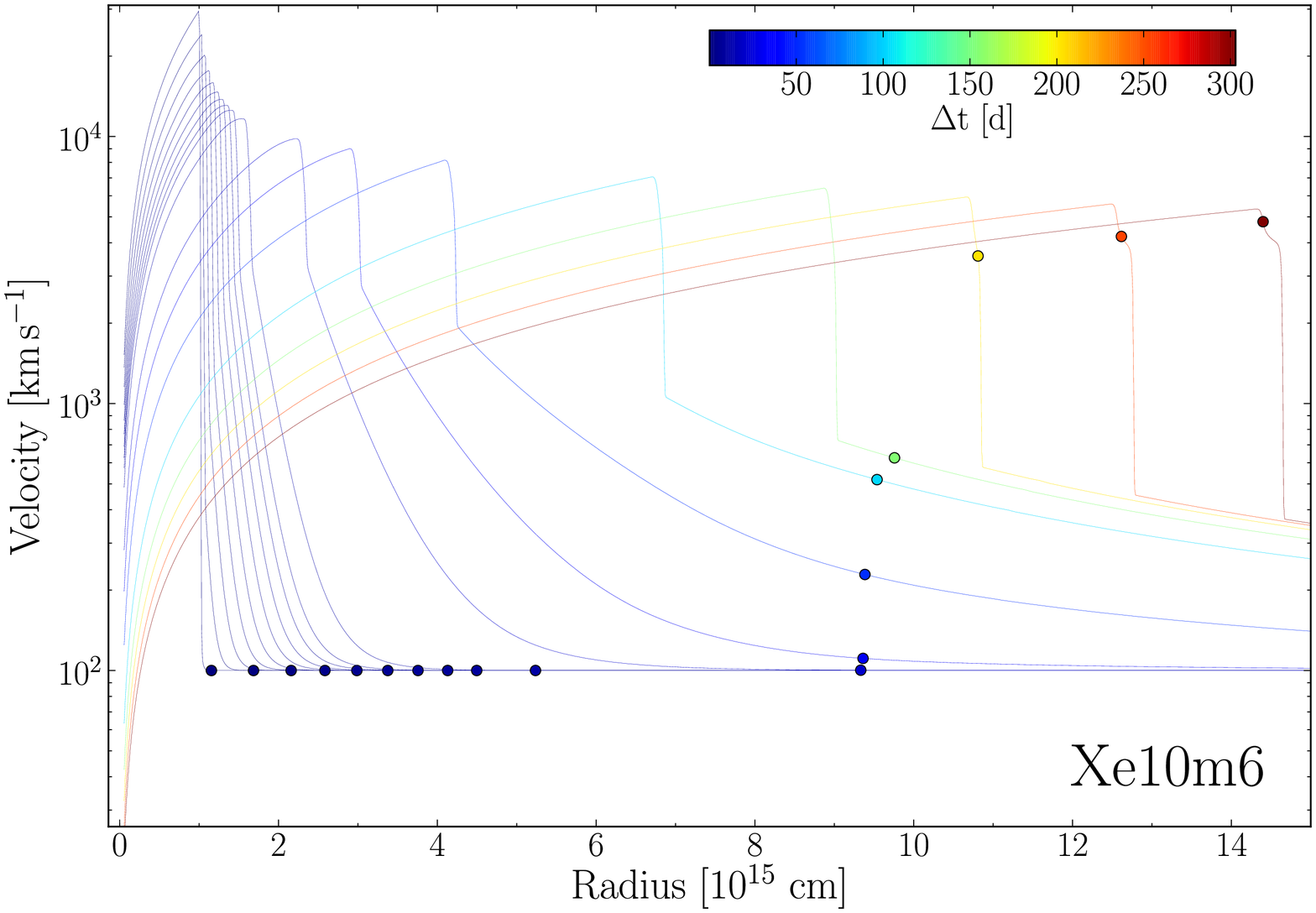,width=8.5cm}
\epsfig{file=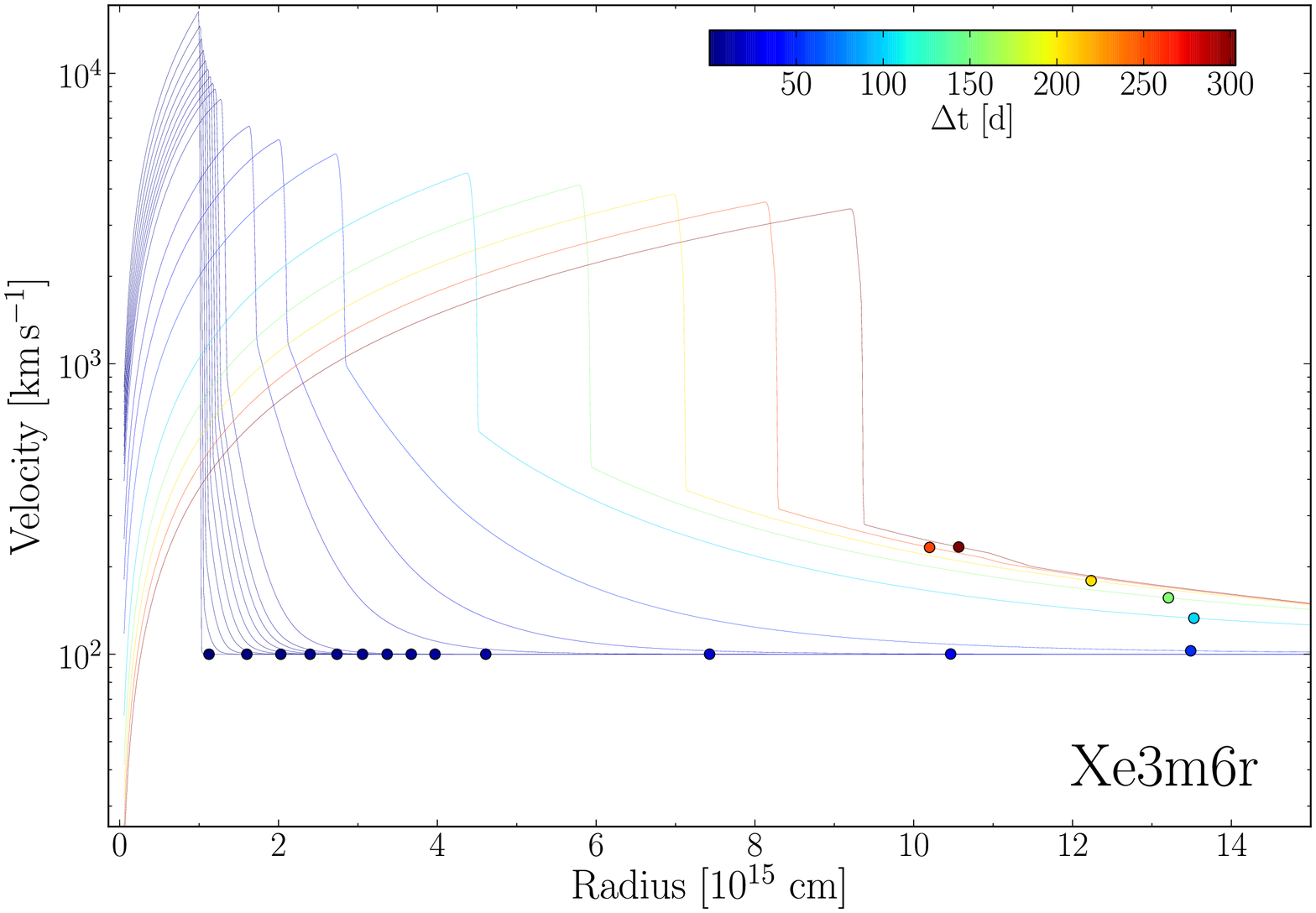,width=8.5cm}
\epsfig{file=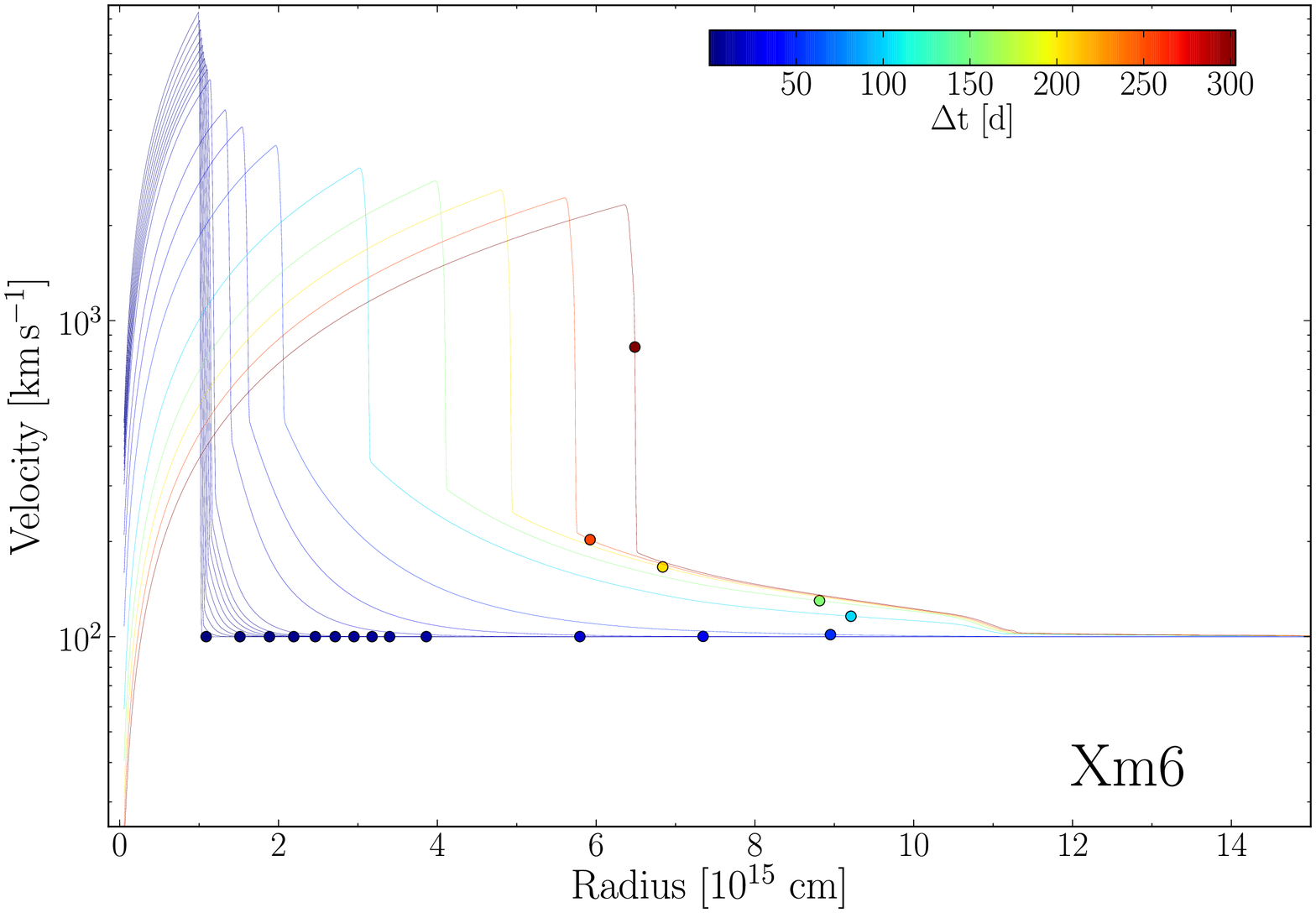,width=8.5cm}
\caption{Velocity profiles from the \heracles\ simulations for models Xe10m6 (top), Xe3m6r (middle), and Xm6 (bottom).
The epochs shown correspond to 0.01,1, 2, 3, 4, 5, 6, 7, 8, 10, 20, 30, 50, 100, 150, 200, 250, and 300\,d
after the onset of the interaction. Notice the pre-acceleration that takes place ahead of the shock.
\label{fig_evol_vel_dep}
}
\end{figure}

The pre-acceleration of the CSM by the radiation from the shock depends on the adopted configuration.
In model X, we obtained a modest enhancement in CSM velocity from 100 to 200\,\kms.
This pre-acceleration is stronger when we use a larger ejecta kinetic energy (stronger shock, higher
luminosity) or if the CSM density is increased (there is a delicate balance to get here since a higher CSM
density means a higher opacity and thus more radiation momentum extracted but it also implies
more inertia) --- see Fig.~\ref{fig_evol_vel_dep}.

\subsection{Impact on maximum light spectra}

\begin{figure*}
\epsfig{file=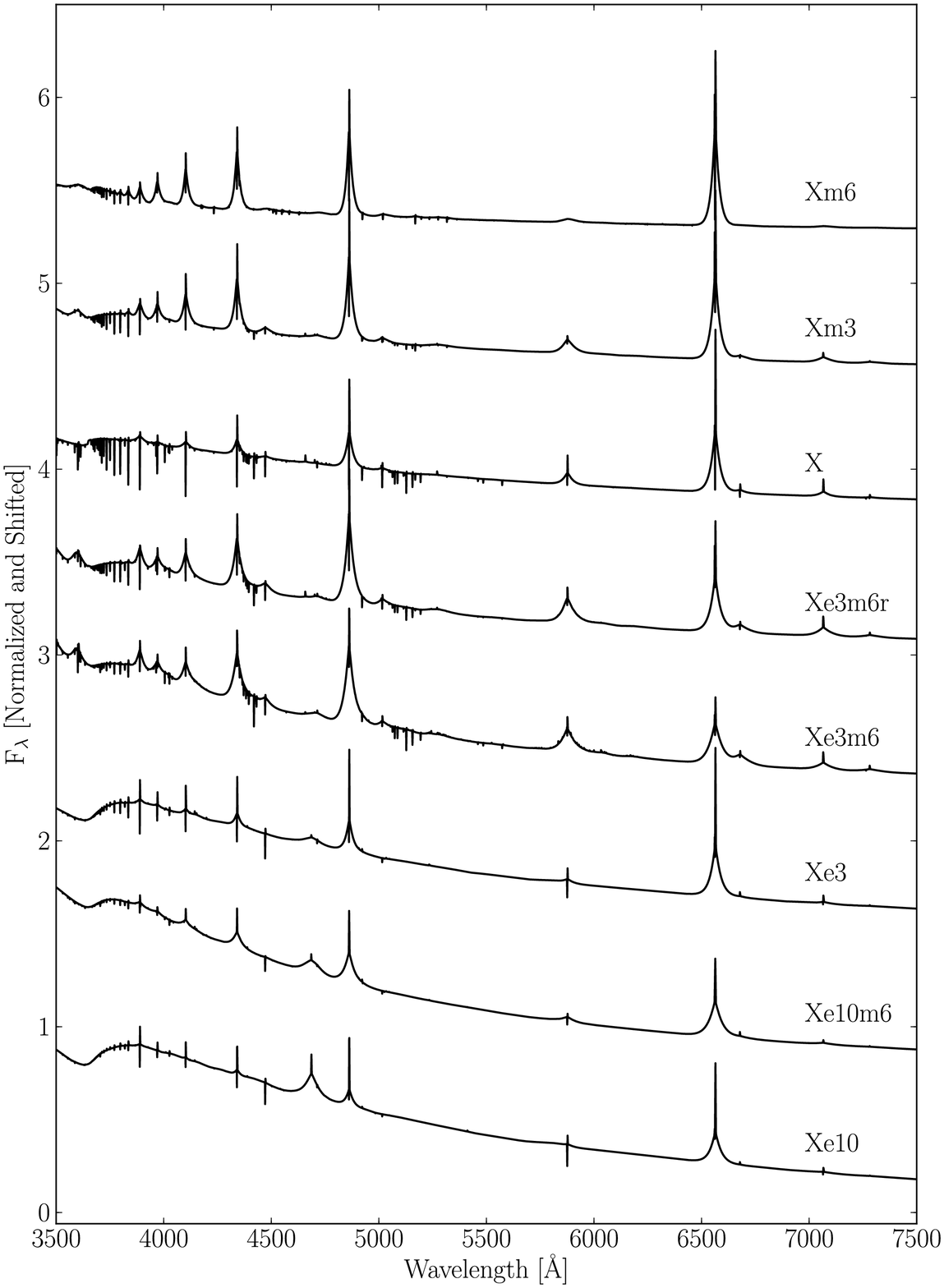,width=8.5cm}
\epsfig{file=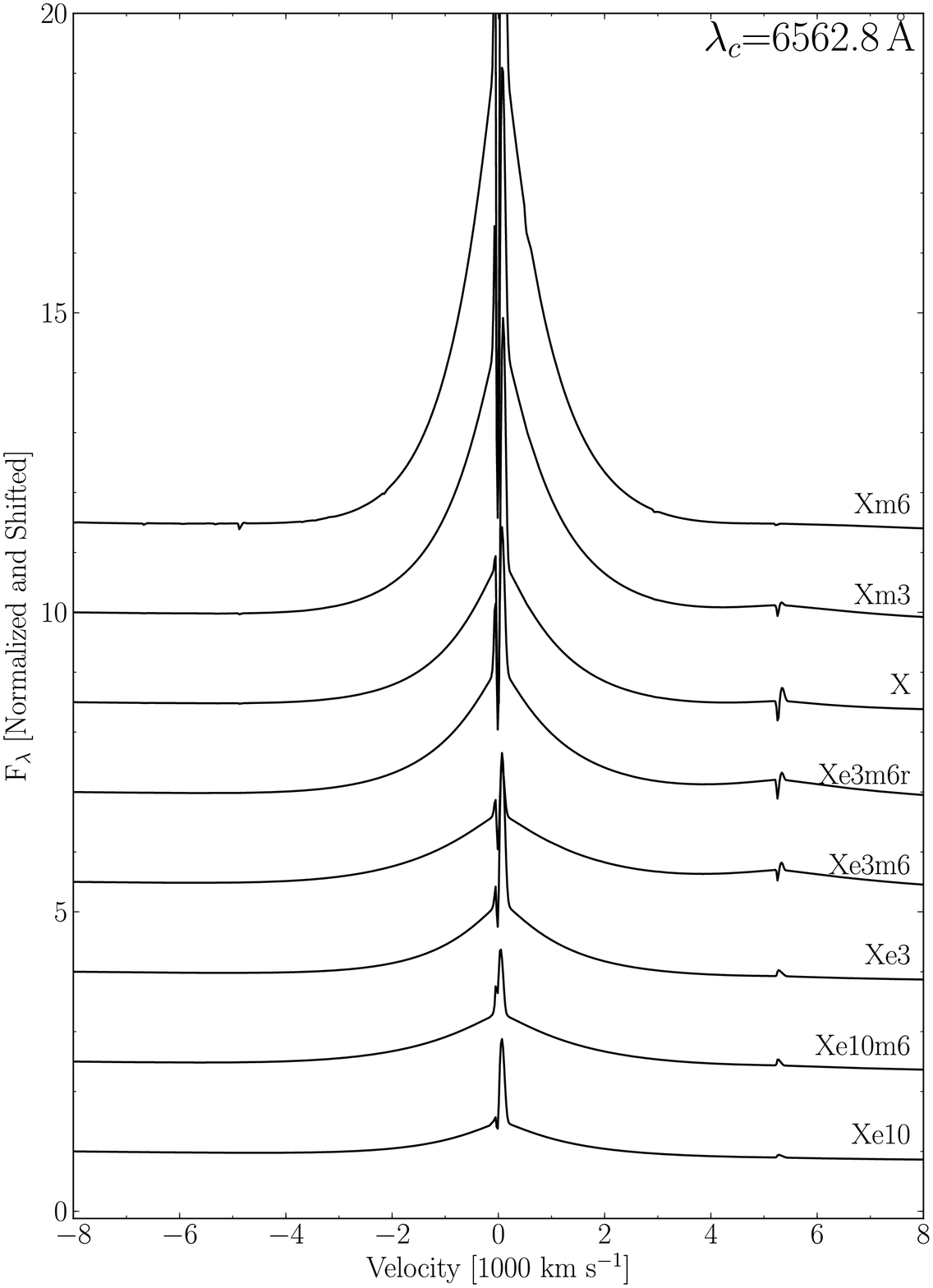,width=8.5cm}
\vspace{-1cm}
\caption{{\it Left:} Maximum light spectra for the interaction models with different ejecta
kinetic energy and/or CSM mass/density (see Table~\ref{tab_mod_sample} for additional
details).
Note the strong shift in colour for interaction models with higher ejecta kinetic energy.
{\it Right:} Same as left, but showing the H$\alpha$ line region, in velocity space.
Interestingly, even for very large CSM densities (and thus mass loss rates),
the maximum extent of the electron-scattering wings is 2000-3000\,\kms.
The much broader wings of H$\alpha$ in SN\,2010jl as time progresses suggest
that they do not stem exclusively from scattering with electrons but must instead contain
a contribution from Doppler-shifted line photons (i.e., emitted from the fast moving material
in the interaction region).
\label{fig_spec_dep}
}
\end{figure*}

   The diversity in bolometric light curves discussed in the preceding section is
also strongly present in the spectra at maximum light (Fig.~\ref{fig_spec_dep}).
For the highest energy explosion, the shift in temperature in the ionised CSM
leads to a major rise in ionisation for hydrogen and helium, producing very blue
and nearly featureless spectra  (whatever blanketing from Fe occurs in the UV and far-UV).
For models with enhanced CSM mass, the temperature/ionisation conditions are comparable
to reference model X, but the line fluxes are increased. The higher the CSM mass, the redder the colour
at peak.

Interestingly, in models with very high CSM mass, although the Balmer line fluxes are much higher,
the extent of the electron scattering wings does not increase beyond a maximum of  $\sim$\,3000\,\kms.
The likely reason is that line photons do not just scatter off free electrons in the CSM --- there is a
non-zero probability that they are absorbed in a bound-free or free-free process leading to photon destruction.
The broader and blue shifted lines seen at late times require that
we see photons blue shifted by expansion. For the present case, the broader blue-shifted component
seen at late times arises from the CDS. We see a blue-shift since the CDS absorbs emission from
the far side of the shell. Thus the H$\alpha$ blueshift seen in
SN\,2010jl \citep{smith_etal_12,zhang_etal_12,fransson_etal_14}
might simply just stem from an optical-depth effect, without any dust intervening in the process.
As the optical depth of the shell declines, we expect the profile to become more symmetric.

\section{Conclusion}
\label{sect_conc}

In this work, we have presented numerical simulations for interacting SNe in which
both the kinetic energy of the explosively-produced inner shell and the mass of the wind-produced
outer shell are very large. We used an inner shell kinetic energy of \foe\ or more, and a CSM mass
of at least 3\,\msun. These simulations correspond to possible configurations for super-luminous SNe IIn.

Our contribution is novel because we post-process our multi-group radiation-hydrodynamics
simulations with radiative transfer tools, to compute non-LTE spectra
and polarisation signatures. With this more global approach, we can study the numerous features
that make super-luminous SNe IIn unique.

 We find that the light curve of a super-luminous SN IIn is composed of several distinct
 phases. Initially, the cold CSM is neutral and transparent to optical photons. As the
 interaction starts, the radiation from the shock drives an ionisation front through the CSM
 and makes it optically thick at all wavelengths.
 This first phase takes about a week after the onset of the interaction, which corresponds roughly to the
 light crossing time through the shell.
  At the time we record the first emerging photons from the SN IIn, the CSM is therefore already optically
  thick. In our reference model, the photosphere is then located at $\sim$\,7$\times$10$^{15}$\,cm
  and will remain at that radius for months.
  The second phase corresponds to the shock crossing of the optically thick CSM. In our reference model,
  this phases lasts $\gtrsim$\,200\,d. Over this duration, the shock is located below the photosphere,
  at an electron-scattering optical depth of about 15 initially. Radiation from the shock is thus continuously
  released deep in the CSM and comes out on a diffusion time scale.
  This context is analogous to optically thick ejecta influenced by radioactive decay, although the
  shock energy is not all channeled into radiation energy.
  Here, the shock luminosity decreases with time because of the decreasing shock strength
  as more and more CSM is being swept up. Although the CSM hardly expands, the diffusion time scale
  and the optical depth decrease because of the expansion of the interaction region.
  We thus obtain a bell shape morphology for the light curve around maximum.
  This variation is not associated
  by the migration of the photospheric radius, but by the variation in shock luminosity, radiation temperature,
  and CSM/photospheric temperature.
  Optical depth effects on the light curve persist for as long as the shock is located below
  the $\sim$\,7$\times$10$^{15}$\,cm.
  The third phase starts when the shock overtakes this location. The photosphere then follows the dense shell.

  Our multi-group radiation hydrodynamics simulation produces the basic light curve
  morphology of type IIn super-luminous SNe.
  Our reference model was tuned to roughly match the inferred properties
  of SN\,2010jl \citep{fransson_etal_14}, but other simulations with different inner-shell kinetic energy and CSM mass
  encompass a variety of light curves, including extreme events like SN\,2006gy \citep{smith_etal_07a}.

We illustrate the importance of performing {\it multi-group} radiation hydrodynamics. By contrast, {\it grey}
(i.e., one-group) radiation hydrodynamics simulations  do not account adequately
for the opacity of the material and produce a discrepant SN IIn light curve morphology.

Our simulations confirm the earlier work of  \citet{moriya_etal_13a} that the shell shocked model of
\citet{smith_mccray_07} is not physically sound.
The reason is that in interacting SNe, even a massive CSM cannot have a continuum optical depth greater than a few tens.
Because such CSM are by nature extended, the shock-crossing time far exceeds the time it takes the shock luminosity
to diffuse through the optically thick layers. The situation is analogous to shock emergence in an exploding star,
except that here the
configuration is quasi steady state --- it persists for as long as there is optically-thick CSM material ahead
of the shock. Our results, which are in agreement with the independent simulations of
\citet{moriya_etal_13a}, provide a simple and physically consistent explanation for the light curves of type IIn super-luminous SNe.
Overall, the shell-shocked model really belongs to the domain of (successful) core-collapse SNe in which
a radiation-dominated shock crosses a stellar envelope whose optical depth is huge everywhere apart from the surface.

  The non-LTE radiative transfer simulations, based on the radiation hydrodynamics simulations, produce
spectra that are initially blue and redden progressively as the CSM/photosphere cools down after bolometric maximum.
We obtain symmetric emission profiles, broadened by non-coherent scattering with free electrons (of thermal origin),
with wings that extend up to 3000\,\kms\ from line centre, and
together with a narrower P-Cygni profile ($<$\,100\,\kms) forming in the outer cold CSM.
Varying the ejecta kinetic energy or the CSM mass alters the shock luminosity, the mean radiation energy
in the optically thick CSM, and causes clear variations in colour and ionisation.
Varying the CSM density by a factor of 6  changes the strength of recombination lines but does not produce broader
lines at bolometric maximum, despite the enhanced electron-scattering optical depth.
Photons emitted at an optical depth $\tau$ perform $\tau^2$ scatterings before escaping.
If the albedo is 99\%, photons coming from $\tau>$\,10 are likely absorbed within the ionised CSM
and photons do not obtain arbitrarily large frequency shifts from line centre.
Broader electron scattering wings may arise for a greater albedo.

Our simulations indicate that photon absorption is probably efficient only in the Lyman continuum. At low
  photon frequencies, absorptive processes are weak and therefore photon absorption occurs only if the total
  optical depth at the site of emission is large (say above 10) --- in our reference simulation
  with a mass loss rate of 0.1\,\msunyr\
  for the CSM, the maximum CSM continuum $\tau$ is only 15.
   Early on, the optical depth in the CDS is high  enough to ensure complete thermalisation.
   Then, line broadening is probably dominated by non-coherent scattering with thermal electrons.
   However, at later times, thermalisation does not occur at all depths in the CSM.
  In the reference model, this is associated with an increased broadening and blueward asymmetry
  of line profiles with time.
  These features are also present in the observations of SN\,2010jl (e.g., H$\alpha$; Fig.~\ref{fig_spec_halpha}) ---
  this suggests a growing contribution from Doppler broadening, connected with photon emission
  from the fast expanding CDS.
  We believe this hybrid line formation is at the origin of the very large H$\alpha$ line width in SN\,2010jl, as well
  as the blueshift of its peak emission.
  Interestingly, the modest peak luminosities and the absence of broad lines at all times in events like
  SN\,1994W \citep{chugai_etal_04,dessart_etal_09} or SN\,2011ht \citep{humphreys_etal_12} are perplexing
  because the emission from the CDS should be seen once the CSM becomes transparent.

  To complement our radiation hydrodynamics simulations with \heracles\ and our non-LTE radiative transfer simulations with \cmfgen,
  which assume spherical symmetry, we present some polarisation simulations for aspherical but axially symmetric configurations.
    Our results for a prolate morphology with a pole-to-equator density ratio of $\sim$\,3 yield a maximum
  $\lesssim$\,2\% continuum polarisation compatible with the value observed for SN\,2010jl \citep{patat_etal_11}.
  In this context, the asymmetry may stem only from the CSM, although both inner and outer shells may be asymmetric.
  In future work, we will extend the \heracles\ simulations to 2-D to study large scale asymmetries and to 3-D to study
  the stability of the dense shell.

\section*{Acknowledgments}

We thank Ralph Sutherland and Claes Fransson for discussion and comments,
and Nando Patat for providing the spectropolarimetric data for SN\,2010jl published in \citet{patat_etal_11}.
Dessart acknowledges financial support from the European Community through an
International Re-integration Grant, under grant number PIRG04-GA-2008-239184,
and from ``Agence Nationale de la Recherche" grant ANR-2011-Blanc-SIMI-BS56-0007.
Hillier acknowledges support from STScI theory grant HST-AR-12640.01, and NASA theory
grant NNX10AC80G.
This material is based upon work partially supported by the National Science Foundation
under grant number AST-1311993.
Hillier would also like to acknowledge the hospitality and support of the Distinguished Visitor program
at the Research School of Astronomy and Astrophysics (RSAA) at the Australian National University (ANU).
This work was also supported in part by the National Science Foundation under
Grant No. PHYS-1066293 and benefited from the hospitality of the Aspen Center
for Physics.
This work was granted access to the HPC resources of CINES under the
allocations c2013046608 and c2014046608 made by GENCI (Grand Equipement National
de Calcul Intensif).
This work also utilised computing resources of the M\'esocentre SIGAMM,
hosted by the Observatoire de la C\^ote d'Azur, Nice, France.

\label{lastpage}


\begin{thebibliography}{60}
\expandafter\ifx\csname natexlab\endcsname\relax\def\natexlab#1{#1}\fi

\bibitem[{{Anderson} {et~al.}(2014){Anderson}, {Dessart}, {Gutierrez}, {Hamuy},
  {Morrell}, {Phillips}, {Folatelli}, {Stritzinger}, {Freedman},
  {Gonz{\'a}lez-Gait{\'a}n}, {McCarthy}, {Suntzeff}, \&
  {Thomas-Osip}}]{anderson_etal_14}
{Anderson}, J.~P., {Dessart}, L., {Gutierrez}, C.~P., {Hamuy}, M., {Morrell},
  N.~I., {Phillips}, M., {Folatelli}, G., {Stritzinger}, M.~D., {Freedman},
  W.~L., {Gonz{\'a}lez-Gait{\'a}n}, S., {McCarthy}, P., {Suntzeff}, N., \&
  {Thomas-Osip}, J. 2014, \mnras, 441, 671

\bibitem[{{Arnett}(1982)}]{arnett_82}
{Arnett}, W.~D. 1982, \apj, 253, 785

\bibitem[{{Barkat} {et~al.}(1967){Barkat}, {Rakavy}, \&
  {Sack}}]{barkat_etal_67}
{Barkat}, Z., {Rakavy}, G., \& {Sack}, N. 1967, Physical Review Letters, 18,
  379

\bibitem[{{Blondin} {et~al.}(1996){Blondin}, {Lundqvist}, \&
  {Chevalier}}]{blondin_etal_96}
{Blondin}, J.~M., {Lundqvist}, P., \& {Chevalier}, R.~A. 1996, \apj, 472, 257

\bibitem[{{Brown} \& {McLean}(1977)}]{brown_mclean_77}
{Brown}, J.~C. \& {McLean}, I.~S. 1977, \aap, 57, 141

\bibitem[{{Busche} \& {Hillier}(2005)}]{BH05_2D}
{Busche}, J.~R. \& {Hillier}, D.~J. 2005, \aj, 129, 454

\bibitem[{{Castor}(1970)}]{castor_70}
{Castor}, J.~I. 1970, \mnras, 149, 111

\bibitem[{{Castor} {et~al.}(1975){Castor}, {Abbott}, \& {Klein}}]{cak}
{Castor}, J.~I., {Abbott}, D.~C., \& {Klein}, R.~I. 1975, \apj, 195, 157

\bibitem[{{Chen} {et~al.}(2014){Chen}, {Woosley}, {Heger}, {Almgren}, \&
  {Whalen}}]{chen_etal_14}
{Chen}, K.-J., {Woosley}, S., {Heger}, A., {Almgren}, A., \& {Whalen}, D.~J.
  2014, \apj, 792, 28

\bibitem[{{Chevalier} \& {Irwin}(2011)}]{chevalier_irwin_11}
{Chevalier}, R.~A. \& {Irwin}, C.~M. 2011, \apjl, 729, L6

\bibitem[{{Chugai}(2001)}]{C01_SN1998S}
{Chugai}, N.~N. 2001, \mnras, 326, 1448

\bibitem[{{Chugai} {et~al.}(2004){Chugai}, {Blinnikov}, {Cumming}, {Lundqvist},
  {Bragaglia}, {Filippenko}, {Leonard}, {Matheson}, \&
  {Sollerman}}]{chugai_etal_04}
{Chugai}, N.~N., {Blinnikov}, S.~I., {Cumming}, R.~J., {Lundqvist}, P.,
  {Bragaglia}, A., {Filippenko}, A.~V., {Leonard}, D.~C., {Matheson}, T., \&
  {Sollerman}, J. 2004, \mnras, 352, 1213

\bibitem[{{Chugai} \& {Danziger}(1994)}]{chugai_danziger_94}
{Chugai}, N.~N. \& {Danziger}, I.~J. 1994, \mnras, 268, 173

\bibitem[{{Dessart} \& {Hillier}(2005{\natexlab{a}})}]{DH05b}
{Dessart}, L. \& {Hillier}, D.~J. 2005{\natexlab{a}}, \aap, 439, 671

\bibitem[{{Dessart} \& {Hillier}(2005{\natexlab{b}})}]{DH05a}
---. 2005{\natexlab{b}}, \aap, 437, 667

\bibitem[{{Dessart} \& {Hillier}(2011)}]{DH11b}
---. 2011, \mnras, 415, 3497

\bibitem[{{Dessart} {et~al.}(2009){Dessart}, {Hillier}, {Gezari}, {Basa}, \&
  {Matheson}}]{dessart_etal_09}
{Dessart}, L., {Hillier}, D.~J., {Gezari}, S., {Basa}, S., \& {Matheson}, T.
  2009, \mnras, 394, 21

\bibitem[{{Dessart} {et~al.}(2013){Dessart}, {Hillier}, {Waldman}, \&
  {Livne}}]{dessart_etal_13b}
{Dessart}, L., {Hillier}, D.~J., {Waldman}, R., \& {Livne}, E. 2013, \mnras,
  433, 1745

\bibitem[{{Dessart} {et~al.}(2010{\natexlab{a}}){Dessart}, {Livne}, \&
  {Waldman}}]{DLW10b}
{Dessart}, L., {Livne}, E., \& {Waldman}, R. 2010{\natexlab{a}}, \mnras, 408,
  827

\bibitem[{{Dessart} {et~al.}(2010{\natexlab{b}}){Dessart}, {Livne}, \&
  {Waldman}}]{DLW10a}
---. 2010{\natexlab{b}}, \mnras, 405, 2113

\bibitem[{{Dopita} {et~al.}(1984){Dopita}, {Cohen}, {Schwartz}, \&
  {Evans}}]{dopita_etal_84}
{Dopita}, M.~A., {Cohen}, M., {Schwartz}, R.~D., \& {Evans}, R. 1984, \apjl,
  287, L69

\bibitem[{{Dubroca} \& {Feugeas}(1999)}]{m1_model}
{Dubroca}, B. \& {Feugeas}, J. 1999, CRAS, 329, 915

\bibitem[{{Fransson} {et~al.}(2014){Fransson}, {Ergon}, {Challis}, {Chevalier},
  {France}, {Kirshner}, {Marion}, {Milisavljevic}, {Smith}, {Bufano},
  {Friedman}, {Kangas}, {Larsson}, {Mattila}, {Benetti}, {Chornock}, {Czekala},
  {Soderberg}, \& {Sollerman}}]{fransson_etal_14}
{Fransson}, C., {Ergon}, M., {Challis}, P.~J., {Chevalier}, R.~A., {France},
  K., {Kirshner}, R.~P., {Marion}, G.~H., {Milisavljevic}, D., {Smith}, N.,
  {Bufano}, F., {Friedman}, A.~S., {Kangas}, T., {Larsson}, J., {Mattila}, S.,
  {Benetti}, S., {Chornock}, R., {Czekala}, I., {Soderberg}, A., \&
  {Sollerman}, J. 2014, \apj, 797, 118

\bibitem[{{Gonz{\'a}lez} {et~al.}(2007){Gonz{\'a}lez}, {Audit}, \&
  {Huynh}}]{gonzalez_etal_07}
{Gonz{\'a}lez}, M., {Audit}, E., \& {Huynh}, P. 2007, \aap, 464, 429

\bibitem[{{Grasberg} \& {Nadezhin}(1986)}]{grasberg_nadezhin_86}
{Grasberg}, E.~K. \& {Nadezhin}, D.~K. 1986, Soviet Astronomy Letters, 12, 68

\bibitem[{{Harries} {et~al.}(1998){Harries}, {Hillier}, \&
  {Howarth}}]{harries_etal_98}
{Harries}, T.~J., {Hillier}, D.~J., \& {Howarth}, I.~D. 1998, \mnras, 296, 1072

\bibitem[{{Hillier}(1991)}]{hillier_91}
{Hillier}, D.~J. 1991, \aap, 247, 455

\bibitem[{{Hillier} \& {Dessart}(2012)}]{HD12}
{Hillier}, D.~J. \& {Dessart}, L. 2012, \mnras, 424, 252

\bibitem[{{Hillier} \& {Miller}(1998)}]{HM98_lb}
{Hillier}, D.~J. \& {Miller}, D.~L. 1998, \apj, 496, 407

\bibitem[{{Hoffman} {et~al.}(2008){Hoffman}, {Leonard}, {Chornock},
  {Filippenko}, {Barth}, \& {Matheson}}]{hoffman_etal_08}
{Hoffman}, J.~L., {Leonard}, D.~C., {Chornock}, R., {Filippenko}, A.~V.,
  {Barth}, A.~J., \& {Matheson}, T. 2008, \apj, 688, 1186

\bibitem[{{Humphreys} {et~al.}(2012){Humphreys}, {Davidson}, {Jones}, {Pogge},
  {Grammer}, {Prieto}, \& {Pritchard}}]{humphreys_etal_12}
{Humphreys}, R.~M., {Davidson}, K., {Jones}, T.~J., {Pogge}, R.~W., {Grammer},
  S.~H., {Prieto}, J.~L., \& {Pritchard}, T.~A. 2012, \apj, 760, 93

\bibitem[{{Klein} \& {Chevalier}(1978)}]{klein_chevalier_78}
{Klein}, R.~I. \& {Chevalier}, R.~A. 1978, \apjl, 223, L109

\bibitem[{{Leonard} {et~al.}(2000){Leonard}, {Filippenko}, {Barth}, \&
  {Matheson}}]{leonard_etal_00}
{Leonard}, D.~C., {Filippenko}, A.~V., {Barth}, A.~J., \& {Matheson}, T. 2000,
  \apj, 536, 239

\bibitem[{{Margutti} {et~al.}(2014){Margutti}, {Milisavljevic}, {Soderberg},
  {Chornock}, {Zauderer}, {Murase}, {Guidorzi}, {Sanders}, {Kuin}, {Fransson},
  {Levesque}, {Chandra}, {Berger}, {Bianco}, {Brown}, {Challis},
  {Chatzopoulos}, {Cheung}, {Choi}, {Chomiuk}, {Chugai}, {Contreras}, {Drout},
  {Fesen}, {Foley}, {Fong}, {Friedman}, {Gall}, {Gehrels}, {Hjorth}, {Hsiao},
  {Kirshner}, {Im}, {Leloudas}, {Lunnan}, {Marion}, {Martin}, {Morrell},
  {Neugent}, {Omodei}, {Phillips}, {Rest}, {Silverman}, {Strader},
  {Stritzinger}, {Szalai}, {Utterback}, {Vinko}, {Wheeler}, {Arnett},
  {Campana}, {Chevalier}, {Ginsburg}, {Kamble}, {Roming}, {Pritchard}, \&
  {Stringfellow}}]{margutti_etal_14}
{Margutti}, R., {Milisavljevic}, D., {Soderberg}, A.~M., {Chornock}, R.,
  {Zauderer}, B.~A., {Murase}, K., {Guidorzi}, C., {Sanders}, N.~E., {Kuin},
  P., {Fransson}, C., {Levesque}, E.~M., {Chandra}, P., {Berger}, E., {Bianco},
  F.~B., {Brown}, P.~J., {Challis}, P., {Chatzopoulos}, E., {Cheung}, C.~C.,
  {Choi}, C., {Chomiuk}, L., {Chugai}, N., {Contreras}, C., {Drout}, M.~R.,
  {Fesen}, R., {Foley}, R.~J., {Fong}, W., {Friedman}, A.~S., {Gall}, C.,
  {Gehrels}, N., {Hjorth}, J., {Hsiao}, E., {Kirshner}, R., {Im}, M.,
  {Leloudas}, G., {Lunnan}, R., {Marion}, G.~H., {Martin}, J., {Morrell}, N.,
  {Neugent}, K.~F., {Omodei}, N., {Phillips}, M.~M., {Rest}, A., {Silverman},
  J.~M., {Strader}, J., {Stritzinger}, M.~D., {Szalai}, T., {Utterback}, N.~B.,
  {Vinko}, J., {Wheeler}, J.~C., {Arnett}, D., {Campana}, S., {Chevalier}, R.,
  {Ginsburg}, A., {Kamble}, A., {Roming}, P.~W.~A., {Pritchard}, T., \&
  {Stringfellow}, G. 2014, \apj, 780, 21

\bibitem[{{Mauerhan} {et~al.}(2013){Mauerhan}, {Smith}, {Filippenko},
  {Blanchard}, {Blanchard}, {Casper}, {Cenko}, {Clubb}, {Cohen}, {Fuller},
  {Li}, \& {Silverman}}]{mauerhan_etal_13}
{Mauerhan}, J.~C., {Smith}, N., {Filippenko}, A.~V., {Blanchard}, K.~B.,
  {Blanchard}, P.~K., {Casper}, C.~F.~E., {Cenko}, S.~B., {Clubb}, K.~I.,
  {Cohen}, D.~P., {Fuller}, K.~L., {Li}, G.~Z., \& {Silverman}, J.~M. 2013,
  \mnras, 430, 1801

\bibitem[{{Moriya} {et~al.}(2013{\natexlab{a}}){Moriya}, {Blinnikov},
  {Baklanov}, {Sorokina}, \& {Dolgov}}]{moriya_etal_13b}
{Moriya}, T.~J., {Blinnikov}, S.~I., {Baklanov}, P.~V., {Sorokina}, E.~I., \&
  {Dolgov}, A.~D. 2013{\natexlab{a}}, \mnras, 430, 1402

\bibitem[{{Moriya} {et~al.}(2013{\natexlab{b}}){Moriya}, {Blinnikov},
  {Tominaga}, {Yoshida}, {Tanaka}, {Maeda}, \& {Nomoto}}]{moriya_etal_13a}
{Moriya}, T.~J., {Blinnikov}, S.~I., {Tominaga}, N., {Yoshida}, N., {Tanaka},
  M., {Maeda}, K., \& {Nomoto}, K. 2013{\natexlab{b}}, \mnras, 428, 1020

\bibitem[{{Niemela} {et~al.}(1985){Niemela}, {Ruiz}, \&
  {Phillips}}]{niemela_etal_85}
{Niemela}, V.~S., {Ruiz}, M.~T., \& {Phillips}, M.~M. 1985, \apj, 289, 52

\bibitem[{{Ofek} {et~al.}(2014){Ofek}, {Zoglauer}, {Boggs}, {Barri{\'e}re},
  {Reynolds}, {Fryer}, {Harrison}, {Cenko}, {Kulkarni}, {Gal-Yam}, {Arcavi},
  {Bellm}, {Bloom}, {Christensen}, {Craig}, {Even}, {Filippenko},
  {Grefenstette}, {Hailey}, {Laher}, {Madsen}, {Nakar}, {Nugent}, {Stern},
  {Sullivan}, {Surace}, \& {Zhang}}]{ofek_etal_14}
{Ofek}, E.~O., {Zoglauer}, A., {Boggs}, S.~E., {Barri{\'e}re}, N.~M.,
  {Reynolds}, S.~P., {Fryer}, C.~L., {Harrison}, F.~A., {Cenko}, S.~B.,
  {Kulkarni}, S.~R., {Gal-Yam}, A., {Arcavi}, I., {Bellm}, E., {Bloom}, J.~S.,
  {Christensen}, F., {Craig}, W.~W., {Even}, W., {Filippenko}, A.~V.,
  {Grefenstette}, B., {Hailey}, C.~J., {Laher}, R., {Madsen}, K., {Nakar}, E.,
  {Nugent}, P.~E., {Stern}, D., {Sullivan}, M., {Surace}, J., \& {Zhang}, W.~W.
  2014, \apj, 781, 42

\bibitem[{{Owocki}(2015)}]{owocki_14}
{Owocki}, S.~P. 2015, in Astrophysics and Space Science Library, Vol. 412,
  Astrophysics and Space Science Library, ed. J.~S. {Vink}, 113

\bibitem[{{Owocki} {et~al.}(2004){Owocki}, {Gayley}, \&
  {Shaviv}}]{owocki_etal_04}
{Owocki}, S.~P., {Gayley}, K.~G., \& {Shaviv}, N.~J. 2004, \apj, 616, 525

\bibitem[{{Pastorello} {et~al.}(2013){Pastorello}, {Cappellaro}, {Inserra},
  {Smartt}, {Pignata}, {Benetti}, {Valenti}, {Fraser}, {Tak{\'a}ts}, {Benitez},
  {Botticella}, {Brimacombe}, {Bufano}, {Cellier-Holzem}, {Costado}, {Cupani},
  {Curtis}, {Elias-Rosa}, {Ergon}, {Fynbo}, {Hambsch}, {Hamuy}, {Harutyunyan},
  {Ivarson}, {Kankare}, {Martin}, {Kotak}, {LaCluyze}, {Maguire}, {Mattila},
  {Maza}, {McCrum}, {Miluzio}, {Norgaard-Nielsen}, {Nysewander}, {Ochner},
  {Pan}, {Pumo}, {Reichart}, {Tan}, {Taubenberger}, {Tomasella}, {Turatto}, \&
  {Wright}}]{pastorello_etal_13}
{Pastorello}, A., {Cappellaro}, E., {Inserra}, C., {Smartt}, S.~J., {Pignata},
  G., {Benetti}, S., {Valenti}, S., {Fraser}, M., {Tak{\'a}ts}, K., {Benitez},
  S., {Botticella}, M.~T., {Brimacombe}, J., {Bufano}, F., {Cellier-Holzem},
  F., {Costado}, M.~T., {Cupani}, G., {Curtis}, I., {Elias-Rosa}, N., {Ergon},
  M., {Fynbo}, J.~P.~U., {Hambsch}, F.-J., {Hamuy}, M., {Harutyunyan}, A.,
  {Ivarson}, K.~M., {Kankare}, E., {Martin}, J.~C., {Kotak}, R., {LaCluyze},
  A.~P., {Maguire}, K., {Mattila}, S., {Maza}, J., {McCrum}, M., {Miluzio}, M.,
  {Norgaard-Nielsen}, H.~U., {Nysewander}, M.~C., {Ochner}, P., {Pan}, Y.-C.,
  {Pumo}, M.~L., {Reichart}, D.~E., {Tan}, T.~G., {Taubenberger}, S.,
  {Tomasella}, L., {Turatto}, M., \& {Wright}, D. 2013, \apj, 767, 1

\bibitem[{{Patat} {et~al.}(2011){Patat}, {Taubenberger}, {Benetti},
  {Pastorello}, \& {Harutyunyan}}]{patat_etal_11}
{Patat}, F., {Taubenberger}, S., {Benetti}, S., {Pastorello}, A., \&
  {Harutyunyan}, A. 2011, \aap, 527, L6

\bibitem[{{Schlegel}(1990)}]{schlegel_90}
{Schlegel}, E.~M. 1990, \mnras, 244, 269

\bibitem[{{Schulte-Ladbeck} {et~al.}(1991){Schulte-Ladbeck}, {Nordsieck},
  {Taylor}, {Nook}, {Bjorkman}, {Magalhaes}, \&
  {Anderson}}]{schulte_ladbeck_etal_91}
{Schulte-Ladbeck}, R.~E., {Nordsieck}, K.~H., {Taylor}, M., {Nook}, M.~A.,
  {Bjorkman}, K.~S., {Magalhaes}, A.~M., \& {Anderson}, C.~M. 1991, \apj, 382,
  301

\bibitem[{{Smith} {et~al.}(2007){Smith}, {Li}, {Foley}, {Wheeler}, \& {et
  al.}}]{smith_etal_07a}
{Smith}, N., {Li}, W., {Foley}, R.~J., {Wheeler}, J.~C., \& {et al.} 2007,
  \apj, 666, 1116

\bibitem[{{Smith} \& {McCray}(2007)}]{smith_mccray_07}
{Smith}, N. \& {McCray}, R. 2007, \apjl, 671, L17

\bibitem[{{Smith} {et~al.}(2012){Smith}, {Silverman}, {Filippenko}, {Cooper},
  {Matheson}, {Bian}, {Weiner}, \& {Comerford}}]{smith_etal_12}
{Smith}, N., {Silverman}, J.~M., {Filippenko}, A.~V., {Cooper}, M.~C.,
  {Matheson}, T., {Bian}, F., {Weiner}, B.~J., \& {Comerford}, J.~M. 2012, \aj,
  143, 17

\bibitem[{{Sollerman} {et~al.}(1998){Sollerman}, {Cumming}, \&
  {Lundqvist}}]{sollerman_etal_98}
{Sollerman}, J., {Cumming}, R.~J., \& {Lundqvist}, P. 1998, \apj, 493, 933

\bibitem[{{Stathakis} \& {Sadler}(1991)}]{stathakis_sadler_91}
{Stathakis}, R.~A. \& {Sadler}, E.~M. 1991, \mnras, 250, 786

\bibitem[{{Stoll} {et~al.}(2011){Stoll}, {Prieto}, {Stanek}, {Pogge},
  {Szczygie{\l}}, {Pojma{\'n}ski}, {Antognini}, \& {Yan}}]{stoll_etal_11}
{Stoll}, R., {Prieto}, J.~L., {Stanek}, K.~Z., {Pogge}, R.~W., {Szczygie{\l}},
  D.~M., {Pojma{\'n}ski}, G., {Antognini}, J., \& {Yan}, H. 2011, \apj, 730, 34

\bibitem[{{Turatto} {et~al.}(1993){Turatto}, {Cappellaro}, {Danziger},
  {Benetti}, {Gouiffes}, \& {della Valle}}]{turatto_etal_93}
{Turatto}, M., {Cappellaro}, E., {Danziger}, I.~J., {Benetti}, S., {Gouiffes},
  C., \& {della Valle}, M. 1993, \mnras, 262, 128

\bibitem[{{van Marle} {et~al.}(2010){van Marle}, {Smith}, {Owocki}, \& {van
  Veelen}}]{van_marle_etal_10}
{van Marle}, A.~J., {Smith}, N., {Owocki}, S.~P., \& {van Veelen}, B. 2010,
  \mnras, 407, 2305

\bibitem[{{Vaytet} {et~al.}(2011){Vaytet}, {Audit}, {Dubroca}, \&
  {Delahaye}}]{vaytet_etal_11}
{Vaytet}, N.~M.~H., {Audit}, E., {Dubroca}, B., \& {Delahaye}, F. 2011, \jqsrt,
  112, 1323

\bibitem[{{Wang} {et~al.}(2001){Wang}, {Howell}, {H{\"o}flich}, \&
  {Wheeler}}]{wang_etal_01}
{Wang}, L., {Howell}, D.~A., {H{\"o}flich}, P., \& {Wheeler}, J.~C. 2001, \apj,
  550, 1030

\bibitem[{{Whalen} {et~al.}(2013){Whalen}, {Even}, {Lovekin}, {Fryer},
  {Stiavelli}, {Roming}, {Cooke}, {Pritchard}, {Holz}, \&
  {Knight}}]{whalen_etal_13}
{Whalen}, D.~J., {Even}, W., {Lovekin}, C.~C., {Fryer}, C.~L., {Stiavelli}, M.,
  {Roming}, P.~W.~A., {Cooke}, J., {Pritchard}, T.~A., {Holz}, D.~E., \&
  {Knight}, C. 2013, \apj, 768, 195

\bibitem[{{Williams} {et~al.}(2014){Williams}, {Dessart}, {Hoffman}, {Huk},
  {Leonard}, {Milne}, {Smith}, \& {Smith}}]{williams_etal_14}
{Williams}, G.~G., {Dessart}, L., {Hoffman}, J.~L., {Huk}, L.~N., {Leonard},
  D.~C., {Milne}, P., {Smith}, N., \& {Smith}, P.~S. 2014, in American
  Astronomical Society Meeting Abstracts, Vol. 223, 354.22

\bibitem[{{Woosley} {et~al.}(2007){Woosley}, {Blinnikov}, \&
  {Heger}}]{woosley_etal_07}
{Woosley}, S.~E., {Blinnikov}, S., \& {Heger}, A. 2007, \nat, 450, 390

\bibitem[{{Zel'dovich} \& {Raizer}(1967)}]{ZR67}
{Zel'dovich}, Y.~B. \& {Raizer}, Y.~P. 1967, {Physics of shock waves and
  high-temperature hydrodynamic phenomena}

\bibitem[{{Zhang} {et~al.}(2012){Zhang}, {Wang}, {Wu}, {Chen}, {Chen}, {Liu},
  {Huang}, {Liang}, {Zhao}, {Lin}, {Wang}, {Dennefeld}, {Zhang}, {Zhai}, {Wu},
  {Fan}, {Zou}, {Zhou}, \& {Ma}}]{zhang_etal_12}
{Zhang}, T., {Wang}, X., {Wu}, C., {Chen}, J., {Chen}, J., {Liu}, Q., {Huang},
  F., {Liang}, J., {Zhao}, X., {Lin}, L., {Wang}, M., {Dennefeld}, M., {Zhang},
  J., {Zhai}, M., {Wu}, H., {Fan}, Z., {Zou}, H., {Zhou}, X., \& {Ma}, J. 2012,
  \aj, 144, 131

\end{thebibliography}
\end{document}